%% file: paper.tex
\documentclass[12pt]{article}

\usepackage{adjustbox}
\usepackage{amsmath}
\usepackage{amssymb}
\usepackage{amsthm}
\usepackage{arydshln}
\usepackage{authblk}
\usepackage{bbm}
\usepackage{bm}
\usepackage{caption}
\usepackage{color}
\usepackage{colortbl}
\usepackage{graphicx}
\usepackage{hyperref}
\usepackage{mdframed}
\usepackage{multirow}
\usepackage{natbib}
\usepackage{setspace}
\usepackage{textcomp}
\usepackage{xcolor}

\setcitestyle{round}
\hypersetup{
  colorlinks=true,
  linkcolor=blue,
  filecolor=blue,
  urlcolor=blue,
  citecolor=blue
}

\newcommand{\blind}{0}

\newcommand{\Gau}{\mathrm{Gau}}

\newcommand{\Sigmavec}{\boldsymbol{\Sigma}}

\newcommand{\Avec}{\mathbf{A}}
\newcommand{\Cvec}{\mathbf{C}}

\newcommand{\Ivec}{\mathbf{I}}
\newcommand{\Ovec}{\mathbf{O}}
\newcommand{\Qvec}{\mathbf{Q}}
\newcommand{\Rvec}{\mathbf{R}}

\newcommand{\Vvec}{\mathbf{V}}
\newcommand{\Wvec}{\mathbf{W}}
\newcommand{\Xvec}{\mathbf{X}}
\newcommand{\Yvec}{\mathbf{Y}}
\newcommand{\Zvec}{\mathbf{Z}}

\newcommand{\zerovec}{\bm{0}}
\newcommand{\alphavec}{\bm{\alpha}}
\newcommand{\betavec}{\bm{\beta}}
\newcommand{\deltavec}{\bm{\delta}}

\newcommand{\gammavec}{\bm{\gamma}}
\newcommand{\kappavec}{\bm{\kappa}}

\newcommand{\omegavec}{\bm{\omega}}
\newcommand{\zetavec}{\bm{\zeta}}

\newcommand{\fvec}{\mathbf{f}}
\newcommand{\gvec}{\mathbf{g}}
\newcommand{\hvec}{\mathbf{h}}
\newcommand{\svec}{\mathbf{s}}
\newcommand{\uvec}{\mathbf{u}}

\newcommand{\xvec}{\mathbf{x}}
\newcommand{\yvec}{\mathbf{y}}

\newcommand{\nospellcheck}[1]{#1}

\date{}

\begin{document}

\title{\bf GeoWarp: Warped spatial processes for inferring subsea sediment properties}
\author{
  Michael Bertolacci${}^{1,2}$\footnote{Contact: michael.bertolacci@uwa.edu.au},
  Andrew Zammit-Mangion${}^2$,
  Juan~Valderrama~Giraldo${}^3$,
  Michael O'Neill${}^3$,
  Fraser Bransby${}^3$, and
  Phil~Watson${}^3$

  \vspace{0.3cm}
  
  ${}^1$ Department of Mathematics and Statistics, The University of Western Australia \\
  ${}^2$ School of Mathematics and Applied Statistics, University of Wollongong \\ \vspace{0.5cm}
  ${}^3$ Oceans Graduate School, The University of Western Australia
}
\maketitle

\begin{abstract}
  For offshore structures like wind turbines, subsea infrastructure, pipelines, and cables, it is crucial to quantify the properties of the seabed sediments at a proposed site. However, data collection offshore is costly, so analysis of the seabed sediments must be made from measurements that are spatially sparse. Adding to this challenge, the structure of the seabed sediments exhibits both nonstationarity and anisotropy. To address these issues, we propose GeoWarp, a hierarchical spatial statistical modeling framework for inferring the 3-D geotechnical properties of subsea sediments. GeoWarp decomposes the seabed properties into a region-wide vertical mean profile (modeled using B-splines), and a nonstationary 3-D spatial Gaussian process. Process nonstationarity and anisotropy are accommodated by warping space in three dimensions and by allowing the process variance to change with depth. We apply \mbox{GeoWarp} to measurements of the seabed made using cone penetrometer tests (CPTs) at six sites on the North West Shelf of Australia. We show that GeoWarp captures the complex spatial distribution of the sediment properties, and produces realistic 3-D simulations suitable for downstream engineering analyses. Through cross-validation, we show that GeoWarp has predictive performance superior to other state-of-the-art methods, demonstrating its value as a tool in offshore geotechnical engineering.
\end{abstract}

\noindent%

\section{Introduction}
\label{sec:introduction}

The design of offshore energy projects, including wind turbines, platforms, anchors, subsea infrastructure, pipelines, and cables, requires estimates of key properties of the three-dimension (3-D) sediment structure below the seabed. Such estimates are usually based on a mix of data and subjective engineering judgment \citep{baecher2021}, and the subsequent design process must account for the challenges posed by both strong and weak sediments under different load conditions \citep{houlsby2016}. Overly conservative designs that handle a wide range of sediment conditions are costly, while designs that target an overly narrow range can lead to failure. Consequently, both the natural variability and the uncertainty associated with the seabed sediments can dictate the construction methodology, scheduling, cost, and even the overall viability of an offshore project.

Obtaining seabed data involves costly geotechnical surveys, sometimes reaching into the millions of dollars \citep{randolphetal2005}. These financial constraints restrict data collection to a modest number of locations within the project site. As a result, the 3-D sediment properties must be inferred from sparse data, a challenging task given the sediment variability. The  global shift toward renewable energy projects, which span larger areas, further complicates this process, both increasing costs and adding complexity to the inference tasks \citep{cooketal2022}.  There is therefore a growing need for methods that can effectively characterize seabed sediment natural variability and uncertainty from sparse data \citep{phoonetal2022}. Such methods would ensure that designs considered during the project planning phase are neither too risky nor too conservative.

\begin{figure}
  \begin{center}
    \includegraphics{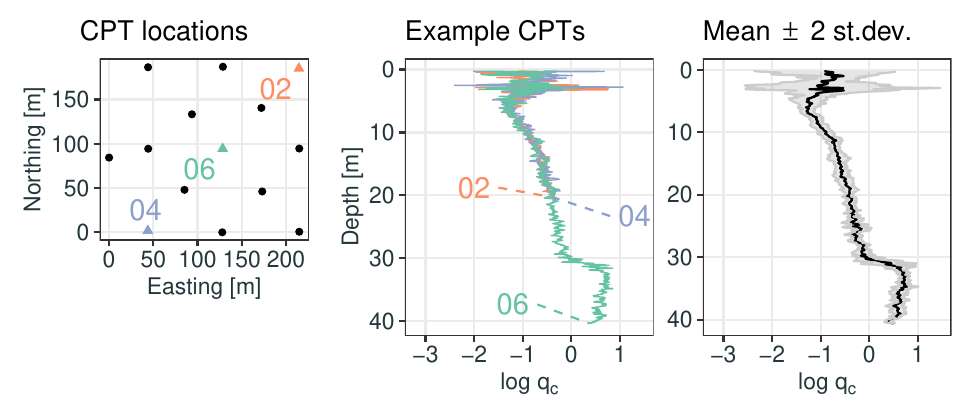}
  \end{center}
  \caption{
    The horizontal locations of 14 CPTs taken at the A2 site (left panel; see Section~\ref{sec:application} for the details of the site), the values of $\log q_c$ for three example CPTs (middle panel; the original unit before transformation is MPa), and the mean $\pm$ 2 st.dev.\@ of $\log q_c$ across all 14 CPTs within  0.1m depth bins (right panel). Labels and colors identify the example CPTs. The locations and values of all 14 CPTs are given in Figures~\ref{fig:map_a} and \ref{fig:datasets_a} of Section~\ref{sec:additional_figures} of the supplementary material, respectively.
  }
  \label{fig:summary_a2}
\end{figure}

Cone penetration tests (CPTs) are favored in offshore geotechnical surveys for their cost-effectiveness and efficiency \citep{lunneetal2011}. During a CPT, a conical-tipped cylindrical instrument is pushed into the seabed, returning near-continuous cone-tip resistance readings (denoted by $q_c$) in megapascals (MPa). These readings provide an indication of sediment strength \citep{lunneetal2011}. Figure~\ref{fig:summary_a2} displays the natural logarithm of $q_c$ against the depth below the seabed from 14 CPTs collected at a site we call A2 on the North West Shelf (NWS) of Australia (one of six sites we analyze in Section~\ref{sec:application}). Log values are adopted since log-normal distributions are recognized as a reasonable statistical model for sediment (and geological) properties \citep{lacasse1994,griffithsetal2009}. The figure's left panel show the CPT locations, the middle shows $\log q_c$ profiles from three example CPTs, and the right summarizes the mean and variability of $\log q_c$ within each 0.1 m depth bin. With vertical measurements every 1-2 cm and the closest pair of CPTs spaced 45 m apart, the data are simultaneously dense vertically and sparse horizontally.

Geotechnical site characterization focuses on sediment stratification and variability \citep{phoonetal2022}. At the A2 site, the upper 5 m exhibits strong but variable sediment; beyond 5 m, a more homogeneous and weaker layer emerges, with a stronger layer beyond 32 m. There is a general trend of increasing cone resistance with depth. As discussed above, the fluctuations of sediment strength inform the design ranges for seabed infrastructure. Characterization methods must therefore account for vertical nonstationarity \citep{phoonetal2003}, horizontal changes in sediment property and layer depth \citep[e.g.,][]{lietal2020}, and anisotropy, as horizontal dependencies are generally stronger than vertical ones \citep{zhuzang2013}. Computational efficiency also comes into play with large CPT datasets, ranging from 4,265 to 35,538 measurements in our study (see Section~\ref{sec:application}).

Historically, sediment characterization methods have leaned on subjective engineering judgment \citep{lacasse1988}, but the advent of larger geotechnical datasets has spurred a shift toward data-driven approaches in both research and practice \citep{phoonetal2022}. Many such methods use spatial random field theory \citep{fenton1999}, with assumptions either of process stationarity \citep[e.g.,][]{fenton1999,uziellietal2005}, or of piecewise stationarity, where the process is segmented in the vertical direction and each segment is modeled as an independent stationary process \citep[e.g.,][]{caowang2013,lietal2020}. More recent models characterize 3-D sediment properties using basis-function representations, which allow for ``mild nonstationarity'' and which help address computational challenges inherent in geotechnical data analysis.

The methods using basis functions differ in their choice of basis functions, the way in which surface smoothness is constrained when estimating the sediment properties, and the model-fitting procedure. \citet{zhaowang2020} and \citet{lyuetal2023a} propose using Bayesian compressive sensing \citep[BCS; ][]{jietal2008} with a set of cosine basis functions defined over a gridded discretization of the 3-D volume of the site. Computational efficiency is assured by limiting the number of basis functions, while model sparsity and smoothness are enforced through the prior. Markov chain Monte Carlo algorithms are used to infer the BCS model parameters. \citet{shukuphoon2021,shukuphoon2023} on the other hand consider both piecewise-constant basis vectors and Legendre polynomials in their geotechnical lasso (GLasso) method, with sparsity enforced through the statistical lasso \citep{tibshirani1996}, and with smoothness guided by a regularizer. The high-dimensional parameter space is managed using specialized optimization techniques.

While the existing methods for sediment characterization perform well in some circumstances, they have substantial limitations that restrict their adoption by the geotechnical engineering community. The BCS method tends to produce overly smooth predictions, a common limitation of low-rank methods \citep{stein2014}, and it does not accommodate complex patterns of nonstationarity. Predictions using BCS also appear to be overly sensitive to the choice of basis functions. Furthermore, while the random field and BCS methods quantify the uncertainty in 3-D sediment properties, GLasso provides only point predictions. Methods that do not decompose the process using basis functions assume stationarity or vertical segmentation, and offer a poor characterization of the spatial heterogeneity of the sediment's 3-D structure.

To address these shortcomings, we introduce GeoWarp, a new method for geotechnical site characterization that uses a 3-D extension and adaptation of the hierarchical warped spatial processes of \citet{zammitmangionetal2022} to model seabed sediment properties from CPT data. GeoWarp divides sediment properties into a region-wide vertical mean profile (using B-splines) and a deviation process (using a nonstationary 3-D spatial Gaussian process). Nonstationarity is addressed by three-dimensional spatial warping and depth-variable deviation process variance. Evaluation with CPT data from six sites on Australia's North West Shelf demonstrates GeoWarp's capability to capture site-specific properties and shows its improved performance over other methods, including the BCS method.

GeoWarp contributes to the current state-of-the-art in geotechnical site characterization. It is among the few methods that produce a model-based, continuously-indexed full prediction distribution over the 3-D site, and it has been validated with field data from multiple real sites. Compared to other methods, GeoWarp also better accommodates nonstationarity in the sediment structure, resulting in improved predictions and uncertainty quantification. GeoWarp is also a full-rank method (unlike the BCS and GLasso) that addresses computational challenges using the Vecchia approximation \citep{vecchia1988}. These attributes make GeoWarp suitable for risk management and cost optimization in geotechnical engineering practice and research.

Section~\ref{sec:geowarp_model} details the spatial hierarchical model that underpins GeoWarp. Section~\ref{sec:inference_computation} discusses parameter inference and prediction. Section~\ref{sec:application} applies GeoWarp to CPT data from six sites on the NWS of Australia, comparing its probabilistic predictions to other methods. Section~\ref{sec:conclusion} concludes the paper, highlighting the strengths and limitations of our method and suggesting future avenues for development.

\section{The GeoWarp hierarchical spatial model}
\label{sec:geowarp_model}

Consider a bounded spatial region of geotechnical interest $S \subset \mathbb{R}^D$, where $D = 1$ corresponds to a transect with a single horizontal coordinate, and $D = 2$ corresponds to an areal study region with two horizontal coordinates such as eastings and northings. Let $Y(\svec, h)$ for $\svec \in S$ and $h \in [0, h_\mathrm{max}]$ be an unknown field over the region $S$ that varies both in space, $\svec$, and depth, $h$, where $h_\mathrm{max} > 0$ is the maximum depth of interest. For example, $Y(\svec, h)$ could represent the logarithm of the sediment's resistance to a cone penetrometer at location $\svec$ and depth $h$.

Measurements are made at a set of horizontal locations, $\svec_1, \ldots, \svec_M \in S$, $M > 0$. At each location, the measurement yields a depth-referenced vector, which we denote by $\Zvec_i \equiv (Z_{i, 1}, \ldots, Z_{i, N_i})'$ for $i = 1, \ldots, M$ and $N_i > 0$. We model $Z_{i, j}$ for $i = 1, \ldots, M$ and $j = 1, \ldots, N_i$ as $Z_{i, j} = Y(\svec_i, h_{i, j}) + \epsilon_{i, j}$, where $h_{i, j}$ and $\epsilon_{i, j}$ are the depth and error of the measurement, respectively. We assume that $\epsilon_{i, j} \sim \Gau(0, \sigma_\epsilon^2)$, where $\sigma_\epsilon^2 > 0$ is unknown and has an inverse Gamma prior distribution, $\sigma_\epsilon^2 \sim \mathrm{IG}(a_\epsilon, b_\epsilon)$, with fixed shape and scale hyperparameters $a_\epsilon, b_\epsilon > 0$. For the application to CPT data in Section~\ref{sec:application}, we take $Z_{i, j}$ to be the logarithm of the measured cone-tip resistance ($\log q_c$); the additive error $\epsilon_{i, j}$ may then be interpreted as contributing to multiplicative error on the original scale.

GeoWarp models the unknown field $Y(\svec, h)$ as the sum of a region-wide vertical mean-profile process, $\mu(h)$, and a deviation process, $\delta(\svec, h)$:
\begin{equation}
  Y(\svec, h) = \mu(h) + \delta(\svec, h),
  \label{eqn:y_process}
\end{equation}
where $\svec \in S$ and $h \in [0, h_\mathrm{max}]$. This model decomposes the unknown process $Y(\svec, h)$ into a component $\mu(h)$ that is shared among all horizontal locations, and a component $\delta(\svec, h)$ that describes the deviation of the process from $\mu(h)$ at location $\svec$ and depth $h$. The vertical-only mean profile captures changes to the mean strength of the sediment with depth; these changes are evident in the $\log q_c$ data shown in Figure~\ref{fig:summary_a2} for the site A2, as well as for the other sites shown in Figures~\ref{fig:datasets_a} and \ref{fig:datasets_b} in Section~\ref{sec:additional_figures} of the supplementary material.

The decomposition in \eqref{eqn:y_process} has precedents in the literature. \citet{yoshidaetal2021} also decomposed the sediment properties into what they termed trend and random processes. Similarly, spatial functional data analysis \citep[spatial FDA; see][for a review]{martinezhernandezgenton2020} methods, which deal with curves that vary over space, often have a mean curve component \citep[e.g.,][]{gromenkoetal2012}. The main difference between \mbox{GeoWarp} and spatial FDA is the modeling of the small-scale variation process, $\delta(\svec, h)$. Most spatial FDA methods model this term through a truncated basis-function representation, while GeoWarp uses a full-rank, parametric, and highly flexible covariance model (see Section~\ref{sec:deviation_process}), thus circumventing known limitations of fixed-rank methods \citep{stein2014}.

\subsection{Mean-profile process}
\label{sec:mean_profile}

We model the vertical mean profile process $\mu(h)$ as the sum of an intercept, a depth-linear trend, and a basis-function representation:
\begin{equation}
  \mu(h) = \alpha_0 + \alpha_1 h + \sum_{k = 1}^{K_\beta} \phi_k^\mu(h) \beta_k,
  \quad
  h \in [0, h_\mathrm{max}],
  \label{eqn:mu_process}
\end{equation}
where $K_\beta \geq 0$ is the number of basis functions, $\phi_1^\mu(h), \ldots, \phi_{K_\beta}^\mu(h)$ are basis functions over $h \in [0, h_\mathrm{max}]$, and $\alpha_0, \alpha_1, \beta_1, \ldots, \beta_{K_\beta}$ are unknown coefficients. The basis functions in \eqref{eqn:mu_process} are B-splines of order 4 defined by knots spaced at intervals $\Delta_\mu > 0$, where for simplicity we choose $\Delta_\mu$ such that $h_\mathrm{max} / \Delta_\mu$ is an integer \citep{deboor1977}. We place three knots before and after the target depths to reduce boundary effects, so the knots span from $-3 \Delta_\mu$ to $h_\mathrm{max} + 3 \Delta_\mu$, and therefore $K_\beta = h_\mathrm{max} / \Delta_\mu + 7$.

The model in \eqref{eqn:mu_process} includes a depth-linear trend because the measured sediment strength tends to increase with depth due to the overlying sediment layers exerting more pressure on the underlying sediment layers. The basis-function representation captures deviations from this trend, due to the layering of the sediment, that can either increase or decrease the sediment strength. Both of these features can be seen in Figure~\ref{fig:summary_a2} for the A2 site: the depth-average of $\log q_c$ tends to increase smoothly with depth, except near the surface (where it is stronger), and around 30 m depth (where it increases rapidly).

To the coefficients $\alphavec \equiv (\alpha_0, \alpha_1)'$ and $\betavec \equiv (\beta_1, \ldots, \beta_{K_\beta})'$ we assign independent multivariate Gaussian priors $\alphavec \sim \mathrm{Gau}(\zerovec_2, \sigma_\alpha^2 \Ivec_2)$ and $\betavec \sim \mathrm{Gau}(\zerovec_{K_\beta}, \Sigmavec_\beta)$, where $\zerovec_P$ is a vector of zeros of length $P > 0$, $\Ivec_P$ is the $P \times P$ identity matrix, $\sigma_\alpha^2 > 0$ is chosen so that the prior over $\alphavec$ is largely uninformative, and $\Sigmavec_\beta$ is a $K_\beta \times K_\beta$ covariance matrix, potentially also unknown. The matrix $\Sigmavec_\beta$ is structured to induce smoothing in the basis-function coefficients, which in turn induces smoothing in the estimate of $\mu(h)$; we discuss two possible choices for $\Sigmavec_\beta$ in Section~\ref{sec:spline_coefficient_prior} of the supplementary material.

The representation in \eqref{eqn:mu_process} can be considered as a fixed rank Gaussian process \citep[see, e.g.,][]{stein2008,cressiejohannesson2008}. As an alternative, \citet{yoshidaetal2021} used a full-rank Gaussian process for the vertical trend in their model. Both models are quite flexible in a 1-D context, but the fixed-rank model in \eqref{eqn:mu_process} is computationally convenient because its parameters can be marginalized out during inference (see Section~\ref{sec:inference_computation}).

\subsection{Deviation process}
\label{sec:deviation_process}

We model the deviation process, $\delta(\cdot, \cdot)$, as a Gaussian stochastic process with expectation zero and nonstationary covariance function given by
\begin{equation}
  \mathrm{Cov}(\delta(\svec_1, h_1), \delta(\svec_2, h_2))
  = \sqrt{\sigma_\delta^2(h_1) \sigma_\delta^2(h_2)}
    \mathcal{M}_\nu(d),
    \enskip \svec_1, \svec_2 \in S,
    \enskip h_1, h_2 \in [0, h_\mathrm{max}],
  \label{eqn:deviation_covariance}
\end{equation}
where $\sigma_\delta^2(h) > 0$ for $h \in [0, h_\mathrm{max}]$, $d \equiv ||\gvec(\svec_1, h_1) - \gvec(\svec_2, h_2)||_2$ is the Euclidean distance between $(\svec_1, h_1)$ and $(\svec_2, h_2)$ after warping by the function $\gvec(\cdot, \cdot)$, and $\mathcal{M}_\nu(d)$ is the \nospellcheck{Mat\'{e}rn} correlation function. We model the logarithm of the depth-varying variance $\sigma_\delta^2(h)$ using B-splines; full details are in Section~\ref{sec:depth_varying_variance}. The warping function $\gvec(\cdot, \cdot)$ is an injective mapping from $\mathbb{R}^{D + 1}$ to $\mathbb{R}^{D + 1}$, described further in Section~\ref{sec:warping_function}.

The \nospellcheck{Mat\'{e}rn} correlation function is equal to $\mathcal{M}_\nu(d) \equiv \frac{2^{1 - \nu}}{\Gamma(\nu)} \left( \sqrt{2 \nu} d \right)^\nu \mathcal{K}_\nu(\sqrt{2 \nu} d)$ for $d > 0$, $\nu > 0$, where $\Gamma(\cdot)$ is the gamma function and $\mathcal{K}_\nu(\cdot)$ is the modified Bessel function of the second kind \citep[see, e.g.,][]{guttorpgneiting2006}. The smoothness of the process increases with the parameter $\nu$; the case $\nu = 1 / 2$ corresponds to the exponential correlation function, while the case $\nu \rightarrow \infty$ corresponds to the squared exponential (or Gaussian) correlation function. Both \citet{chingetal2019} and \citet{yoshidaetal2021} studied this correlation function in the geotechnical context and found that the flexibility to control the smoothness was useful when modeling sediment properties. In Section~\ref{sec:application}, we find that $\nu = 3 / 2$ is a reasonable choice for the data we analyze.

The covariance model in \eqref{eqn:deviation_covariance} is chosen to reflect the features of the seabed. As seen in Figure~\ref{fig:summary_a2}, the variance of the sediment strength can change with depth \citep[this is also the case with land soil, and the model of][for clay content in land soils, also includes depth-varying variances]{ortonetal2016}. Nonstationarity of variances in the vertical dimension is common in other 3-D applications too; for example, both \citet{nguyenetal2017} and \citet{salvanayun2022} found this feature necessary when modeling atmospheric and 3-D oceanographic data, respectively. The warping function in \eqref{eqn:deviation_covariance} allows the scales of dependence to change with depth, and is important to characterize several of the sites we study in Section~\ref{sec:application}. The warping function also accounts for anisotropy, which is an expected feature of seabed sediments \citep{watsonetal2019}. 

Similar to \eqref{eqn:deviation_covariance}, \citet{berildfuglstad2023} proposed a nonstationary 3-D model, which they built using stochastic partial differential equations \citep[SPDEs;][]{lindgrenetal2011}. Their approach is very flexible but grapples with complex parameterization of nonstationarity. Instead, GeoWarp simplifies by focusing primarily on depth-dependent nonstationarity, which is appropriate for seabed sediments.

\subsubsection{Depth-varying variance}
\label{sec:depth_varying_variance}

To ensure that the depth-varying variance $\sigma_\delta^2(h)$ is positive, we assign a basis-function model to its logarithm. The basis-function model uses the same B-spline basis functions of order 4 as the mean-profile process in \eqref{eqn:mu_process}, though with a different number of knots and with the depth-linear term omitted. We model it as
\begin{equation}
  \log \sigma_\delta^2(h) = \eta + \sum_{k = 1}^{K_\zeta} \phi_k^\sigma(h) \zeta_k,
  \label{eqn:sigma_process}
\end{equation}
where $K_\zeta \geq 0$ is the number of basis functions, $\phi_1^\sigma(h), \ldots, \phi_{K_\zeta}^\sigma(h)$ are basis functions over $h \in [0, h_\mathrm{max}]$, and $\eta, \zeta_1, \ldots, \zeta_{K_\zeta}$ are unknown coefficients. As in Section~\ref{sec:mean_profile}, we use B-splines of order 4 as basis functions, and we place the knots at intervals $\Delta_\sigma > 0$, with three knots before and after the target depths. To $\eta$ and $\zetavec \equiv (\zeta_1, \ldots, \zeta_{K_\zeta})'$ we assign independent Gaussian priors $\eta \sim \mathrm{Gau}(0, \sigma_\eta^2)$ and $\zetavec \sim \mathrm{Gau}(\zerovec_{K_\zeta}, \Sigmavec_\zeta)$, where $\sigma_\eta^2 > 0$ is set so that the prior over $\eta$ is uninformative, and $\Sigmavec_\zeta$ is an unknown $K_\zeta \times K_\zeta$ covariance matrix structured to induce smoothness in $\log \sigma_\delta^2(h)$. We discuss two choices for $\Sigmavec_\zeta$ in Section~\ref{sec:spline_coefficient_prior} of the supplementary material; in our experiments we use an exponential correlation function with unknown length scale $\ell_\zeta > 0$ and unknown variance $\sigma_\zeta^2 > 0$.

\subsubsection{Warping function}
\label{sec:warping_function}

GeoWarp follows \citet{zammitmangionetal2022} in using a composition of simple injective warpings called warping units to model $\gvec(\cdot, \cdot)$. To simplify the notation, let $\uvec \equiv (\svec', h)'$ for $\svec \in S$ and $h \in [0, h_\mathrm{max}]$. We define
\begin{equation}
  \gvec(\uvec) = (\gvec_G \circ \cdots \circ \gvec_1)(\uvec),
  \label{eqn:warping_composition}
\end{equation}
where $\gvec_l : \mathbb{R}^{D + 1} \to \mathbb{R}^{D + 1}$ for $l = 1, \ldots, G$ are the injective warping units, each of which is unknown and governed by its own parameters. Since each warping unit in the composition is injective, the overall warping $\gvec(\cdot)$ is injective too. \citet{zammitmangionetal2022} found that the compositional structure in \eqref{eqn:warping_composition} can model complex warpings, and considered several types of warping units. In this work, we consider two types of warping: axial warping, and geometric warping, applied in that order.

\noindent\emph{Axial warping units:} An axial warping unit (AWU) rescales a single input coordinate using a monotonically increasing (and hence injective) mapping. The AWU for coordinate $d \in \{ 1, \ldots, D + 1 \}$ is given by $\fvec_d(\uvec) \equiv (u_1, \ldots, u_d^*(u_d), \ldots, u_{D + 1})'$, where $u_i$ is the $i$th element of $\uvec$, and $u_d^*(u_d)$ is the warped coordinate. Since the AWUs only affect a single coordinate at a time, they can be applied in any order without changing the overall warping. We use a Bernstein polynomial basis expansion of order $L_d > 0$ to construct $u_d^*(u_d)$. This basis can reproduce a linear mapping exactly (representative of no warping), easily accommodates constraints such as monotonicity, and is flexible enough to reproduce complex warpings \citep{mckaycurtisghosh2011}. The expansion is given by
\begin{equation}
  u_d^*(u_d) \equiv \sum_{l = 1}^{L_d} \psi_{l, L_d}(u_d^\dagger) \lambda_{d, l},
  \quad d = 1, \ldots, D + 1,
  \label{eqn:bernstein_warping}
\end{equation}
where $L_d > 0$; $\lambda_{d, 1}, \ldots, \lambda_{d, L_d}$ are unknown coefficients; $\psi_{l, L_d}(x) \equiv \binom{L_d}{l} x^l (1 - x)^{L_d - l}$, $l = 1, \ldots, L_d$, is the $l$th Bernstein basis polynomial of degree $L_d$; $u_d^\dagger = (u_d - u_d^-) / (u_d^+ - u_d^-)$; and $u_d^-$ and $u_d^+$ are the infimum and supremum of coordinate $d$, respectively (thus, $u_d^\dagger \in [0, 1])$. In Section~\ref{sec:application} we use $L_d = 2$ for the horizontal dimensions $d = 1, \ldots, D$, corresponding to a linear warping, and $L_{D + 1} = 20$ for the vertical coordinate. This choice is made because the sediment properties are more variable vertically than horizontally, as seen in Figure~\ref{fig:summary_a2}.

The representation in \eqref{eqn:bernstein_warping} is slightly unconventional in that it omits the Bernstein basis function corresponding to $l = 0$. This is a necessary and sufficient condition to ensure that $u_d^*(0) = 0$, which in our setting serves as a reference point to make the warpings comparable between fits to different datasets. To enforce a monotonically increasing mapping, we follow \citet{mckaycurtisghosh2011}, who showed that a necessary and sufficient condition for \eqref{eqn:bernstein_warping} to increase monotonically is that $\lambda_{d, 1} < \ldots < \lambda_{d, L_d}$. To ensure this, we reparameterize the $\lambda$s as $\lambda_{d, l} = \sum_{l' = 1}^l \gamma_{d, l'}$ for $l = 1, \ldots, L_d$, where $\gamma_{d, 1}, \ldots, \gamma_{d, L_d}$ are positive and unknown increments. In Section~\ref{sec:application}, we use different choices for the prior distribution on the increments depending on the dimension $d$. For the horizontal dimensions $d = 1, \ldots, D$, for which $L_d = 2$, the only increment $\gamma_{d, 1}$ corresponds to the inverse of a standard length scale parameter. Therefore to $\gamma_{d, 1}$ we assign an inverse-uniform prior such that $\gamma_{d, l}^{-1}$ is uniformly distributed over the interval $[(\gamma_d^+)^{-1}, (\gamma_d^-)^{-1}]$, where $\gamma_d^+ > \gamma_d^- > 0$ are fixed to physically-motivated upper and lower bounds on the horizontal length scale. For the vertical dimension, we assign independent gamma prior distributions $\gamma_{D + 1, l} \sim \mathrm{Ga}(a_{D + 1}^\gamma, b_{D + 1}^\gamma)$, $l = 1, \ldots, L_{D + 1}$, where $a_{D + 1}^\gamma > 0$ and $b_{D + 1}^\gamma > 0$ are fixed such that the prior distribution on $\gamma_{D + 1, l}$ is uninformative. The uninformative prior for $\gamma_{D + 1, l}$ without upper and lower bounds is motivated by fact that the vertical length scales are well-identified by the vertically-dense data.

The Bernstein basis expansion AWU in \eqref{eqn:bernstein_warping} has two advantages over the AWU introduced by \citet{zammitmangionetal2022}, which models the warping as a sum of sigmoidal basis functions. First, our AWU admits the linear warping as a special case, achieved when $\gamma_{d, i}$, $i = 1, \ldots, L_d$, are all equal to each other. Second, the modeler need only choose the number of basis functions $L_d$, whereas in the AWUs of \citet{zammitmangionetal2022} there is an additional tuning parameter which determines the shape of the basis functions.

\noindent\emph{Geometric warping unit:} We specify a geometric warping unit as $\hvec(\uvec) = \Rvec \uvec$, where $\Rvec$ is an unknown upper triangular $(D + 1) \times (D + 1)$ matrix, constrained such that the matrix $\Avec \equiv \Rvec' \Rvec$ is a correlation matrix; $\Rvec$ is therefore the Cholesky factor of $\Avec$. Let $\uvec_1, \uvec_2 \in S \times [0, h_\mathrm{max}]$. If $\hvec(\cdot)$ were the only warping unit in our warping function, then $||\hvec(\uvec_1) - \hvec(\uvec_2)||_2 = ||\Rvec (\uvec_1 - \uvec_2)||_2 = \sqrt{(\uvec_1 - \uvec_2)' \Avec (\uvec_1 - \uvec_2)}$, so that the Euclidean distance in \eqref{eqn:deviation_covariance} can be interpreted as a Mahalanobis distance. The diagonal elements of the matrix $\Avec$ therefore determine the relative scaling of the different dimensions; for $\Avec$, a correlation matrix, these elements are fixed to one, so that the scaling comes from the AWUs. In fact, while the geometric warping unit induces geometric anisotropy \citep{shengelfand2019} in the deviation process, it does not induce process nonstationary; nonstationarity is introduced in our models through the AWUs. The prior distribution we assign to $\Rvec$ is described in Section~\ref{sec:geometric_warping_unit_prior} of the supplementary material.

\noindent\emph{Combined effect of axial and geometric warping units:} The overall warping in GeoWarp is as follows: the AWUs are applied first, followed by the geometric warping unit, resulting in the warping $\gvec(\uvec) = (\hvec \circ \fvec_{D + 1} \circ \cdots \circ \fvec_1)(\uvec)$. The squared-distance after warping between the pairs $\uvec_1 \equiv (u_{1, 1}, \ldots, u_{1, D + 1})'$ and $\uvec_2 \equiv (u_{2, 1}, \ldots, u_{2, D + 1})'$, both in $S \times [0, h_\mathrm{max}]$, is therefore given by $||\gvec(\uvec_1) - \gvec(\uvec_2)||_2^2 = (\uvec_1^* - \uvec_2^*)' \Avec (\uvec_1^* - \uvec_2^*)$, where $\uvec_i^* = (u_1^*(u_{i, 1}), \ldots, u_{D + 1}^*(u_{i, D + 1}))'$ for $i = 1, 2$.

\subsection{Summary of the GeoWarp hierarchical model}
\label{sec:model_summary}

Let $\Zvec \equiv (Z_{1, 1}, Z_{1, 2}, \ldots, Z_{M, N_M})'$ and $\deltavec \equiv (\delta(\svec_1, h_{1, 1}), \delta(\svec_1, h_{1, 2}), \ldots, \delta(\svec_M, h_{M, N_M}))'$ be vectors of length $N \equiv \sum_{i = 1}^M N_i$ containing all the measurements and their associated realizations of $\delta(\cdot, \cdot)$, respectively. The data model may be written in vector form as
\begin{equation}
  (\Zvec \mid \omegavec, \kappavec, \gammavec, \Rvec, \sigma_\epsilon^2)
    \sim \mathrm{Gau}(\Xvec \omegavec, \Sigmavec_Z),
  \quad \sigma_\epsilon^2
    \sim \mathrm{IG}(a_\epsilon, b_\epsilon),
  \label{eqn:data_model_vectorised}
\end{equation}
where $\omegavec \equiv (\alphavec', \betavec')'$; $\kappavec \equiv (\eta, \zetavec')'$; $\gammavec \equiv (\gamma_{1, 1}, \gamma_{1, 2}, \ldots, \gamma_{D + 1, L_{D + 1}})'$; $\Xvec \equiv (\xvec_{1, 1}, \xvec_{1, 2}, \ldots, \xvec_{M, N_M})'$ is an $N \times (K_\beta + 2)$ matrix in which $\xvec_{i, j} \equiv (1, h_{i, j}, \phi_1^\mu(h_{i, j}), \ldots, \phi_{K_\beta}^\mu(h_{i, j}))'$ for $i = 1, \ldots, M$ and $j = 1, \ldots, N_i$; and  $\Sigmavec_Z \equiv \Sigmavec_\delta + \sigma_\epsilon^2 \Ivec_N$ is an $N \times N$ covariance matrix, where $\Sigmavec_\delta \equiv \mathrm{Var}(\deltavec \mid \kappavec, \gammavec, \Rvec)$ is the covariance matrix of the deviation process, described below.

The hierarchical model for the parameters that govern the mean-profile process (Section~\ref{sec:mean_profile}) is
\begin{equation}
  (\omegavec \mid \sigma_\beta^2)
    \sim \mathrm{Gau}(\zerovec_{K_\beta + 2}, \Sigmavec_\omega),
  \quad
  \sigma_\beta^2
    \sim \mathrm{IG}(a_\beta, b_\beta),
\end{equation}
where $\Sigmavec_\omega \equiv \mathrm{bdiag}(\sigma_\alpha^2 \Ivec_2, \Sigmavec_\beta)$ is a block diagonal $(K_\beta + 2) \times (K_\beta + 2)$ covariance matrix, with $\Sigmavec_\beta$ discussed in Section~\ref{sec:spline_coefficient_prior} of the supplementary material.

The deviation process (Section~\ref{sec:deviation_process}) enters \eqref{eqn:data_model_vectorised} through the matrix $\Sigmavec_\delta$, the entries of which are given by \eqref{eqn:deviation_covariance}, which combines the depth-varying variances, governed by $\kappavec$, and the axial and geometric warping units, governed by $\gammavec$ and $\Rvec$, respectively (see Section~\ref{sec:warping_function}). The hierarchical model for these parameters is
\begin{equation}
  \begin{gathered}
    (\kappavec \mid \ell_\zeta, \sigma_\zeta^2)
      \sim \mathrm{Gau}(\zerovec_{K_\zeta + 1}, \Sigmavec_\kappa),
    \quad \ell_\zeta
      \sim \text{Half-N}(\sigma_\ell^2),
      \quad \sigma_\zeta^2
      \sim \mathrm{IG}(a_\zeta, b_\zeta), \\
    \gamma_{d, 1} \sim \text{Inv-Unif}(\gamma_d^-, \gamma_d^+),
      \enskip d = 1, \ldots, D, \\
    \gamma_{D + 1, 1}, \ldots, \gamma_{D + 1, L_{D + 1}} \sim \mathrm{Ga}(a_{D + 1}^\gamma, b_{D + 1}^\gamma), \\
    \Rvec \sim \text{LKJ-R}(\rho_R),
  \end{gathered}
\end{equation}
where $\Sigmavec_\kappa \equiv \mathrm{bdiag}(\sigma_\eta^2, \Sigmavec_\zeta)$ is a block diagonal $(K_\zeta + 1) \times (K_\zeta + 1)$ covariance matrix, $\text{Half-N}(\sigma_\ell^2)$ denotes the half-normal distribution with parameter $\sigma_\ell^2$, $\text{Inv-Unif}(\gamma_d^-, \gamma_d^+)$ denotes the inverse-uniform distribution over the interval $[\gamma_d^-, \gamma_d^+]$, and where $\text{LKJ-R}(\rho_R)$ denotes a modified version of the Lewandowski--Kurowicka--Joe prior distribution with parameter $\rho_R$; see Sections~\ref{sec:geometric_warping_unit_prior} and~\ref{sec:spline_coefficient_prior} in the supplementary material for more details. Excluding $\omegavec$, which is marginalized out during inference in Section~\ref{sec:inference_computation}, the total number of parameters in the model is $K_\zeta + D(D + 1)/2 + \sum_{d = 1}^{D + 1} L_d + 11$. In our application to CPT data in Section~\ref{sec:application}, our model choices lead to a total of 74 parameters.

\section{Inference and prediction}
\label{sec:inference_computation}

We now consider inference and prediction with the GeoWarp model. CPT measurements are usually taken at fine vertical intervals of 1--2 cm and, of the datasets we consider in Section~\ref{sec:application}, the smallest and largest have 4,265 and 35,538 data points, respectively. This is too large for exact posterior inference, so we instead perform calculations approximately using techniques based on the Vecchia approximation \citep{vecchia1988}. Section~\ref{sec:vecchia_approximation} summarizes the Vecchia approximation and explains how it can be adapted to CPT data, Section~\ref{sec:parameter_inference} details how we make inference on GeoWarp's parameters, and Section~\ref{sec:prediction} shows how we predict the sediment properties at unobserved locations.

\subsection{Vecchia approximation}
\label{sec:vecchia_approximation}

Consider an $n$-dimensional random vector, $\Vvec \equiv (V_1, \ldots, V_n)'$, for which $\Vvec \sim \mathrm{Gau}(\zerovec, \Qvec_V^{-1})$, where $\Qvec_V^{-1}$ is an $n \times n$ covariance matrix with its $(i, j)$th entry given by $\mathrm{Cov}(V_i, V_j)$. In the general case, evaluating the density of $\Vvec$ requires $O(n^3)$ operations, which is prohibitive for large $n$. The Vecchia approximation $\tilde{p}(\Vvec)$ of $p(\Vvec)$ is done by first factorizing $p(\Vvec)$ as $p(\Vvec) = p(V_1) \prod_{i = 2}^n p(V_i \mid V_{i - 1}, \ldots, V_1)$, and then replacing the full conditional distributions in this expression with ones that condition on just a subset of each element's predecessors in $\Vvec$. This yields $\tilde{p}(\Vvec) \equiv p(V_1) \prod_{i = 2}^n p(V_i \mid V_{p_{i, 1}} \ldots, V_{p_{i, J_i}})$, where $\mathcal{P}_i \equiv \{ p_{i, 1}, \ldots, p_{i, J_i} \}$ for $i = 2, \ldots, n$ is a set of size $J_i$ containing integers between 1 and $i - 1$. The set $\mathcal{P}_i$ is called the parent set (or neighbor set) of $V_i$. The quality of the approximation depends on the ordering of the elements in the vector $\Vvec$, as well as on the choice of the parent sets, both of which can be chosen arbitrarily; \citet{guinness2018} showed that, when these considerations are managed carefully, the approximation is sufficiently accurate in the spatial context for the purpose of both parameter estimation and prediction.

\citet{guinness2018} showed that the density $\tilde{p}(\Vvec)$ is a multivariate Gaussian density, $\tilde{p}(\Vvec) \sim \mathrm{Gau}(\zerovec, \tilde{\Qvec}_V^{-1})$, where $\tilde{\Qvec}_V^{-1}$ is an $n \times n$ covariance matrix. The matrix $\tilde{\Qvec}_V$ is a sparse matrix, and \citet{guinness2021} described efficient methods to calculate quantities involving $\tilde{\Qvec}_V$ using a single pass through the vector $\Vvec$, including $|\tilde{\Qvec}_V|$, $\tilde{\Qvec}_V^{1 / 2} \yvec$, and $\tilde{\Qvec}_V \yvec$, where $\yvec$ is an $n$-dimensional vector. These operations (and hence the density $\tilde{p}(\Vvec)$) can be computed in $O(n m^3)$ operations, where $m \equiv \max \{ J_i : i = 2, \ldots, n \}$. For large $n$ and small $m$ (the number of parents is typically chosen to be between 10 and 50), many fewer computations are required when using $\tilde{\Qvec}_V$ instead of $\Qvec_V$. In Section~\ref{sec:parameter_inference} we use this method to construct an approximation $\tilde{\Sigmavec}_Z^{-1}$ of $\Sigmavec_Z^{-1}$ (defined in Section~\ref{sec:model_summary}), and in Section~\ref{sec:prediction} we construct an approximate prediction distribution.

When the elements of $\Vvec$ are spatially indexed, the most common scheme for picking the parent set $\mathcal{P}_i$, $i = 1, \ldots, n$, is to use the $m$ locations nearest to $i$ among its predecessor locations, where $m$ is chosen to balance computation time and statistical efficiency. This scheme does not work well for CPT data because the locations of the data are vertically dense and horizontally sparse, so that the nearest neighbors of a point are almost always from the same CPT. This nearest neighbor strategy would therefore lead to an approximation that largely neglects horizontal dependencies between CPTs. To address this issue, for each location we instead pick from among the location's predecessors half of the parents to be the closest according to the Euclidean distance over the full coordinate space $(\svec', h)' \in \mathbb{R}^d \times [0, h_\mathrm{max}]$ (that is, using the conventional nearest neighbor scheme described above), and half of the parents to be the points from other CPTs that are closest in depth. We use a random ordering of the measurements, which was found to work well by \citet{guinness2018}. An example of this scheme applied to 2-D data with $m = 10$ is shown in Figure~\ref{fig:parent_example} in Section~\ref{sec:additional_figures} of the supplementary material.

\subsection{Parameter inference}
\label{sec:parameter_inference}

For computational efficiency when estimating the parameters of GeoWarp, we marginalize $\omegavec$ in \eqref{eqn:data_model_vectorised} to yield $(\Zvec \mid \kappavec, \gammavec, \Rvec, \sigma_\epsilon^2) \sim \mathrm{Gau}(\zerovec_N, \Xvec \Sigmavec_\omega \Xvec' + \Sigmavec_Z)$. The logarithm of the density of the corresponding marginalized posterior distribution is given by
\begin{multline}
  \log p(\kappavec, \gammavec, \Rvec, \sigma_\epsilon^2, \sigma_\beta^2, \ell_\zeta, \sigma_\zeta^2 \mid \Zvec)
  =
    \log p(\Zvec \mid \kappavec, \gammavec, \Rvec, \sigma_\epsilon^2)
    + \log p(\kappavec \mid \ell_\zeta, \sigma_\zeta^2)
    + \log p(\gammavec)
    + \log p(\Rvec) \\
    + \log p(\sigma_\epsilon^2)
    + \log p(\sigma_\beta^2)
    + \log p(\ell_\zeta)
    + \log p(\sigma_\zeta^2)
    + \mathrm{const.},
  \label{eqn:model_posterior_density}
\end{multline}
where ``$\mathrm{const.}$'' is an arbitrary constant that does not depend on the parameters. The term $\log p(\Zvec \mid \kappavec, \gammavec, \Rvec, \sigma_\epsilon^2)$ in \eqref{eqn:model_posterior_density} is given by
\begin{equation}
  \log p(\Zvec \mid \kappavec, \gammavec, \Rvec, \sigma_\epsilon^2)
  = -\frac{1}{2} \left[
      N \log(2\pi)
      + \log |\Xvec \Sigmavec_\omega \Xvec' + \Sigmavec_Z|
      + \Zvec' (\Xvec \Sigmavec_\omega \Xvec' + \Sigmavec_Z)^{-1} \Zvec
    \right] .
  \label{eqn:marginal_log_likelihood}
\end{equation}
The quantities $\log |\Xvec \Sigmavec_\omega \Xvec' + \Sigmavec_Z|$ and $(\Xvec \Sigmavec_\omega \Xvec' + \Sigmavec_Z)^{-1}$ in \eqref{eqn:marginal_log_likelihood} can be recast using a matrix-determinant lemma and the Sherman--Morrison--Woodbury matrix identity \citep{hendersonsearle1981} as $\log |\Xvec \Sigmavec_\omega \Xvec' + \Sigmavec_Z| = \log |\Ovec| - \log |\Sigmavec_\omega| - \log |\Sigmavec_Z|$ and $(\Xvec \Sigmavec_\omega \Xvec' + \Sigmavec_Z)^{-1} = \Sigmavec_Z^{-1} - \Sigmavec_Z^{-1} \Xvec \Ovec^{-1} \Xvec' \Sigmavec_Z^{-1}$, where $\Ovec \equiv \Sigmavec_\omega^{-1} + \Xvec' \Sigmavec_Z^{-1} \Xvec$. The matrices $\Sigmavec_\omega$ and $\Ovec$ are relatively small, but for most CPT datasets the matrix $\Sigmavec_Z$ is too large to use directly. Using the Vecchia approximation described in Section~\ref{sec:vecchia_approximation}, we instead approximate its inverse, $\Sigmavec_Z^{-1}$, as $\tilde{\Sigmavec}_Z^{-1}$. We then use $\tilde{\Sigmavec}_Z^{-1}$ in \eqref{eqn:marginal_log_likelihood} to define a computationally-tractable approximate log-density, $\log \tilde{p}(\Zvec \mid \kappavec, \gammavec, \Rvec, \sigma_\epsilon^2)$, which is substituted into \eqref{eqn:model_posterior_density} to define an approximate log posterior, $\log \tilde{p}(\kappavec, \gammavec, \Rvec, \sigma_\epsilon^2, \sigma_\beta^2, \ell_\zeta, \sigma_\zeta^2 \mid \Zvec)$.

The parameters of GeoWarp are inferred using Markov chain Monte Carlo (MCMC), which yields samples from the approximate posterior distribution of $\kappavec$, $\gammavec$, $\Rvec$, $\sigma_\epsilon^2$, $\sigma_\beta^2$, $\ell_\zeta$ and $\sigma_\zeta^2$. The parameter $\omegavec$ can then be sampled from its full conditional distribution. We implemented the computation of the approximate log likelihood using the R and Stan programming languages \citep[][respectively]{rcore2022,stan2022}, and for MCMC used the No-U-Turn Sampler \citep[NUTS,][]{hoffmangelman2014} implemented in Stan. An alternative to MCMC with reduced computational cost is to find the maximum \emph{a posteriori} (MAP) estimates of the parameters; we do this in the cross-validation in Section~\ref{sec:cross_validation}. We explore the differences between the MAP estimates and MCMC inferences in Section~\ref{sec:discussion}.

\subsection{Prediction at unsampled locations}
\label{sec:prediction}

We now discuss how GeoWarp predicts the process $Y(\svec, h)$, $\svec \in S, h \in [0, h_\mathrm{max}]$, at $N_\mathrm{pred} > 0$ unsampled locations and depths. Let $\Yvec^* \equiv (Y(\svec_1^*, h_1^*), \ldots, Y(\svec_{N_\mathrm{pred}}^*, h_{N_\mathrm{pred}}^*))'$ be the vector of unknowns, where the locations and depths to be predicted are $\svec_i^* \in S$ and $h_i^* \in [0, h_\mathrm{max}]$ for $i = 1, \ldots, N_\mathrm{pred}$. The prediction distribution of $\Yvec^*$ given the data and the parameters is $p(\Yvec^* \mid \Zvec, \omegavec, \kappavec, \gammavec, \Rvec, \sigma_\epsilon^2)$. This is a multivariate Gaussian distribution whose direct manipulation is computationally infeasible. Therefore, to perform predictions, we approximate the distribution of $\Yvec^*$ using the Vecchia approximation described in Section~\ref{sec:vecchia_approximation}, yielding the approximate prediction distribution $\tilde{p}(\Yvec^* \mid \Zvec, \omegavec, \kappavec, \gammavec, \Rvec, \sigma_\epsilon^2)$, from which it is easy to draw samples of $\Yvec^*$. We then draw one sample from $\tilde{p}(\Yvec^* \mid \Zvec, \omegavec, \kappavec, \gammavec, \Rvec, \sigma_\epsilon^2)$ for each MCMC sample of the unknown parameters $\omegavec, \kappavec, \gammavec, \Rvec$ and $\sigma_\epsilon^2$. Details of the prediction distribution are given in Section~\ref{sec:prediction_approximation} of the supplementary material.

In some cases, such as in cross-validation, one needs to predict a vector of CPT measurements, $\Zvec^* \equiv (Z_1^*, \ldots, Z_{N_\mathrm{pred}}^*)'$, where $Z_i^* \equiv Y(\svec_i^*, h_i^*) + \epsilon_i^*$ and $\epsilon_i^* \sim \mathrm{Gau}(0, \sigma_\epsilon^2)$ for $i = 1, \ldots, N_\mathrm{pred}$. Samples of $\Zvec^*$ are produced by first sampling $\Yvec^*$ at the measurement locations, and then adding Gaussian white noise with variance $\sigma_\epsilon^2$ to each element.

\section{Application to CPT data from the North West Shelf of Australia}
\label{sec:application}

\begin{figure}[t!]
  \begin{center}
   \includegraphics[width=\textwidth]{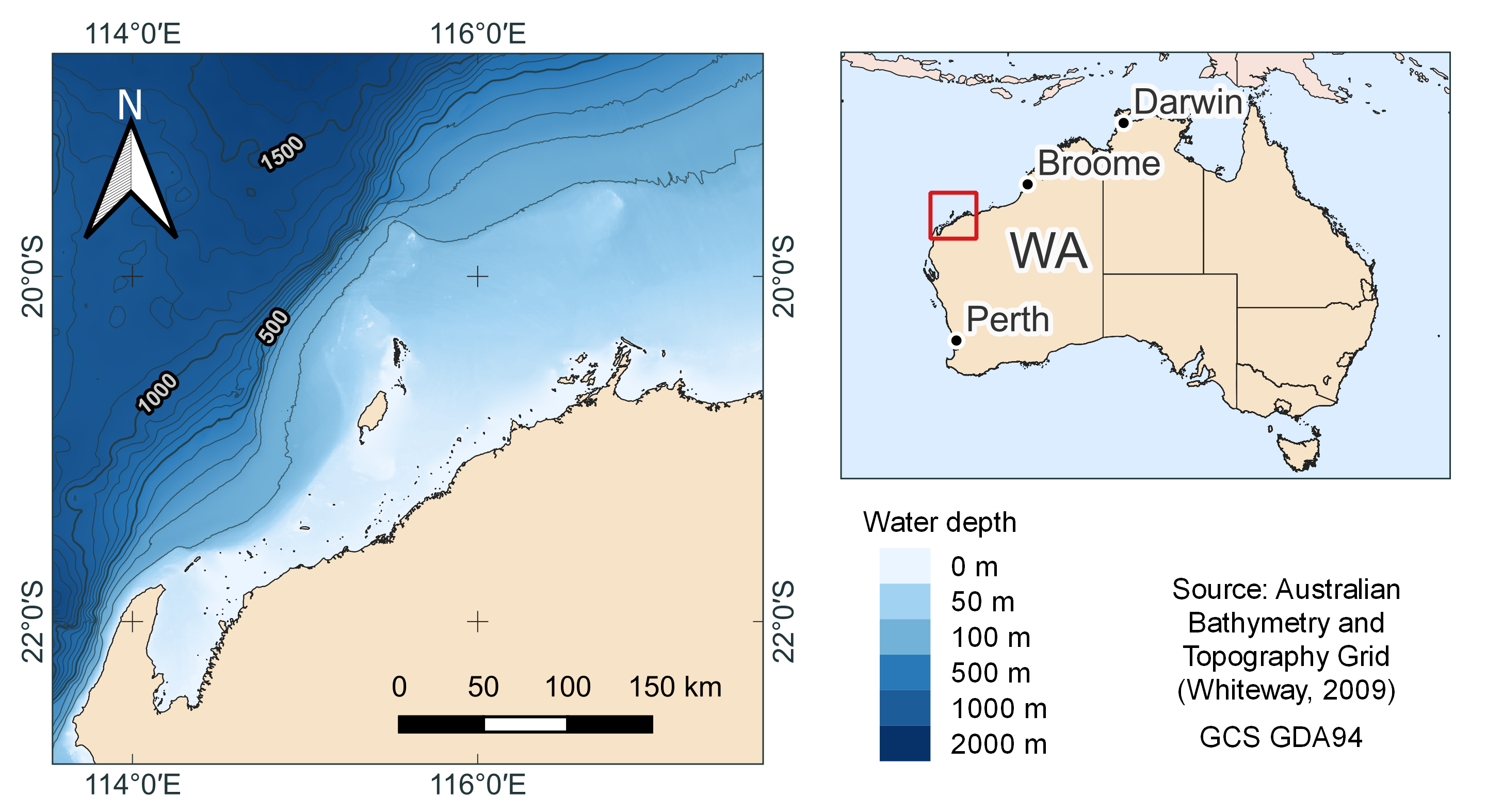}
  \end{center}
  \caption{
    Bathymetric map of Australia's North West Shelf. The exact locations of the fields used in this study are not shown for confidentiality reasons.
  }
  \label{fig:region_map}
\end{figure}

The North West Shelf (NWS) of Australia, depicted in Figure~\ref{fig:region_map} has been the site of significant offshore infrastructure developments over the past 30 years. The seabed in this region is abundant in carbonate sediments, characterized by very high and anisotropic spatial variability \citep{hoganetal2017,watsonetal2019}, which presents challenges in site characterization.

We apply GeoWarp to CPT data from six sites on the NWS, grouped into two fields. Field A consists of sites A1, A2, and A3, while Field B includes sites B1, B2, and B3. The exact locations of the fields have been anonymized for confidentiality reasons. We have also constructed two 2-D datasets, labeled A2-T and B2-T, constructed from five CPTs that lie along a horizontal transect at sites A2 and B2, respectively. These cases are examined as geotechnical investigations sometimes occur along transects, such as prior to laying subsea pipelines or installing cables. All sites except B3 have measurements down to 41 meters ($h_\mathrm{max} = 41$ m); the data for B3 go down to 22 meters. More details on the sites, including maps and plots of all the CPT data, are given in Section~\ref{sec:detailed_cpt_data} of the supplementary material.

The following sections provide further details of our study. Section~\ref{sec:model_fits_and_predictions} examines \mbox{GeoWarp's} fits at the six 3-D sites and two 2-D transects. Section~\ref{sec:cross_validation} evaluates the predictive capability of the model using cross-validation. Finally, Section~\ref{sec:discussion} discusses the results.

\subsection{Model fits and predictions}
\label{sec:model_fits_and_predictions}

Some choices must be made to tailor GeoWarp to a data set. The choices we make for this data set are detailed in Section~\ref{sec:model_for_nws_data} of the supplementary material, which gives the model configuration, the prior distributions, and the number of parents to use in the Vecchia approximation. With these choices, we used the GeoWarp software to generate MCMC samples of the parameters for each site. We also fitted the model using MAP estimation, which we contrast to MCMC in Section~\ref{sec:discussion}. Details of the MAP and MCMC fitting procedures as well as run times are given in Section~\ref{sec:fitting_procedure} of the supplementary material.

\subsubsection{Estimated sediment characteristics}
\label{sec:estimated_sediment_characteristics}

The posterior mean estimates of the mean profile, $\mu(h)$, $h \in [0, h_\mathrm{max}]$, and of the standard deviation of the deviation process, $\sqrt{\sigma_\delta^2(h)}$ for the six 3-D sites and the two 2-D sites are shown in the left and middle panels of Figure~\ref{fig:vertical_profiles}, respectively. Figure~\ref{fig:vertical_profiles_full} in the supplementary material expands on these plots by showing the estimated 95\% posterior intervals for these quantities. To visualize the combined effect of the horizontal and vertical axial warping units and the geometric warping unit, for each MCMC sample we calculate iso-correlation contours which indicate where the correlation with a given point is equal to 0.5; points inside the contour have correlation larger than 0.5, while points outside have correlation smaller than 0.5. The following plots in the supplementary material visualize the distribution over these contours: Figure~\ref{fig:isocorrelation_horizontal} shows the distribution for the easting--northing plane (the contours in this plane do not vary with depth); Figure~\ref{fig:isocorrelation_vertical_easting} shows the contours at various depths in the easting--depth plane; and Figure~\ref{fig:isocorrelation_vertical_northing} shows the contours at various depths in the northing--depth plane. The posterior distribution for each iso-correlation contour is visualized over a grid, where each grid cell is colored according to the probability that the contour passes it. The following are some observations on the characteristics of the fits.

\begin{figure}
  \begin{center}
    \includegraphics{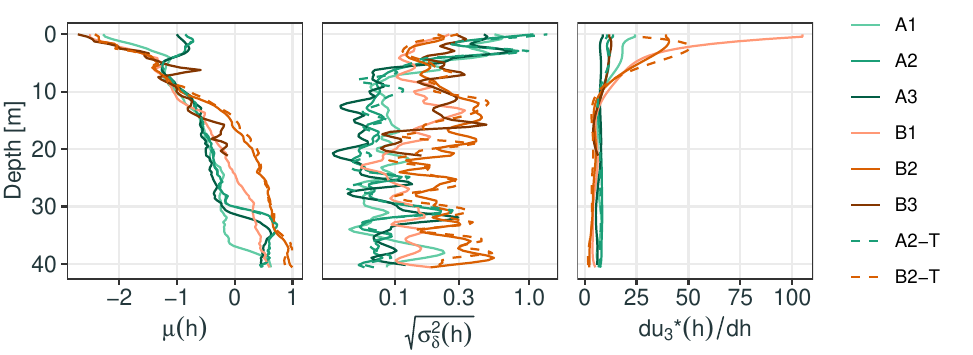}
  \end{center}

  \caption{
    Posterior mean estimates of the mean profile ($\mu(h)$, left), vertical standard deviation profiles ($\sqrt{\sigma_\delta^2(h)}$, middle, log scale), and the derivative of the vertical warping ($\textrm{d}u_3^*(h) / \textrm{d}h$, right), for each of the six 3-D sites (solid lines) and the two 2-D sites (dashed lines).
  }
  \label{fig:vertical_profiles}
\end{figure}

\emph{The mean profile increases with depth:} As expected, the mean profile increases with depth, justifying the inclusion of $h$ as a covariate in \eqref{eqn:mu_process}.

\emph{Considerable variability in the top 5 m}: The standard deviation of the profile is high in the first 5 m for A-field sites, but it is relatively low at these depths for the B-field sites. At the B-field sites and at the site A1 the mean profile is at its lowest for the top 5 m, while for A2 and A3 the mean profile starts relatively high. This variability is also clearly seen in Figure~\ref{fig:residuals_plot}; this plot shows the residuals of the model fits at each site, calculated by subtracting the posterior mean estimate of the mean profile from the measurements at each CPT.

\emph{Uncertainty around layer changes}: At depths between 10 m and 15 m, the mean profile for B2 increases rapidly, and there is a corresponding increase in $\sigma_\delta^2(h)$. From Figure~\ref{fig:datasets_b}, this appears to correspond to a sudden layer change, where some of the CPTs transition to higher values over a short space. The increased value of $\sigma_\delta^2(h)$ in this region may therefore represent the horizontal uncertainty in the depth of the layer change. Similar co-occurring increases in the mean profile and the variance occur around depth 32 m for the A2 and A3 sites, and at around 38 m for the A1 site. The increased variability around these depths is also apparent in Figure~\ref{fig:residuals_plot}.

\emph{Varied horizontal length scales}: The posterior distributions of the horizontal iso-correlation contours in Figure~\ref{fig:isocorrelation_horizontal} vary substantially between sites. At A1, B1, and B2, the horizontal dependencies are small (dropping to 0.5 correlation by 25 meters) relative to the distance between CPTs (at least 22 m apart). At A2, the estimated dependence in the northing direction is around 50 meters, which is long relative to the sizes of the sites. At A3, B1, B2, and B3 there is also considerable uncertainty in the iso-correlation contours, suggesting that the horizontally-sparse data provide only weak information about horizontal dependencies.

\emph{Limited easting--depth and northing--depth rotation near the surface}: The iso-correlation contours in Figure~\ref{fig:isocorrelation_vertical_easting} and Figure~\ref{fig:isocorrelation_vertical_northing} show that the geometric warping unit has induced little rotation in the easting--depth and northing--depth planes near the surface. This suggests that features in the deviation process largely do not occur at an incline near the surface.

\emph{Shorter correlation lengths in the top 10 m}: The posterior mean estimates of the derivative of the vertical warping, $\textrm{d}u_3^*(h) / \textrm{d}h$, are shown for the six sites in the right-most panel of Figure~\ref{fig:vertical_profiles}, and their posterior intervals are shown in Figure~\ref{fig:vertical_profiles_full} in the supplementary material. The derivative is shown because this quantity is inversely related to the vertical correlation of the deviation process: the larger (smaller) the derivative, the weaker (stronger) the spatial dependence. The most striking feature of the warping function is the large derivative in the top 10 m at the B1 and B2 sites (and to a lesser degree at A1). This indicates a short correlation length at these depths. Deeper down, the correlation length increases. The remaining sites A2, A3, and B3 have relatively constant derivatives, indicating that the length scales do not vary much with depth.

\emph{The estimates for A2-T and B2-T are similar to those for A2 and B2, respectively}: The estimated vertical profile characteristics for the 2-D sites A2-T and B2-T are similar to those for the full 3-D sites (A2 and B2, respectively) from which they were constructed. This is reassuring, and indicates that GeoWarp can be used with both 2-D and 3-D data.

\subsubsection{Prediction at unknown locations}
\label{sec:prediction_results}

We now use the methods described in Section~\ref{sec:prediction} to predict the log sediment resistance $Y(\cdot, \cdot)$ over a 3-D grid covering the A2 site, and over a 2-D grid covering the B2-T site. The predictions were computed using the Vecchia approximation with 100 parents (see Section~\ref{sec:vecchia_approximation}).

\begin{figure}[t]
  \begin{center}
    \includegraphics[width=0.5\linewidth]{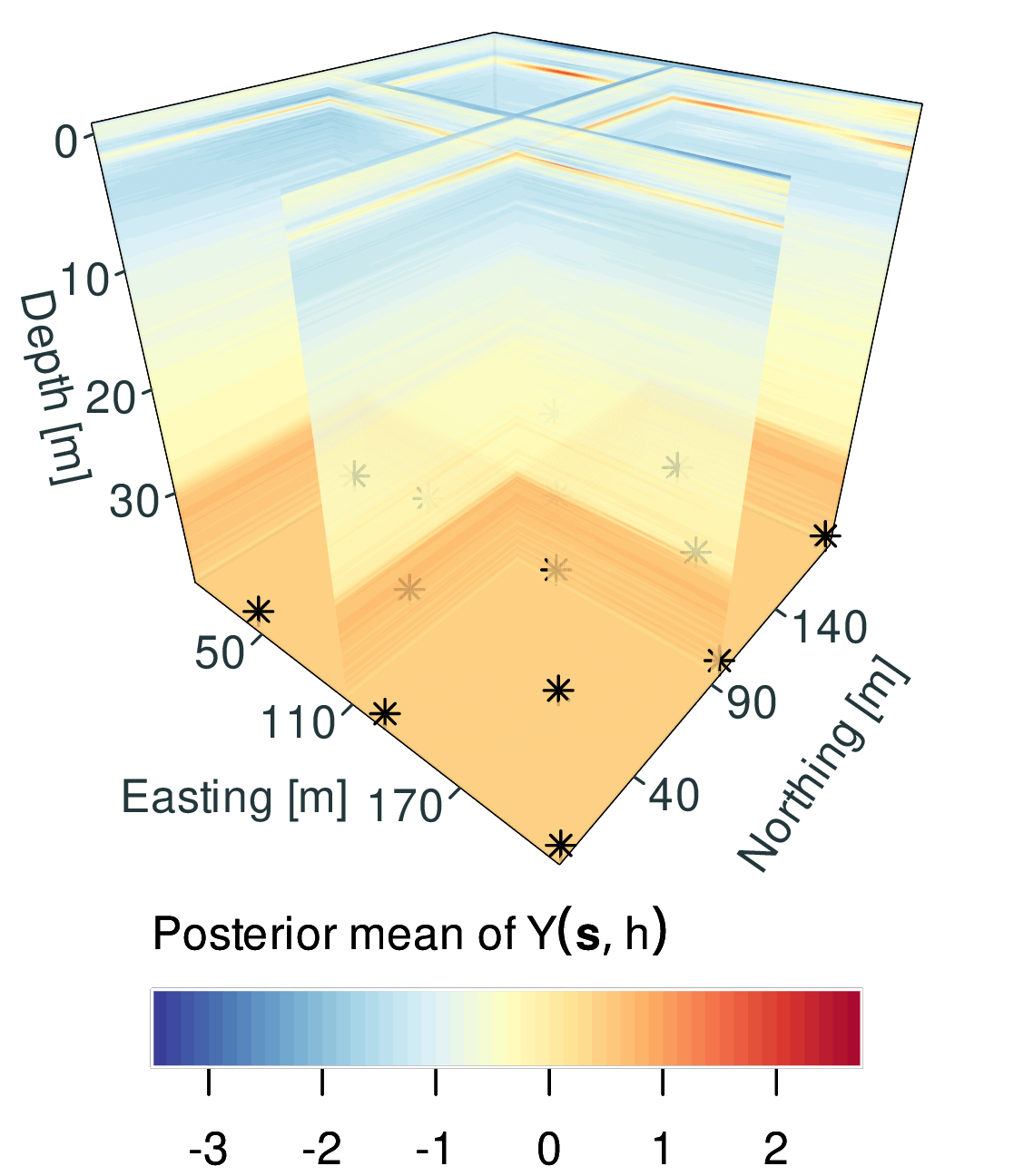}%
    \includegraphics[width=0.5\linewidth]{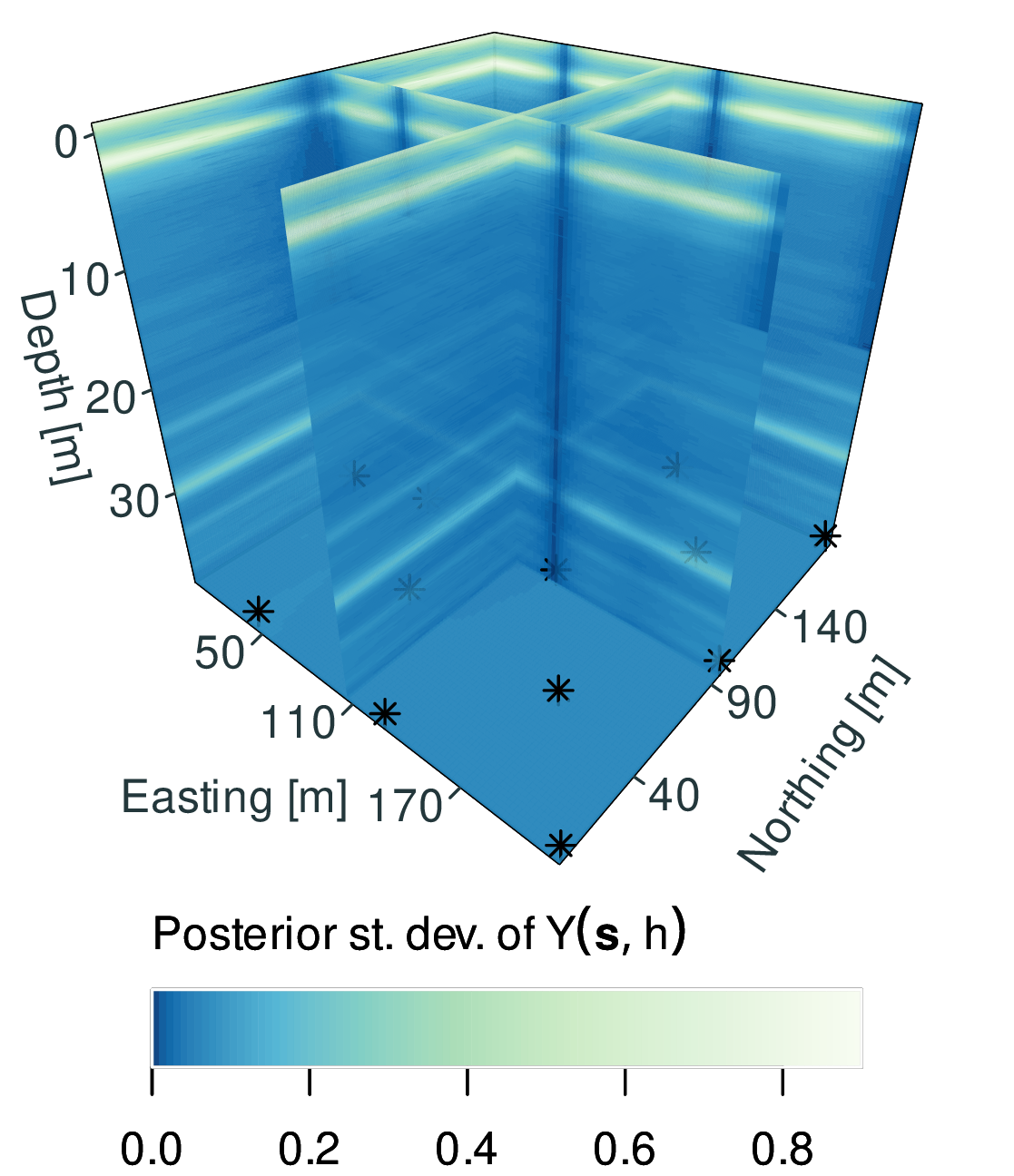}
  \end{center}
  \caption{
    Predictions over a 3-D grid at the A2 site. The left panel shows the posterior mean of $Y(\cdot, \cdot)$, while the right panel shows the posterior standard deviation. Points mark the horizontal locations of the CPTs performed at the site.
  }
  \label{fig:predicted_a2}
\end{figure}

The posterior mean and standard deviation of the A2 predictions are depicted in Figure~\ref{fig:predicted_a2}. Several of the features of the vertical profile discussed in Section~\ref{sec:estimated_sediment_characteristics} are visible here, including the increased variance in the top 5 m of the profile, as well as the steady increase with depth in the mean of $Y(\cdot, \cdot)$. The standard deviation is smaller at locations close to CPTs.

\begin{figure}[t]
  \begin{center}
    \includegraphics[width=16.5cm]{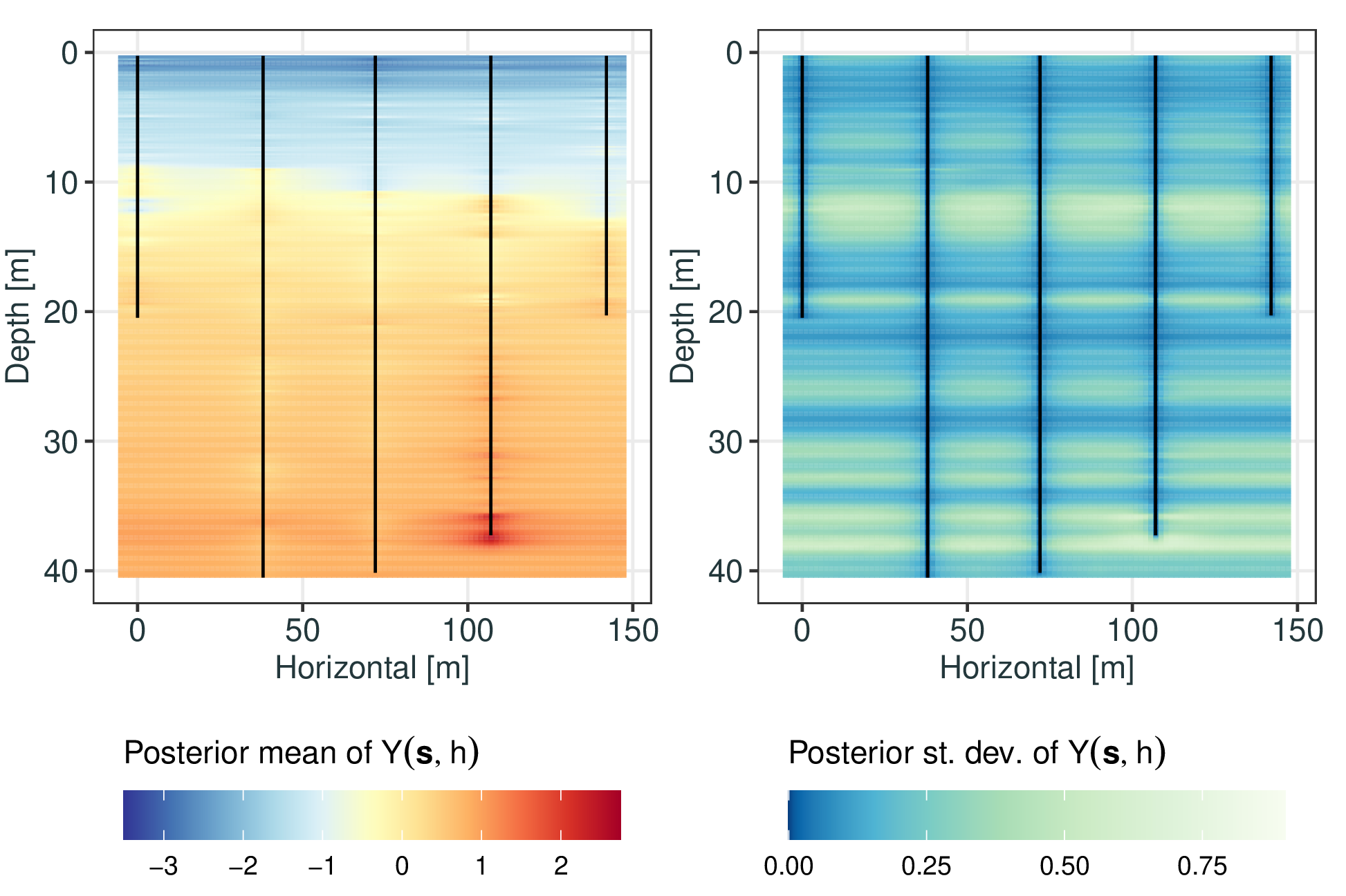}
  \end{center}
  \caption{
    Predictions over a 2-D grid at the B2-T site. Lines show the locations and depths of the CPTs. The left panel shows the posterior mean of $Y(\cdot, \cdot)$, while the right panel shows the posterior standard deviation.
  }
  \label{fig:predicted_b2t}
\end{figure}

Figure~\ref{fig:predicted_b2t} shows the 2-D predictions for the B2-T site. The most striking feature in this plot is the spike in $Y(\cdot, \cdot)$ at the bottom of the CPT near horizontal coordinate 110 m. The predicted spike increases below the maximum depth of the CPT, which is largely due to the combination of vertical dependence in the prediction process, and the linear depth trend. After a few more meters, the prediction reverts back to the estimated mean profile $\mu(h)$. The other striking feature, also discussed earlier in Section~\ref{sec:estimated_sediment_characteristics}, is the increased posterior variance at the layer transition that occurs between depths 10 m and 15 m.

In addition to producing posterior summaries, a major advantage of using a full stochastic model is the ability to make realistic posterior simulations of the sediment properties at any desired location. These can be used as inputs to the design process, as they embody the full posterior distribution of the sediment properties. Posterior simulations also provide a check on the realism of the model; simulations should look similar to the observed data. Figure~\ref{fig:posterior_samples} in Section~\ref{sec:additional_figures} of the supplementary material shows single samples from the posterior predictive distribution of GeoWarp for the A1 and B1 sites, made using the Vecchia approximation with 100 parents. The left panels show slices of the full 3-D sample of $Y(\cdot, \cdot)$ for a grid covering the site, while the right panels show a single vertical sample of $\log q_c$ (equal to $Y(\cdot, \cdot)$ plus Gaussian white noise with variance $\sigma_\epsilon^2$) together with CPTs from the site. The samples reproduce the features of the data, including the increased variance near the surface for A1 and the shorter correlation length near the surface for B1.

\subsection{Cross-validation study}
\label{sec:cross_validation}

The predictions and simulations discussed in the previous section show that the model in Section~\ref{sec:geowarp_model} is able to accurately reproduce the features of CPT data, but it remains to assess how effective the model is at prediction at unsampled locations. To this end, we perform a cross-validation study \citep[see, e.g.,][]{stone1974} by iteratively withholding entire CPTs at each site and using the remaining CPTs at that site to estimate the model parameters and predict the withheld data; this is repeated for every CPT at every site. Parameters are estimated using MAP estimation due to the high computational cost of repeated runs of MCMC. The predictions are compared to the withheld CPT, and the performance is summarized using metrics described below. To focus the metrics on interpolation performance, we only calculate the metrics for depths in the withheld CPTs for which there is at least one other CPT at the site that reaches that depth.

We compare our method to three others. The first is ordinary linear regression with depth as the covariate, a common starting-point for geotechnical analysis \citep[e.g.,][]{fenton1999,uziellietal2005}. The second is Bayesian compressive sensing (BCS) as implemented in the software package \nospellcheck{Analytics of Sparse Spatial Data} \citep[\nospellcheck{ASSD-BCS};][]{lyuetal2023b}. The third and final method we compare to is a pseudo-model in the form of 0.1m depth-binned statistics.  This method corresponds closely to a common practice in geotechnical analysis where empirical summaries of the CPTs are used to characterize a site. Details of how we configured the BCS method and how the depth-binned statistics were constructed are given in Section~\ref{sec:other_methods_configuration} of the supplementary material.

We assess the quality of the prediction and prediction uncertainties on the withheld data using the mean-squared error (MSE), the continuous ranked probability score (CRPS), the interval score of the 95\% prediction interval (Int05), and the Dawid--Sebastiani score (DSS); see \citet{gneitingraftery2007} for more details on these scores. The scores are calculated by comparing the marginal predictive distribution for each measurement in the withheld CPT, and we take the average score to be a summary of the performance. We also calculate a pairwise variant of the DSS (DSS2), by averaging the bivariate DSS \citep[][]{gneitingraftery2007} of each vertically-adjacent pair of measurements.

The MSE only assesses the quality of the point predictions, while the remaining scores take into account the prediction uncertainty. DSS2 further takes into account correlations in the predictions, which are available for methods like GeoWarp that produce the full predictive distribution. For all scores, a lower value corresponds to better performance.

Table~\ref{tab:cross_validation_metrics} shows the cross-validated scores for the different sites and models. The column marked `All' gives the average performance across all CPTs at all sites. DSS and DSS2 are omitted for the binned statistics approach because it does not produce a full prediction distribution. Since the scores are subject to random variation, for each site and metric we assess whether the difference between the average score of the best-performing (lowest-scoring) variant and the other models is significant using one-sided two-sample $t$-tests at $\alpha = 0.05$. The best performing model, and those that are not statistically significantly different from it, are indicated in bold. We also plot the cross-validation predictions in Figures~\ref{fig:cv_predictions_a1} to \ref{fig:cv_predictions_b2_t} in Section~\ref{sec:additional_figures} of the supplementary material for the sites A1, A2, A3, B1, B2, B3, A2-T, and B2-T, respectively.

GeoWarp gives the best or equivalent-best performance in all metrics at all but one site, B2-T. When averaged across all measurements (the `All' column), GeoWarp performs the best for all metrics. The linear model is usually the worst performing, and the performance of BCS is typically between those of GeoWarp and the linear model.

The binned statistics approach performs well overall, particularly for the MSE, where it gives equivalent performance to GeoWarp for A1, A3, B1, and B3. The binned statistics approach does not incorporate horizontal dependencies, so the similar performance between GeoWarp and the binned statistics approach suggests that the inclusion of the vertical mean profile process in GeoWarp is the most important feature for predictive performance at these sites; we investigate this more in Section~\ref{sec:discussion}.

At B2-T the linear model outperforms GeoWarp for Int05 and DSS. Inspection of the actual predictions in Figure~\ref{fig:cv_predictions_b2_t} shows that the poor performance at these sites is due to an outlier CPT: CPT 11 for B2-T. Its outlier status can be seen clearly in the residuals for site B2-T in Figure~\ref{fig:residuals_plot}; only two other CPTs go to the same depth as CPT 11, and those two are similar to each other. Therefore, when CPT 11 is omitted in the cross-validation, the inferred variances $\sigma_\delta^2(h)$ are too small, which affects both the prediction mean and uncertainty. One solution to this problem is to assume constant variance with depth; we consider this further in Section~\ref{sec:discussion}.

\begin{table}[t]
  \caption{
    Cross-validated prediction scores. The scores for the different sites are shown in each column, and the `All' column gives the score averaged over all measurements. The best (lowest-scoring) model for each site and metric is marked in bold, as are those models that are not statistically different from the best (determined using one-sided two-sample $t$-tests at $\alpha = 0.05$).
  }
  \begin{center}
    \footnotesize
    \setlength\tabcolsep{5.5pt}
    \vspace{-0.4cm}
    \input{figures/cv-metrics-table}
  \end{center}

  \label{tab:cross_validation_metrics}
\end{table}

\subsection{Discussion}
\label{sec:discussion}

The predictions and simulations in Section~\ref{sec:model_fits_and_predictions} show that the model described in Section~\ref{sec:geowarp_model} is able to capture the features of CPT data, including the shared mean profile within a site and depth-varying variance and scales of correlation. The predictions from the model shown in Figure~\ref{fig:predicted_a2} and Figure~\ref{fig:predicted_b2t} interpolate smoothly between the CPTs, as might be expected, and the prediction uncertainties are highest at locations most distant from observations. We verified with domain experts that the simulations shown in Figure~\ref{fig:posterior_samples} are also reasonable.

The predictions in Section~\ref{sec:model_fits_and_predictions} are obtained using MCMC, which incorporates parameter uncertainty but takes a long time to run. Section~\ref{sec:map_mcmc_comparison} of the supplementary material investigates the use of MAP estimation as a faster alternative to MCMC. The MAP fits are similar to the MCMC fits, with the main difference that MAP tends to suggest longer correlation lengths, and that its prediction standard deviation is around 5\% smaller. This supports MAP estimation as a viable alternative to MCMC, particularly at larger sites where parameter uncertainty is smaller and MCMC is substantially slower.

The cross-validation results in Section~\ref{sec:cross_validation} show that GeoWarp outperforms other methods at most sites. Through an expanded cross-validation study, Section~\ref{sec:parsimonious_versions} of the supplementary material investigates which features of GeoWarp are most important to its performance. We find that quantification of the depth-varying mean profile is important for all aspects of prediction performance, while the depth-varying variance and the warping are important for uncertainty quantification. We also find that the poor performance of GeoWarp at B2-T can be improved by assuming constant variance with depth.

We also assessed GeoWarp's ability to generalize beyond the offshore sites in this study using cross-validation with an onshore dataset. The results, which show that GeoWarp performs well for the onshore site, are given in Section~\ref{sec:jaksa_benchmark} of the supplementary material.

\section{Conclusion}
\label{sec:conclusion}

We have introduced a new Bayesian method, GeoWarp, for 3-D geotechnical sediment characterization that is able to accommodate many important properties of the sediment, such as vertical changes in its mean strength, and heterogeneity in sediment stratigraphy. As described in Section~\ref{sec:geowarp_model}, GeoWarp decomposes the sediment properties into a vertical mean profile process and a deviation process. The deviation process models nonstationarity by warping the 3-D space over which it is defined, and by allowing the variance to change with depth. Computational efficiency of GeoWarp when making inference and performing predictions (described in Section~\ref{sec:inference_computation}) is achieved using the Vecchia approximation. Using data from six sites on the NWS of Australia, we show in Section~\ref{sec:application} that GeoWarp can accommodate different sediment stratigraphies. Through a cross-validation study we show that GeoWarp outperforms most other methods that are typically used for these data.

There are several potential future directions for the GeoWarp framework. The layering of the sediment, which makes the sediment properties vary with depth, is represented in GeoWarp through the combination of the vertical mean-profile process, the variance of the deviation process, and the vertical warping, all of which do not vary horizontally. In Section~\ref{sec:model_fits_and_predictions} it was shown how the uncertainty over the depth of layer transitions is represented in the process variance. This increased variance appears in the prediction uncertainty (see Figure~\ref{fig:predicted_a2}), which is appropriate. In practice, however, the depth at which layer transitions happen varies with horizontal location; see, for example, Figure~\ref{fig:predicted_b2t}, where the depth of the first layer transition appears to vary between 9 m and 13 m over the transect. This feature could be incorporated into GeoWarp by allowing the mean profile, the variance, and the warping, to vary horizontally using, for example, the deep compositional methods outlined by \citet{zammitmangionetal2022}.

Related to the above potential extension, the layering of the sediment in GeoWarp is represented through functions that change continuously with depth. Although these functions are flexible enough to accommodate rapid changes when needed, the underlying physical layer transitions can sometimes be discontinuous. This could be represented in GeoWarp by using piecewise-continuous functions for the mean profile, the variance profile, and the vertical warping function. Estimates of the change points would help understand the stratification of the sediments into different types, one of the fundamental goals of geotechnical analysis. However, such an extension would complicate inference and, given the flexibility of the existing model, may not affect predictions much.


Finally, at some sites it may be helpful to use other sources of data to improve the quality of the estimated geotechnical quantities. One candidate data type is geophysical seismic data, which can be collected more cheaply over a wider area than CPTs and which is often acquired before a geotechnical survey. \citet{chenetal2021} discuss how geophysical data could augment geotechnical analysis, but there are challenges in quantifying the relationship between geophysical and geotechnical data. Nonetheless, incorporating geophysical data into GeoWarp might alleviate limitations we see at sites where geotechnical data is scarce; adapting GeoWarp to accommodate geophysical data is the subject of future research.

\if1\blind
{

\section*{Acknowledgments}

This research is supported by the ARC \nospellcheck{ITRH} for Transforming Energy Infrastructure through Digital Engineering (TIDE), which is led by The University of Western Australia (UWA), delivered with The University of Wollongong and other research partners, and funded by the \nospellcheck{Australian Research Council, INPEX Operations Australia, Shell Australia, Woodside Energy, Fugro Australia Marine, Wood Group Kenny Australia, RPS Group, Bureau Veritas, and Lloyd’s Register Global Technology (grant No. IH200100009)}. Fraser Bransby holds the \nospellcheck{Fugro} Chair in Geotechnics at UWA, whose support is gratefully acknowledged. Phil Watson author leads the Shell Chair in Offshore Engineering research team at UWA, which is supported by Shell Australia.

We are grateful to Professor \nospellcheck{Yu} Wang for providing feedback and assistance in getting the BCS method running on our data sets.

\section*{Conflict of interest}

The authors report there are no competing interests to declare.

}\fi

\bibliography{references}

\section*{Supplementary material}

The remainder of this document contains supplementary material to the main paper. Section~\ref{sec:remaining_details_geowarp} gives extra details on GeoWarp's underlying hierarchical model. Section~\ref{sec:prediction_approximation} gives mathematical details on the approximate prediction distribution used in GeoWarp. Section~\ref{sec:additional_details_application} gives extra details on the application of GeoWarp to cone penetrometer test (CPT) data from Australia's North West Shelf (NWS). Section~\ref{sec:jaksa_benchmark} applies GeoWarp to an onshore benchmark dataset. Finally, Section~\ref{sec:additional_figures} provides additional figures and tables.

\appendix

\section{Remaining details on the GeoWarp model}
\label{sec:remaining_details_geowarp}

Section~\ref{sec:geowarp_model} in the paper describes GeoWarp's hierarchical model. Here, we give the remaining details of the model: Section~\ref{sec:geometric_warping_unit_prior} describes the prior distribution on the geometric warping unit, and Section~\ref{sec:spline_coefficient_prior} describes the prior distribution on the spline coefficients used in both the mean-profile process and the depth-varying variance process.

\subsection{Prior on geometric warping unit}
\label{sec:geometric_warping_unit_prior}

We assign a prior distribution to $\Avec$ and use the change-of-variables formula to obtain the corresponding prior distribution for $\Rvec$. The prior distribution we choose for $\Avec$ is the \nospellcheck{Lewandowski--Kurowicka--Joe} \citep[LKJ; ][]{lewandowskietal2009} prior, which is given by $p(\Avec) \propto |\Avec|^{\rho_R - 1}$, where $\rho_R > 0$ is a fixed and known shape parameter. After change-of-variables, the corresponding prior distribution for $\Rvec$ is given by $p(\Rvec) \propto \prod_{d = 1}^{D + 1} R_{dd}^{D - d + 2 \rho_R - 1}$, where $\{ R_{dd} : d = 1, \ldots, D + 1 \}$ are the elements of the diagonal of $\Rvec$; we denote this distribution by $\Rvec \sim \text{LKJ-R}(\rho_R)$. The value $\rho_R = 1$ corresponds to a uniform distribution over the correlation matrix $\Avec$, while for $\rho_R > 1$ the mode of the distribution is the identity matrix; in Section~\ref{sec:application}, we choose $\rho_R = 6$.

\subsection{Priors for spline coefficients}
\label{sec:spline_coefficient_prior}

The covariance matrices $\Sigmavec_\beta$ and $\Sigmavec_\zeta$ can be chosen to induce additional smoothness in the estimates of $\mu(h)$ and $\log \sigma_\delta^2(h)$, respectively. Below, we discuss $\Sigmavec_\beta$, but the same discussion applies to $\Sigmavec_\zeta$. We let $\Sigmavec_\beta = \sigma_\beta^2 \Cvec_\beta$, where $\sigma_\beta^2 > 0$ is unknown and $\Cvec_\beta$ is a covariance matrix of dimension $K_\beta \times K_\beta$. Denote by $c_{\beta, i, j}$ the $(i, j)$th entry of $\Cvec_\beta$. We consider two possible forms for $\Cvec_\beta$.  The first sets $c_{\beta, i, j} = \min(i, j)$, which corresponds to a random walk over the coefficients \citep{eilersetal1996}. The second sets $c_{\beta, i, j} = \exp(-|i - j| / \ell_\beta)$, where $\ell_\beta > 0$ is unknown, and this corresponds to exponential decay in the correlation between coefficients. To $\sigma_\beta^2$ we assign the prior $\mathrm{IG}(a_\beta, b_\beta)$ and to $\ell_\beta$ we assign the half-normal prior $\ell_\beta \sim \text{Half-N}(\sigma_\ell^2)$, where $a_\beta, b_\beta, \sigma_\ell^2 > 0$ are fixed hyperparameters. The half normal distribution has its mode at zero, and it tends to shrink the estimated length scale towards smaller values. The main difference between random walk and the exponential decay priors is their tendency for mean reversion: the random walk does not revert back to the process mean, while the exponential decay does.

In our application of GeoWarp to CPT data, we find that the random walk prior is more suitable for $\Sigmavec_\beta$, because it induces smoothness in the estimate of the mean profile while allowing it to differ as much as needed from the straight line $\alpha_0 + \alpha_1 h$ (see \eqref{eqn:mu_process}). By contrast, for $\Sigmavec_\zeta$, we find that the exponential decay prior is more suitable. This prior also induces smoothness in the estimate of $\log \sigma_\delta^2(h)$, but does not allow persistent deviations from the mean of the process, $\eta$.

\section{Approximate prediction distribution}
\label{sec:prediction_approximation}

Recall from Section~\ref{sec:prediction} in the manuscript that $\Yvec^* \equiv (Y(\svec_1^*, h_1^*), \ldots, Y(\svec_{N_\mathrm{pred}}^*, h_{N_\mathrm{pred}}^*))'$ is a vector of unknowns, where the locations and depths to be predicted are $\svec_i^* \in S$ and $h_i^* \in [0, h_\mathrm{max}]$ for $i = 1, \ldots, N_\mathrm{pred}$. Let $\Xvec^*$ be an $N_\mathrm{pred} \times (K_\beta + 2)$ matrix defined analogously to $\Xvec$ but for the unsampled locations. Define $\Wvec \equiv (\Zvec', (\Yvec^*)')'$, and let
\begin{equation}
  \Qvec_W \equiv \begin{pmatrix}
    \Qvec_{Z Z} & \Qvec_{Z*} \\
    \Qvec_{*Z} & \Qvec_{**}
  \end{pmatrix}
  \equiv \mathrm{Var}(\Wvec)^{-1}
\end{equation}
be the precision matrix of $\Wvec$, partitioned into submatrices corresponding to $\Zvec$ and $\Yvec^*$. Then the prediction distribution of $\Yvec^*$ given the data and the parameters is a multivariate Gaussian distribution, $(\Yvec^* \mid \Zvec, \omegavec, \kappavec, \gammavec, \Rvec, \sigma_\epsilon^2) \sim \mathrm{Gau}(\tilde{\Yvec}^{*}, \Qvec_{**}^{-1})$, where
\begin{equation}
  \tilde{\Yvec}^*
  \equiv
    \Xvec^* \omegavec - \Qvec_{**}^{-1} \Qvec_{*Z} (\Zvec - \Xvec \omegavec).
  \label{eqn:prediction_equation}
\end{equation}
The matrix $\Qvec_W$ and its submatrices are generally too large to be used directly, so we apply the Vecchia approximation described in Section~\ref{sec:vecchia_approximation} to the distribution of $\Wvec$. This yields a sparse approximation to $\Qvec_W$ and its submatrices \citep[see][for details on how to calculate these approximations]{katzfussetal2020}. Using this approximation we then calculate $\tilde{\Yvec}^*$ via \eqref{eqn:prediction_equation}, and draw samples from the approximate full conditional distribution of $\Yvec^*$. This step depends on the parameters $\omegavec$, $\kappavec$, $\gammavec$, $\Rvec$, and $\sigma_\epsilon^2$, which are unknown. When using MCMC, one sample of $\Yvec^*$ is drawn for each MCMC sample of the parameters. When using MAP estimation, the samples are therefore drawn from the approximate prediction distribution $\tilde{p}(\Yvec^* \mid \Zvec, \hat{\omegavec}, \hat{\kappavec}, \hat{\gammavec}, \hat{\Rvec}, \hat{\sigma}_\epsilon^2)$.

As in Section~\ref{sec:vecchia_approximation}, when doing prediction the choice of ordering and parents for the Vecchia approximation is important. We first order the entries of $\Wvec$ by randomly reordering its sub-vectors, $\Zvec$ and $\Yvec^*$, thereby ensuring that all entries of $\Zvec$ precede those of $\Yvec^*$. We use the same parent scheme for $\Zvec$ as described in Section~\ref{sec:vecchia_approximation}, while the parents for $\Yvec^*$ are split into two halves, the first of which are parents that come from the data, $\Zvec$, and the second of which come from $\Yvec^*$ itself. The $\Zvec$ parents of $\Yvec^*$ are chosen by taking from each CPT the measurements closest in depth to that associated with the entry in $\Yvec^*$. These parents are distributed as evenly as possible across all CPTs. This ensures that both vertical and horizontal dependencies are captured in the approximation. The within-$\Yvec^*$ parents are chosen based on the sparsity pattern of the prediction locations: when they are vertically dense and horizontally sparse (such as when predicting the process at the locations of withheld CPTs) we use the same scheme as for $\Zvec$; when they are regular (such when $\Yvec^*$ is on a grid) we use the standard nearest neighbors scheme.

\section{Additional details of the application to NWS CPT data}
\label{sec:additional_details_application}

Section~\ref{sec:application} in the main paper describes how GeoWarp is applied to CPT data from six sites on Australia's North West Shelf. This section gives additional details on the data (Section~\ref{sec:detailed_cpt_data}), details the configuration of GeoWarp (Section~\ref{sec:model_for_nws_data}), discusses the fitting procedure (Section~\ref{sec:fitting_procedure}), compares the MAP and MCMC estimates of GeoWarp (Section~\ref{sec:map_mcmc_comparison}), details the configuration of the other methods in the cross-validation study (Section~\ref{sec:other_methods_configuration}), and evaluates the cross-validation performance of more parsimonious versions of GeoWarp (Section~\ref{sec:parsimonious_versions}).

\subsection{Detailed description of the data}
\label{sec:detailed_cpt_data}

The sediments at field A are primarily sandy silt, with sporadic, thin layers of cemented calcarenite. Field B consists mainly of high plasticity carbonate muddy silts, silty muds, and sporadic thin lenses of carbonate silty/muddy sands. Horizontal coordinates of the CPTs for fields A and B are provided in Figures~\ref{fig:map_a} and \ref{fig:map_b} in Section~\ref{sec:additional_figures}, shifted so that the coordinate $(0, 0)'$ is at the lower left-hand corner of the each site. We have also constructed two 2-D datasets, labeled A2-T and B2-T, constructed from five CPTs that lie along a horizontal transect at sites A2 and B2, respectively. These cases are examined as geotechnical investigations sometimes occur along transects, such as prior to laying subsea pipelines or installing cables. We excluded the top 25 cm of all CPT data since measurements at these shallow depths are generally less representative of the sediments. The CPT data for fields A and B are presented in Figures~\ref{fig:datasets_a} and \ref{fig:datasets_b} in Section~\ref{sec:additional_figures}.

\subsection{Model for NWS data}
\label{sec:model_for_nws_data}

We describe here the choices made to make the GeoWarp model suitable to the NWS data in Section~\ref{sec:application}. Section~\ref{sec:model_configuration} describes the model configuration. Section~\ref{sec:prior_distributions} describes the prior distributions. Finally, Section~\ref{sec:choice_of_parents} discusses the choice of the number of parents in the Vecchia approximation.

\subsubsection{Model configuration}
\label{sec:model_configuration}

The configuration we use for GeoWarp is given in Table~\ref{tab:geowarp_configuration}. The knots for the splines in the mean profile and the depth-varying process are spaced so as to resolve features at fine scales. We set the smoothness of the deviation process to $\nu = 3 / 2$, a choice we justify below. The AWUs for the horizontal dimensions ($d = 1, 2$) are set to a linear warping ($L_1, L_2 = 2$) because the sites we use have too few horizontal locations to estimate a non-linear warping, and because we expect more variability vertically (a consequence of the layering of the sediments). For the vertical dimension, the choice $L_3 = 20$ allows us to resolve non-linear vertical features at a resolution of roughly 2 m over the 41 m of the CPTs.

To pick a value for $\nu$, we perform an exploratory analysis in which we fit models independently to each of the 70 CPTs across all six sites. At each site we computed the average value of $\log q_c$ in each 0.1 m bin (the same as was done for the binned statistics in Section~\ref{sec:cross_validation}). This is meant as an empirical approximate of the vertical mean-profile process at each site. We then subtracted this average from the $\log q_c$ values in each bin at each CPT, resulting in one residual vector per CPT. We denote this vector and its entries by $\Rvec_i \equiv (R_{i, 1}, \ldots, R_{i, N_i})'$ for $i = 1, \ldots, M$, where each entry $R_{i, j}$ corresponds to a measurement $Z_{i, j}$; $\Rvec_i$ can be considered an estimate of the sum of the measurement errors and the deviation process for the $i$th CPT.

To each $\Rvec_i$, $i = 1, \ldots, M$, and for each value $\nu = 1 / 2$, $3 / 2$, $5 / 2$, and $7 / 2$, we used the maximum likelihood method to independently fit a Gaussian process with zero mean and the following 1-D covariance model:
\begin{equation}
  \mathrm{Cov}(R_{i, j}, R_{i, k}) = \tau_1^2 \mathcal{M}_\nu(\upsilon |h_{i, j} - h_{i, k}|) + \tau_2^2 \mathbbm{1}(j = k),
  \quad j, k = 1, \ldots, N_i,
\end{equation}
where $\tau_1^2, \tau_2^2, \upsilon > 0$ are unknown parameters. These fits were performed using the R software package \nospellcheck{GpGp} \citep{guinessetal2021b}, which uses the methods given by \citet{guinness2018} we also used in Section~\ref{sec:inference_computation} to estimate the full GeoWarp model.

For each fit to each CPT we found the likelihood at the maximum likelihood estimate of the parameters. The values $\nu = 1 / 2$, $3 / 2$, $5 / 2$, and $7 / 2$ gave the highest likelihood 15, 53, 2, 0, and 0 times, respectively, and so we chose the value $\nu = 3 / 2$ for our analyses.

\subsubsection{Prior distributions}
\label{sec:prior_distributions}

The prior distributions we use for GeoWarp are given in Table~\ref{tab:geowarp_configuration}; here we justify these choices.

The prior distributions of the $\gamma$s for the horizontal dimensions are set to inverse-uniform distributions between $\gamma_d^- = 1 / 200$ m$^{-1}$ and $\gamma_d^+ = 1 / 0.5$ m$^{-1}$, corresponding to horizontal length scales between 0.5 m and 200 m. The lower bound rules out unrealistically short-scale dependencies, while the upper bound is large compared to the size of the fields. We find this modestly informative prior to be necessary due to the limited number of horizontal locations at some sites.

For the smallest sites A3 and B3, which have 7 and 5 CPTs, respectively, we further restrict the AWUs to be identical for the easting and northing directions; without this modification, GeoWarp tends to infer an unrealistically large difference between the easting and northing $\gamma$'s, which we attribute to the sparse horizontal data at these sites.

For the AWU for the vertical dimension, the prior distribution on the $\gamma$s for $d = 3$ is uninformative but has a non-zero mode; this is necessary to prevent non-existence of the posterior mode.

For the geometric warping unit, the prior distribution is set to induce shrinkage towards isotropy without excluding realistic patterns of anisotropy through the choice $\rho_R = 6$.

The prior for $\sigma_\epsilon^2$ has 5th and 95th percentiles of 0.1 and 1; this is a diffuse prior for these data. Similarly, the priors on $\sigma_\beta^2$ and $\sigma_\zeta^2$ imply diffuse priors with 5th and 95th percentiles equal to $10^{-6}$ and 100, respectively. We let $\sigma_\ell^2 = 1$ when setting the prior on $\ell_\zeta$.

\subsubsection{Choice of number of parents}
\label{sec:choice_of_parents}

To make an appropriate choice for the number of parents to use with GeoWarp in Section~\ref{sec:application}, we performed MAP estimation for each of the six 3-D sites and the two 2-D sites using a maximum of 25 parent, 50 parents (the value we choose in the main paper), and 100 parents. All other aspects of the model (described in Section~\ref{sec:model_fits_and_predictions}) and the fitting procedure (described in Section~\ref{sec:fitting_procedure}) were unchanged. The fastest and slowest running times for the MAP estimation procedure occur for the A3 and B1 sites, respectively, and the timings are as follows:
\begin{center}
  \begin{tabular}{r|rr}
    \hline \hline
    \# parents & Fastest (A3) & Slowest (B1) \\
    \hline
    25 & 13 minutes & 97 minutes \\
    50 & 19 minutes & 171 minutes \\
    100 & 40 minutes & 451 minutes  \\
    \hline
  \end{tabular}
\end{center}

For all sites and for each choice of the number of parents, Figure~\ref{fig:vertical_profiles_n_parents} shows the estimated mean profile, $\mu(h)$, $h \in [0, h_\mathrm{max}]$ (left-most panel), the estimated standard deviation of the deviation process, $\sqrt{\sigma_\delta^2(h)}$ (middle panel), and the estimated derivative of the vertical warping, $\textrm{d}u_3^*(h) / \textrm{d}h$ (right-most panel). The derivative of the warping is shown because this quantity is inversely related to the vertical correlation of the deviation process. All quantities shown are very similar between the fits with 25 parents, 50 parents, and 100 parents.

To characterize the dependencies in each fit, Figure~\ref{fig:isocorrelation_horizontal_n_parents} shows iso-correlation contours in the easting--northing plane, Figure~\ref{fig:isocorrelation_vertical_n_parents_easting} shows iso-correlation contours in the easting--depth plane at various depths, and Figure~\ref{fig:isocorrelation_vertical_n_parents_northing} shows iso-correlation contours in the northing--depth plane at various depths. Within each site, the contours are very similar between the fits with 25 parents, 50 parents, and 100 parents. The only exception is A1, where the iso-correlation contours are different when 25 parents are used.

Given the similarity in the fits between 50 and 100 parents, and the fact that fitting the model with 50 parents is much faster, we choose to use 50 parents for the analyses in Section~\ref{sec:application}.

\subsection{Fitting procedure details and timings}
\label{sec:fitting_procedure}

The GeoWarp model with the configuration as detailed in Section~\ref{sec:model_for_nws_data} was fitted using both MCMC and MAP estimation to each of the six 3-D sites and the two 2-D sites. To estimate the parameters using MCMC, four MCMC chains were run for each site using the No-U-Turn Sampler (NUTS) implemented in Stan \citep{hoffmangelman2014}. The MCMC chains were also initialized at random, and were run for 2,000 iterations total, the first 1,000 of which were discarded as burn-in. Convergence of the MCMC sampler was confirmed visually using traceplots. To find the MAP estimates, an optimization routine was run 10 times from random initial parameter values in order to avoid local optima; the best estimate was chosen based on the value of the approximate log posterior density.

All routines were run on a computer with two 28-core Intel Xeon Gold 6348 central processing units, and each run of the software (whether for MCMC or optimization), used 14 threads. The fastest and slowest running times for the procedures occur for the A3 and B1 sites, respectively, and the timings are as follows:
\begin{center}
  \begin{tabular}{r|rr}
    \hline \hline
    Method & Fastest (A3) & Slowest (B1) \\
    \hline
    MCMC & 443 minutes & 2218 minutes \\
    MAP & 19 minutes & 171 minutes \\
    \hline
  \end{tabular}
\end{center}

\subsection{Comparison of MAP and MCMC estimates}
\label{sec:map_mcmc_comparison}

Here we contrast the MAP estimates for GeoWarp to those from MCMC. The estimated mean profile, $\mu(h)$, the estimated standard deviation of the deviation process, $\sqrt{\sigma_\delta^2(h)}$, and the estimated derivative of the vertical warping, $\textrm{d}u_3^*(h) / \textrm{d}h$, for both MAP and MCMC are shown in Figure~\ref{fig:vertical_profiles_full}. The estimated MAP and MCMC estimates of all profiles are very similar at all sites, with the exception of the standard deviation at B3. At B3, MAP infers constant variance with depth, while MCMC does not. This suggests that the posterior mode for this quantity is in a region of low posterior volume.

Figure~\ref{fig:isocorrelation_horizontal} shows estimates of iso-correlation contours in the easting--northing plane for both MAP and MCMC for all sites. These are almost identical between MAP and MCMC for sites A1 and A2. By contrast, the MAP length-scale estimates for A3, B1, B2, and B3 are large relative to where the bulk of the posterior mass is. This tendency is likely due to the parameterization of the horizontal AWUs that use inverse length scale parameters $\gamma_{d, 1}$, $d = 1, 2$, for the easting and northing directions, respectively. Figure~\ref{fig:isocorrelation_horizontal} shows that the posterior distributions of the horizontal length scales are relatively flat over their plausible ranges. As a consequence, the modes of the \emph{inverse} length scales occur towards the bottom of their plausible ranges (corresponding to the tops of the plausible length scale ranges). Figure~\ref{fig:gamma_horizontal_marginal_posterior} illustrates this by showing the estimated marginal posterior distribution of $\gamma_{1, 1}$ (constrained to equal $\gamma_{2, 1}$) at site B3.

Figure~\ref{fig:isocorrelation_vertical_easting} and Figure~\ref{fig:isocorrelation_vertical_northing} show estimates of iso-correlation contours for the MAP estimates and for MCMC in the easting--depth and northing--depth planes, respectively, at various depths, and for all sites. Inferences from MAP and MCMC are very similar for all sites, with the exception of B3. This shows that the vertical length scales are generally well-identified by the data, which are dense in the vertical direction. The exception, B3, has longer vertical length scales in the MAP estimates compared to the MCMC estimates at depths above 18 m; these depths are also where the MAP estimate of the depth-varying variance is most different to the MCMC estimates.

We also contrast the MAP and MCMC predictions at unknown locations. We do this for the A2 and B2-T sites, the same sites as were discussed in Section~\ref{sec:prediction_results} of the main paper. We calculated the posterior mean and standard deviation for both methods, and at each location calculated the difference between the posterior means and the ratio between the posterior standard deviations. The results are shown in Figure~\ref{fig:predicted_a2_mcmc} for the A2 site and in Figure~\ref{fig:predicted_b2t_mcmc} for the B2-T site. At A2 the MAP and MCMC predictions are very similar; the only significant difference is that prediction standard deviations are around 5\% smaller for MAP than for MCMC. For B2-T, the predictions are again similar, but the inferred horizontal length scales appear shorter for MCMC than for MAP. This appears in the posterior means as more rapid regression in the MCMC predictions towards the site-wide mean profile, and in the posterior standard deviation as a larger variance in the MCMC predictions when close to the locations of the CPTs. This is consistent with the finding that the MAP estimates of the horizontal length scales tend relatively large. However, the prediction distributions for MAP and MCMC are almost identical away from CPTs, so that the practical impact of this difference in a geotechnical engineering context is expected to be minor.

\subsection{Configuration of other methods in the cross-validation study}
\label{sec:other_methods_configuration}

Predictions for the 0.1m depth-binned statistics are constructed from each bin's observations: the empirical mean of the bin is the point forecast, while the empirical distribution function (ECDF) of the measurements in each bin quantifies the variability.

We apply the BCS method to our data using the authors' software package, \nospellcheck{Analytics of Sparse Spatial Data} based on Bayesian compressive sampling/sensing \citep[\nospellcheck{ASSD-BCS};][]{lyuetal2023b}. For the fits to the 3-D sites, we used an easting--northing--depth grid with a resolution of 8 m $\times$ 8 m $\times$ 0.2 m, while for the 2-D site the horizontal--depth grid is 1 m $\times$ 0.1 m. These grid sizes reflect computational considerations; we were unable to get the \nospellcheck{ASSD-BCS} software to work with finer grid sizes. All other options in the software were left at their default values.

\subsection{Parsimonious versions of GeoWarp}
\label{sec:parsimonious_versions}

The GeoWarp model described in Section~\ref{sec:geowarp_model} has a large number of parameters, which may be difficult to estimate when there are few CPTs, as in the B2-T site. It is also of interest to investigate how the different features of the GeoWarp model impact its predictions. We therefore assess five additional variants of the GeoWarp model in which different features of GeoWarp are enabled or disabled.

The first four model variants constitute a $2 \times 2$ factorial design testing the roles of warping and of depth-varying variances in GeoWarp:
\begin{itemize}
  \item
    \textbf{GeoWarp}: This is the full model described in the text.

  \item
    \textbf{GW-NoWarp}: Here we omit the geometric warping unit, and set all AWUs to a linear warping; this corresponds to a covariance kernel with separate, fixed length scales in the horizontal and vertical directions. Variance is still depth-varying.

  \item
    \textbf{GW-CV}: Here the deviation has constant variance instead of depth-varying variance (hence the suffix CV), which we achieve by setting $K_\zeta = 0$. The warping structure is retained.

  \item
    \textbf{GW-NoWarp-CV}: This variant combines a lack of warping with constant variance. This is therefore a standard stationary kernel with constant variance and separate length scales for each dimension.
\end{itemize}

The final two variants assess the impact of omitting from the deviation process either just horizontal, or both horizontal and vertical dependencies, respectively:
\begin{itemize}
  \item
    \textbf{GW-Vert-CV}: This variant again has constant variance, and the covariance function of its deviation process is
    \begin{equation}
      \mathrm{Cov}(\delta(\svec_1, h_1), \delta(\svec_2, h_2))
      = \sigma_\delta^2
        \mathbbm{1}(\svec_1 = \svec_2)
        \mathcal{M}_\nu(d_h),
      \label{eqn:deviation_vertical_only}
    \end{equation}
    where $\sigma_\delta^2$ is the constant variance inferred for the process, $\mathbbm{1}(\cdot)$ is an indicator function, $d_h \equiv |g_v(h_1) - g_v(h_2)|$, and the warping function $g_v(\cdot)$ applies only to the vertical dimension. Equation~\eqref{eqn:deviation_vertical_only} can be regarded as a limiting case of \eqref{eqn:deviation_covariance} where $S$ is stretched out to the extent that horizontal correlations become negligible. The warping function in this case consists only of an axial warping unit in the vertical dimension.

  \item
    \textbf{GW-WN-CV}: Here we again assume constant variance, and further set the deviation process to correspond to white noise, so that there is no spatial dependence in any dimension.
\end{itemize}
For GW-Vert-CV and GW-WN-CV, which lack horizontal dependencies in the deviation process, the predictive mean is equal to the estimate of the mean-profile process, $\mu(h)$. In this sense these variants are similar to the binned statistics method introduced in Section~\ref{sec:cross_validation} of the main paper.

Table~\ref{tab:cross_validation_metrics_full} expands Table~\ref{tab:cross_validation_metrics} from the main paper to show the performance of the model variants. As before, for each site and metric, the best performing model is marked in bold, as are those whose performance is not significantly different to the best model according to a one-sided two-sample $t$-tests at $\alpha = 0.05$. This table is discussed in Section~\ref{sec:discussion} of the main paper.

At most metrics and sites, GeoWarp has the best or equivalent-best performance relative to the other model variants. For MSE, the best model overall is GW-NoWarp-CV, and GeoWarp is second-best due to under-performing (relative to GW-NoWarp-CV) at one site. GeoWarp is the best overall for the other metrics. This suggests that the warping and depth-varying variances are crucial for uncertainty quantification, but less so for pure point prediction; there is little to distinguish between the GeoWarp variants for point prediction, except for GeoWarp-NoWarp, which performs notably worse.

Among the features of GeoWarp, the most crucial for performance is the mean profile process, as can be seen by the gap in performance between the linear model and the GeoWarp variants across most sites and metrics. This is consistent with the competitive performance of the binned statistics method on many metrics, which also quantifies the depth-varying mean of the data. The warping, depth-varying variances, and full dependency structure are collectively important to performance in most metrics for the sites in the A field, particularly for uncertainty quantification, though less so for those in the B field.

The variants GW-Vert-CV and GW-WN-CV (which lack horizontal dependencies) are competitive for the B field even on metrics that account for uncertainty quantification; this suggests that these sites lack strong horizontal structure beyond the mean profile. For these sites, GW-WN-CV, which lacks even vertical dependencies, is competitive with GW-Vert-CV on most metrics. However, the models with full dependency structures have the advantage that they yield more realistic conditional simulations; these simulations may then be used as inputs to subsequent design processes.

Table~\ref{tab:cross_validation_metrics_full} also suggests that, if a model with depth-varying variance is considered necessary, it is advisable to also include warping. By contrast, model performance is generally reasonable with or without warping under constant variance. This can be seen because the model variant GW-NoWarp, which has depth-varying variances but no warping, performs worst on average (the `All' column) among the GeoWarp variants on all metrics but DSS2, where it is second-last. We found that this is because GW-NoWarp sometimes produces unrealistically large estimates of the depth-varying variance profile.

At B2-T, the site where GeoWarp performed poorly on some metrics, the parsimonious variants with constant variance all performed better than GeoWarp, although the linear model was still best for the Int05 metric. This suggests that, for sites with limited data, constant variance may be a good simplifying assumption to make.

\section{Application of GeoWarp to a benchmark dataset}
\label{sec:jaksa_benchmark}

To evaluate GeoWarp's ability to infer 3-D site structures different from the NWS fields studied in Section~\ref{sec:application}, we performed cross-validation using a dataset of CPTs collected at South Parklands, Adelaide, South Australia. These CPTs were initially collected and analyzed by \citet{jaksa1995}. From the full collection of CPTs collected by \citet{jaksa1995}, \citet{phoonetal2022b} chose 18 CPTs in a 20 m $\times$ 20 m $\times$ 5 m spatial domain to be a benchmark real-world dataset to evaluate methods for data-driven geotechnical site characterization. Figure~\ref{fig:jaksa_data} shows the locations and values of these 18 CPTs. The site characteristics at South Parklands differ substantially from the NWS sites: the site is terrestrial rather than seabed, it is predominantly clay soil instead of compacted silt, it has smaller horizontal distances between CPTs, and it has a shallower depth of 5 m compared to 40 m for the NWS sites. These make it a suitable benchmark for evaluating GeoWarp's ability to generalize to data with different characteristics.

We performed cross-validation with the South Parklands data using the same GeoWarp configuration as for the NWS data, but we removed the knots before and after the target depths in the basis-function representation for $\log \sigma_\delta^2(h)$ in \eqref{eqn:sigma_process}. We omitted these boundary knots because including them led to unrealistic variance estimates near the top and bottom of the column, likely due to insufficient information in the 5 m column to constrain the coefficients outside the boundary. In the cross-validation also included the parsimonious GeoWarp variants described in Section~\ref{sec:parsimonious_versions}.

We calculated the same performance metrics as in Section~\ref{sec:cross_validation}, shown in Table~\ref{tab:cross_validation_metrics_jaksa}. To determine whether any models had equivalent-best performance, we conducted one-sided two-sample $t$-tests at $\alpha = 0.05$ in the same manner as in Section~\ref{sec:parsimonious_versions}, but the best model for each metric was always significantly different to the other models. The full GeoWarp model outperformed all other methods, including the state-of-the-art BCS method, on all metrics except Int05, where the the depth-binned statistics performed the best. Overall, GeoWarp appears to generalize well to the South Parklands data despite the differences in site characteristics.

\newpage

\section{Additional figures and tables}
\label{sec:additional_figures}

\setcounter{figure}{0}
\renewcommand{\theHfigure}{B\arabic{figure}}
\renewcommand{\thefigure}{B\arabic{figure}}
\renewcommand{\thetable}{B\arabic{table}}

\begin{figure}[ht!]
  \begin{center}
    \includegraphics{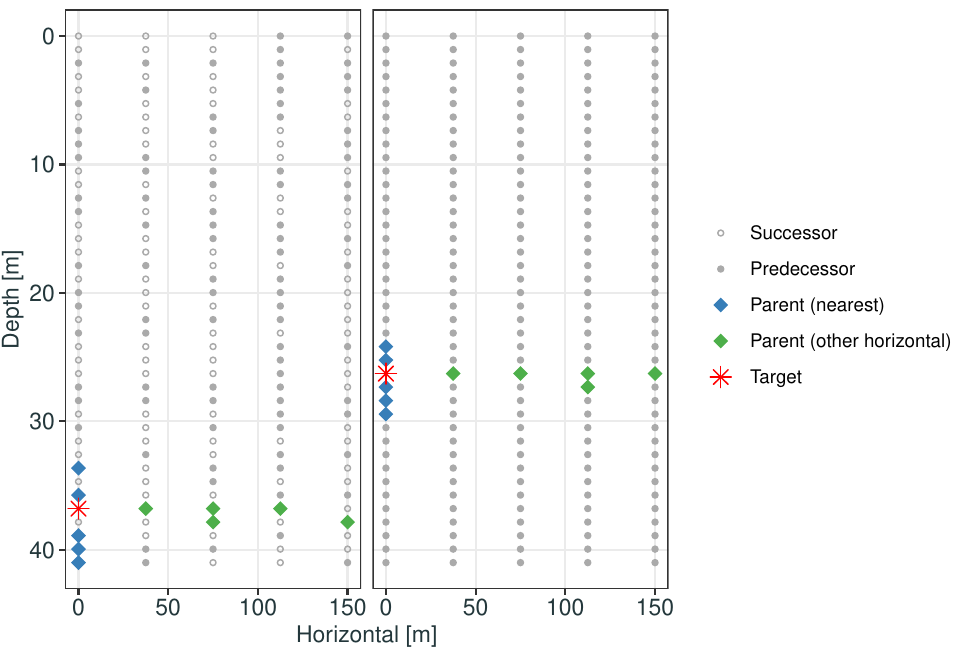}
  \end{center}
  \caption{
    Example of the parent-selection scheme described in Section~\ref{sec:vecchia_approximation} with $m = 10$ when applied to simulated locations of five CPTs along a 2-D transect. The parents, predecessors, and successors for two target points (in red) are shown in the two panels. The parents chosen from their nearest predecessors are in blue, while the parents chosen from predecessors at different horizontal locations are in green.
  }
  \label{fig:parent_example}
\end{figure}

\begin{figure}[ht!]
  \begin{center}
    \includegraphics{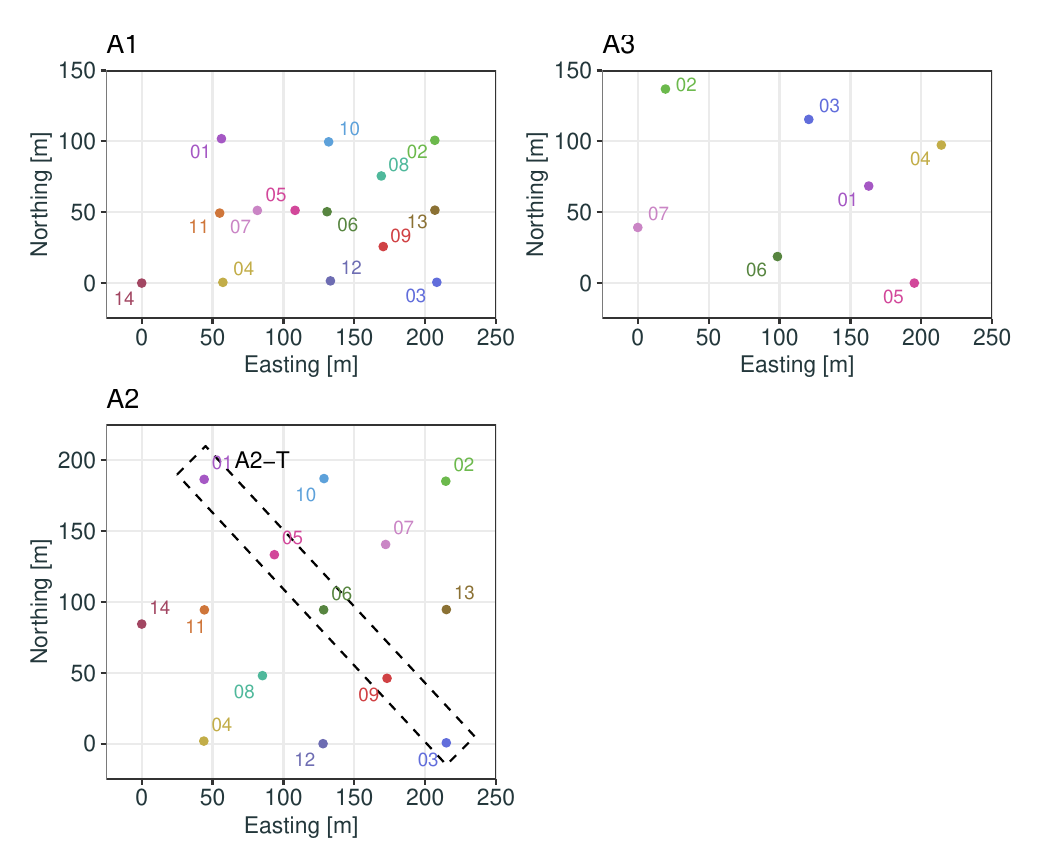}
  \end{center}
  \caption{
    Maps of the horizontal coordinates for each of the CPTs taken at the sites in the A field. The sites from A2 that comprise the 2-D transect site A2-T are shown in the map for A2.
  }
  \label{fig:map_a}
\end{figure}

\begin{figure}[ht!]
  \begin{center}
    \includegraphics{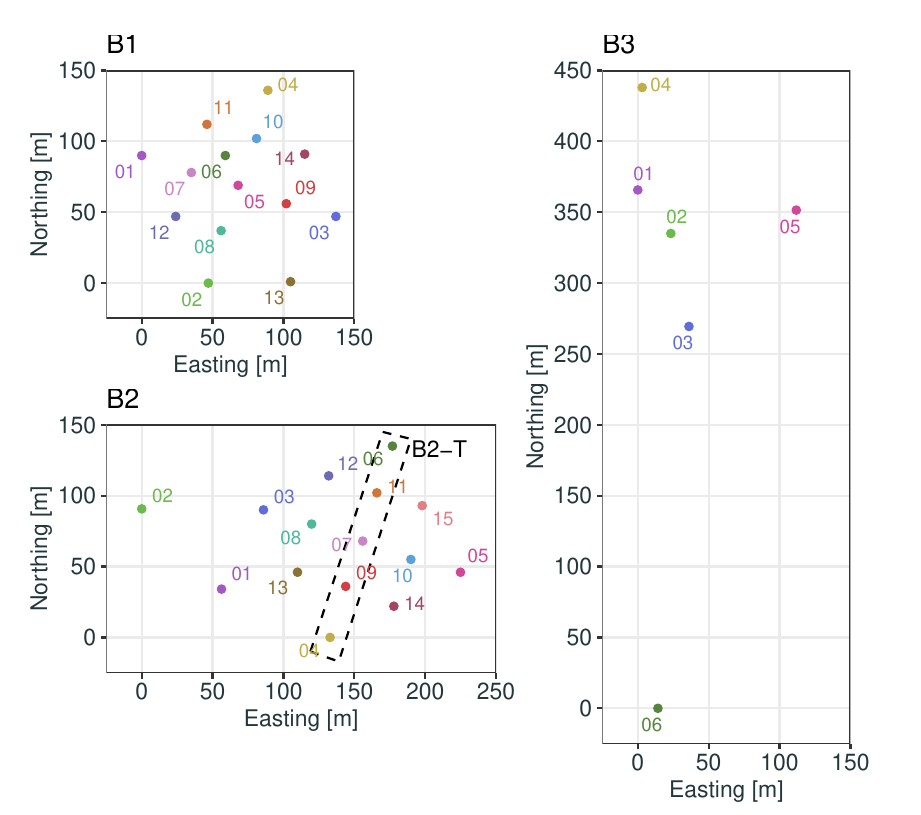}
  \end{center}
  \caption{
    As in Figure~\ref{fig:map_a}, but for the sites from the B field.
  }
  \label{fig:map_b}
\end{figure}

\begin{figure}[ht!]
  \begin{center}
    \includegraphics{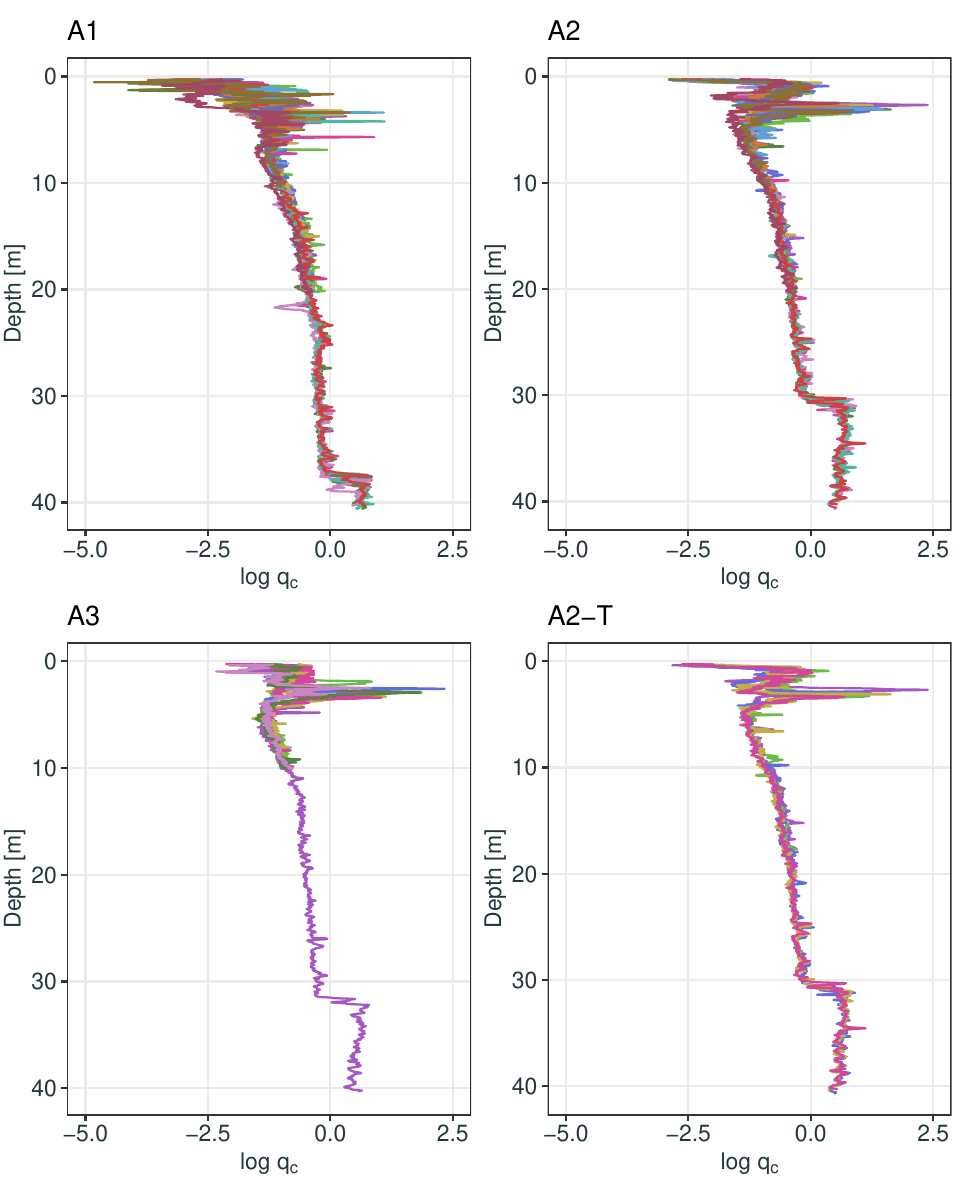}
  \end{center}
  \caption{
    Measurements of $\log q_c$ from CPTs taken at sites in the A field. Each panel shows CPTs for a different site, and labels and colors identify each CPT. A2-T is a special case consisting of five CPTs from the A2 field that lie along a transect line. The colors in this figure match those in Figure~\ref{fig:map_a}.
  }
  \label{fig:datasets_a}
\end{figure}

\begin{figure}[ht!]
  \begin{center}
    \includegraphics{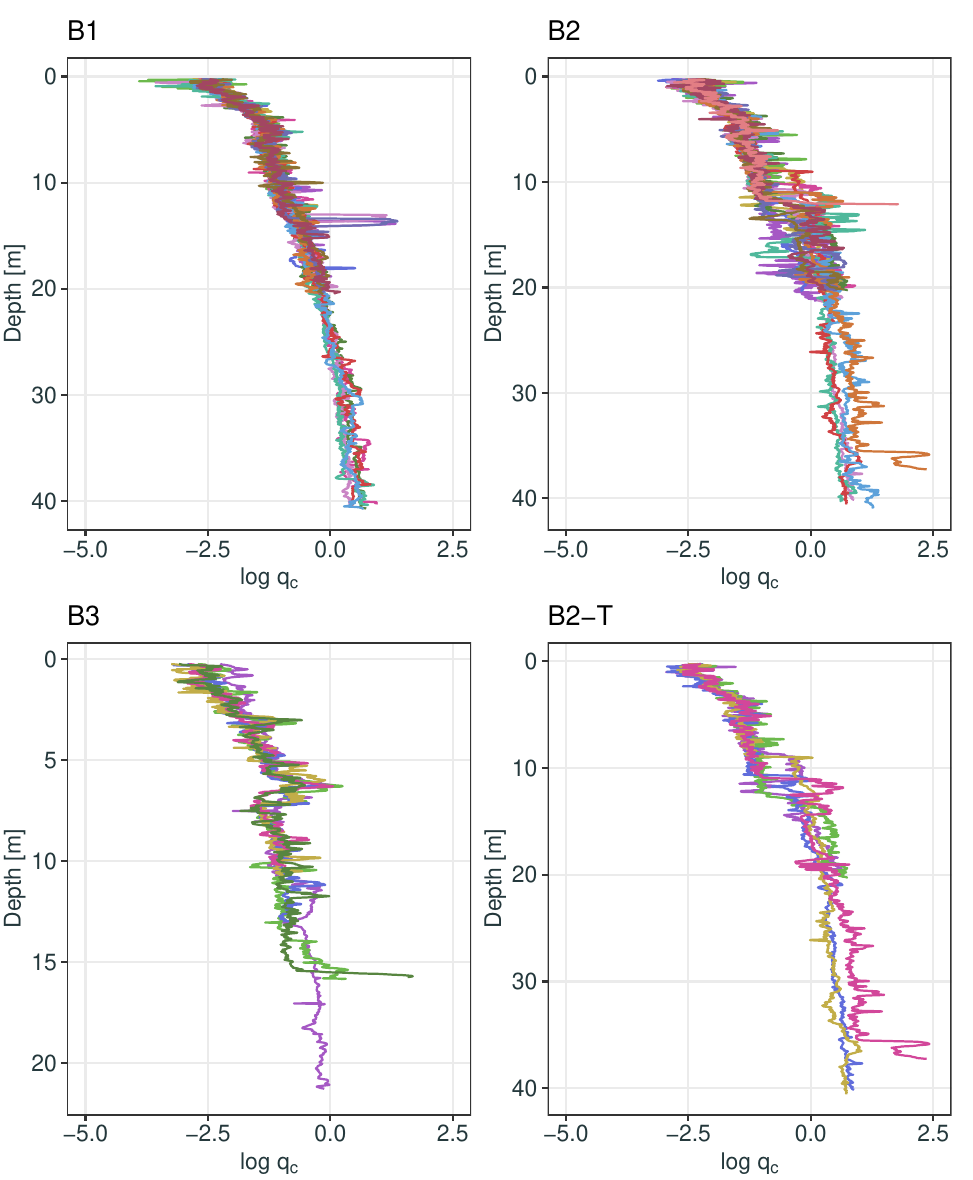}
  \end{center}
  \caption{
    As in Figure~\ref{fig:datasets_a}, but for sites in the B field. B2-T is a special case consisting of five CPTs from the B2 field that lie along a transect line. The colors in this figure match those in Figure~\ref{fig:map_b}.
  }
  \label{fig:datasets_b}
\end{figure}

\begin{figure}[ht!]
  \begin{center}
    \includegraphics{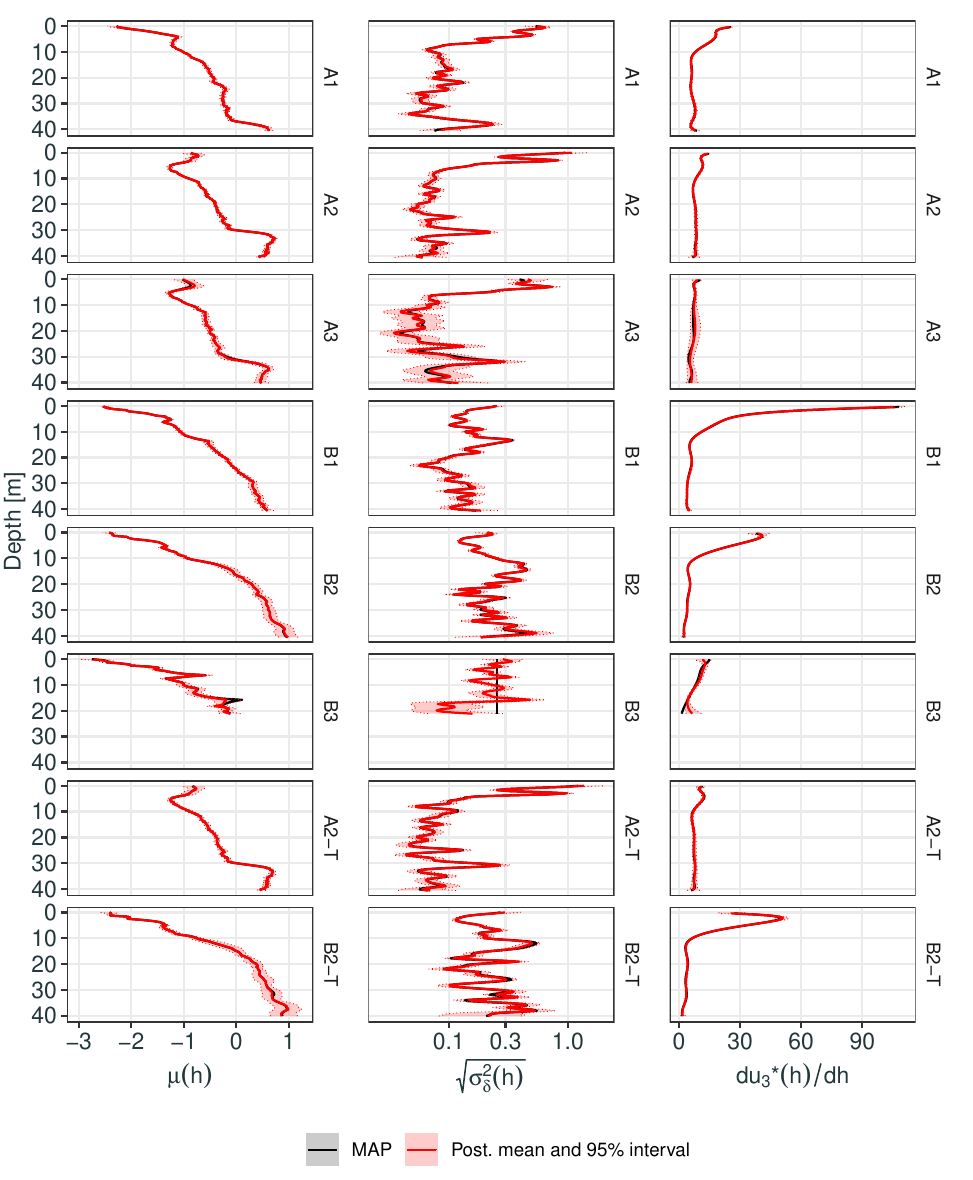}
  \end{center}

  \caption{
    MAP (black) and posterior mean and 95\% interval (red) estimates of the mean profile ($\mu(h)$, left), the vertical standard deviation profiles ($\sqrt{\sigma_\delta^2(h)}$, middle, log scale), and the derivative of the vertical warping ($\textrm{d}u_3^*(h) / \textrm{d}h$, right), for each of the six 3-D sites (first six rows), and the two 2-D sites (last two rows).
  }
  \label{fig:vertical_profiles_full}
\end{figure}

\begin{figure}[ht!]
  \begin{center}
    \includegraphics[width=16.5cm]{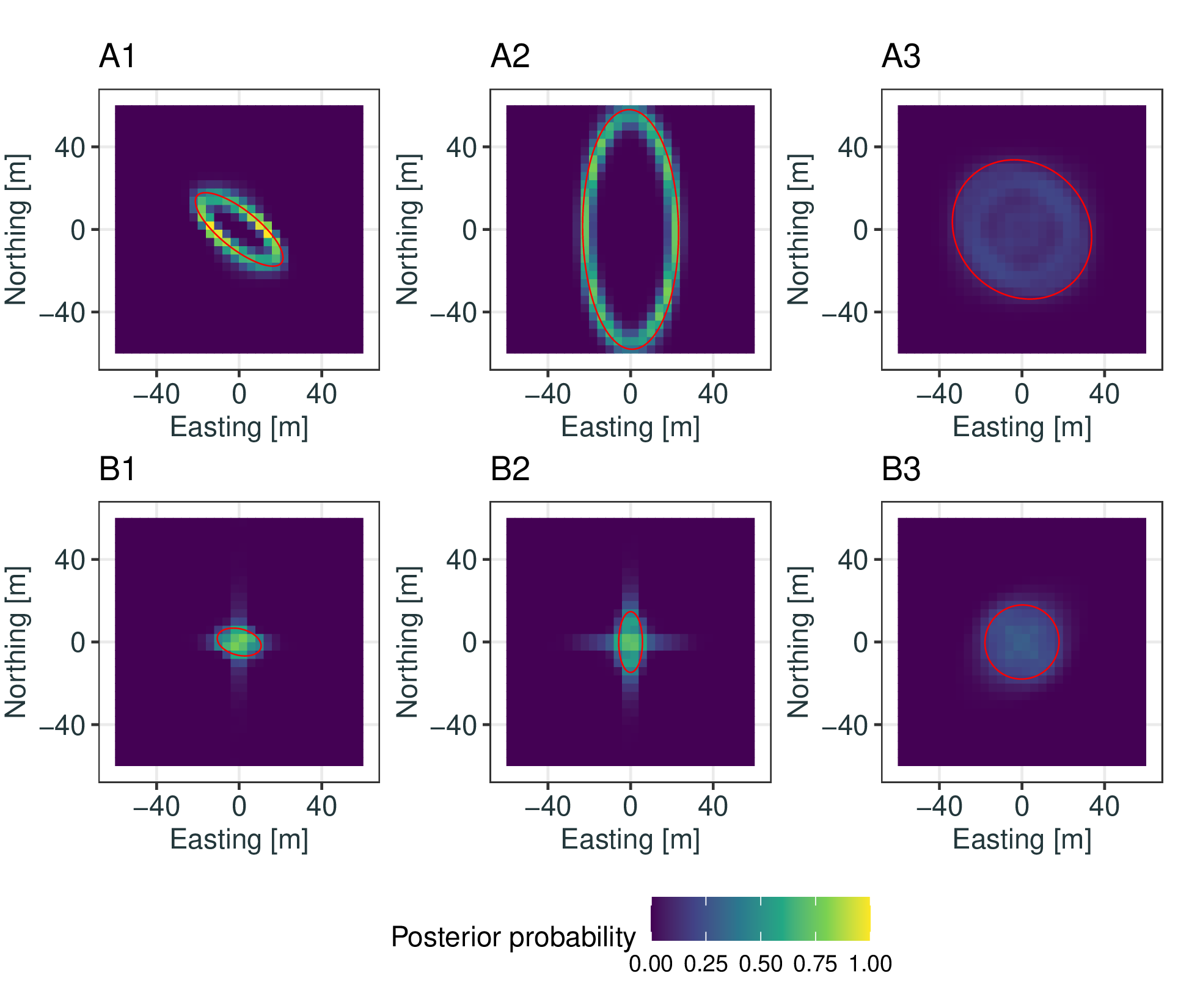}
  \end{center}

  \caption{
    Posterior distributions of iso-correlation contours showing the easting and northing coordinates for which the correlation in the deviation process with the point $(0, 0)'$ is equal to 0.5 at a fixed depth. The posterior distribution is visualized over a 30 $\times$ 30 grid, where the color of each grid cell represents the posterior probability that the iso-correlation contour passes through that cell. The MAP estimate of the iso-correlation contour is also shown in red. Note that these contours do not vary with depth.
  }
  \label{fig:isocorrelation_horizontal}
\end{figure}

\begin{figure}[ht!]
  \begin{center}
    \includegraphics[width=16.5cm]{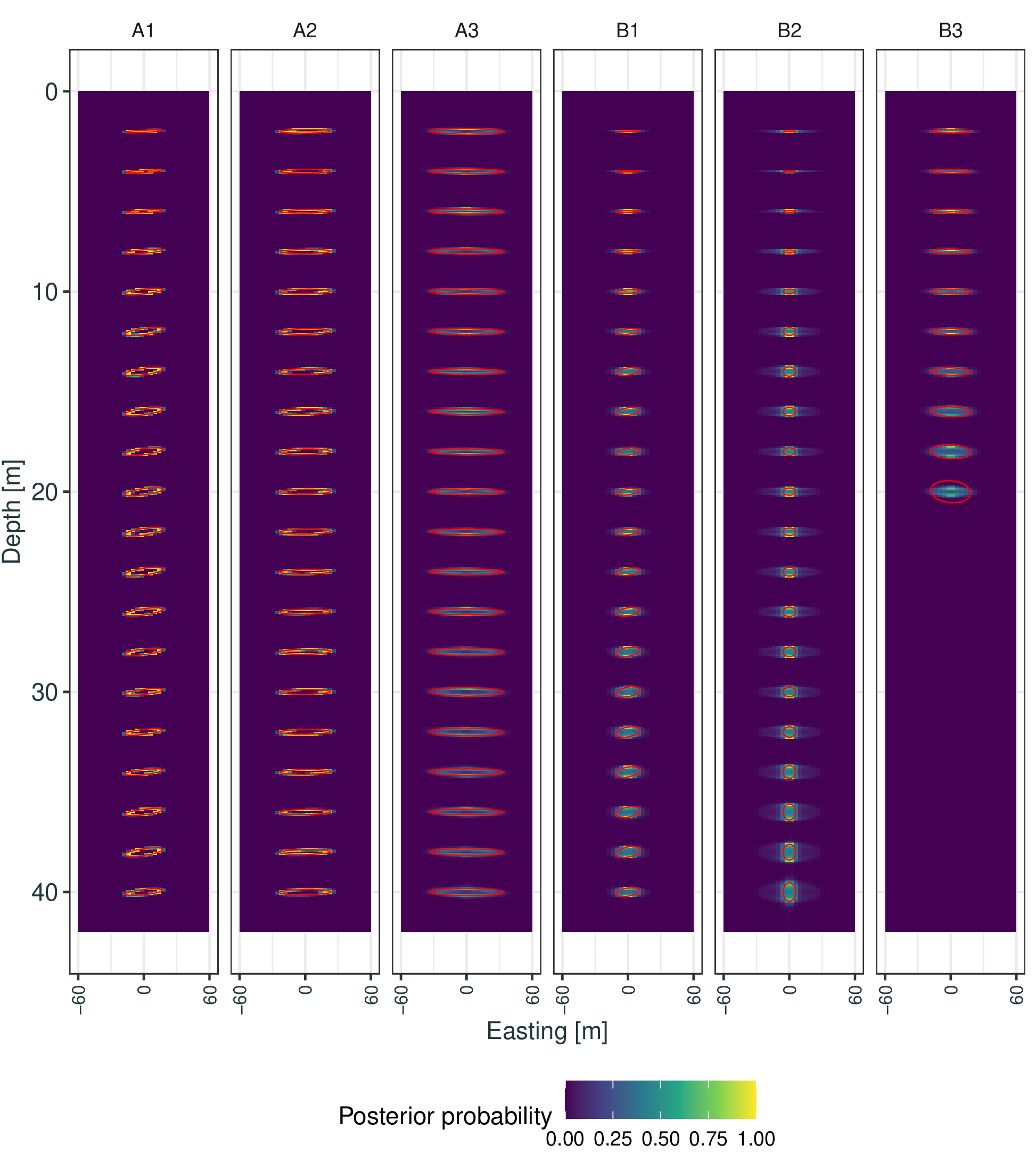}
  \end{center}

  \caption{
    Posterior distributions of iso-correlation contours in the easting--depth plane centered on locations $(0, 0, 2)', \ldots, (0, 0, 40)'$ that show the coordinates for which the correlation in the deviation process with that location is equal to 0.5. Each column shows the contours for one of the six 3-D sites. The posterior distribution is visualized over a 30 $\times$ 500 grid, where the color of each grid cell represents the posterior probability that an iso-correlation contour passes through that cell. Contours are omitted for B3 below 21m, the maximum depth measured at this site. The MAP estimate of the iso-correlation contour is shown in red.
  }
  \label{fig:isocorrelation_vertical_easting}
\end{figure}

\begin{figure}[ht!]
  \begin{center}
    \includegraphics[width=16.5cm]{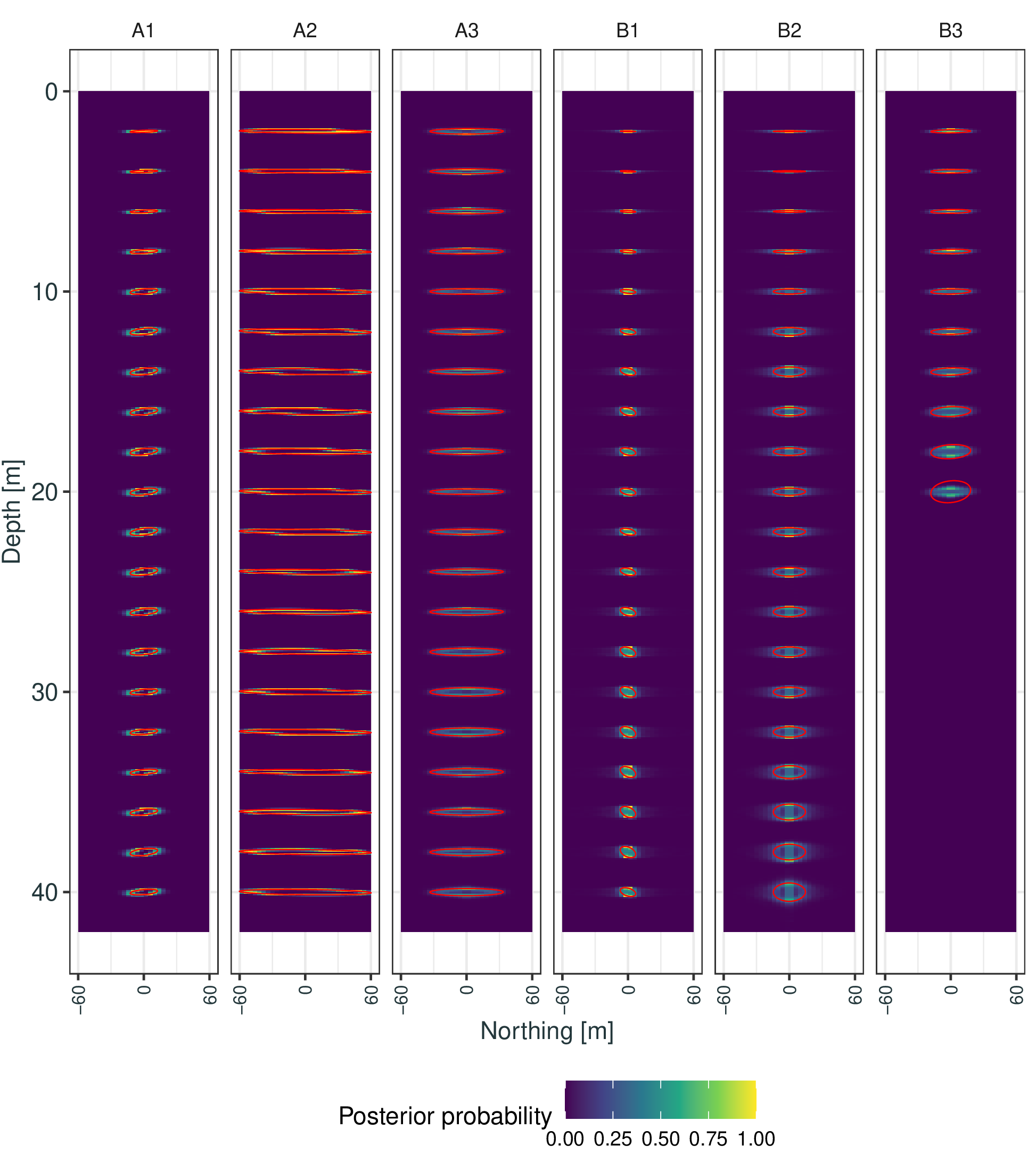}
  \end{center}

  \caption{
    As in Figure~\ref{fig:isocorrelation_vertical_easting}, but showing the posterior distributions of iso-correlation contours in the northing--depth plane.
  }
  \label{fig:isocorrelation_vertical_northing}
\end{figure}

\begin{figure}[ht!]
  \begin{center}
    \includegraphics{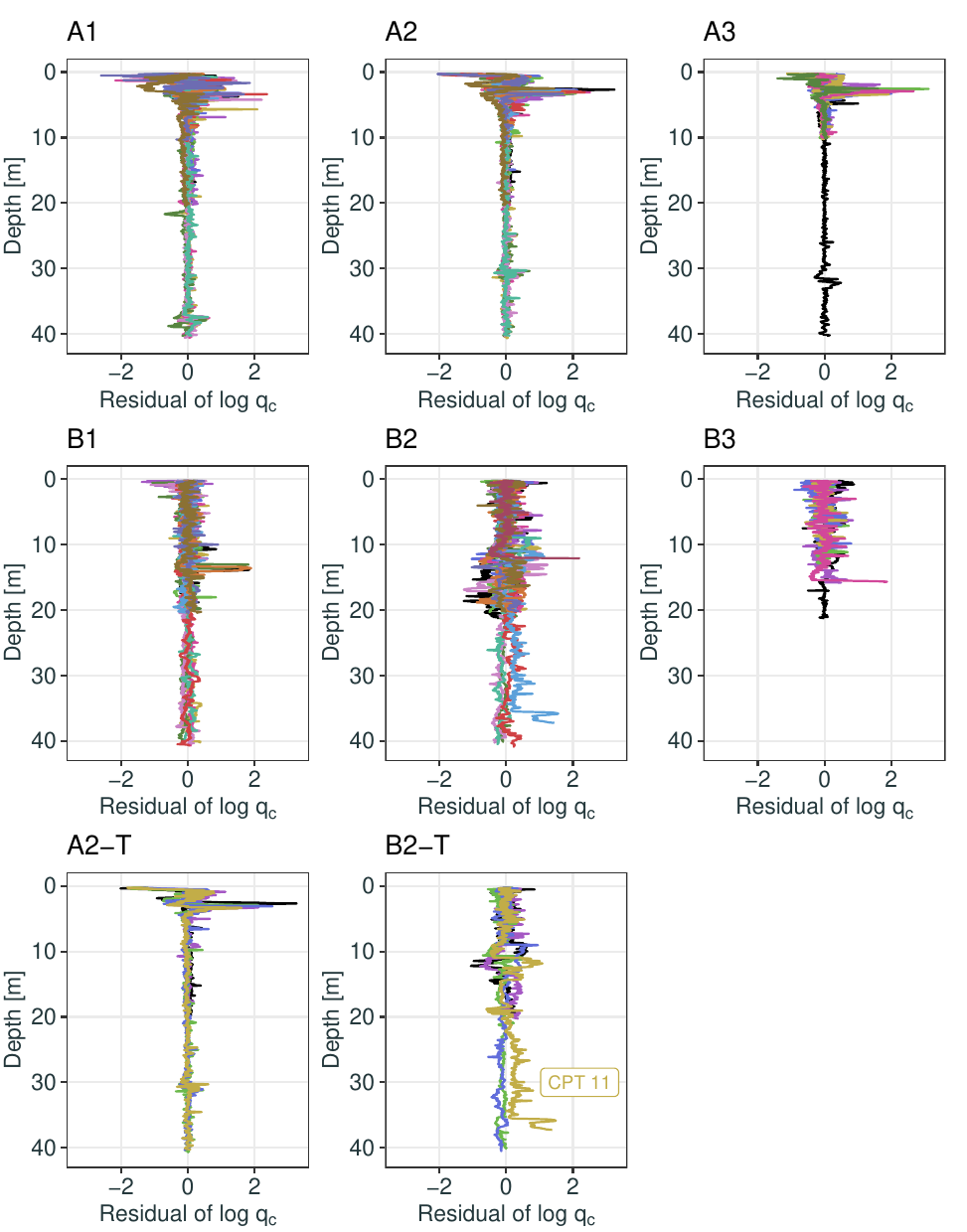}
  \end{center}

  \caption{
    Residuals from the fits to each of the sites, calculated by subtracting the posterior mean estimate of the mean profile from the measurements in each CPT. Each panel shows a different site, and each color corresponds to a different CPT.
  }
  \label{fig:residuals_plot}
\end{figure}

\begin{figure}[ht!]
  \begin{center}
    \includegraphics[width=0.49\linewidth]{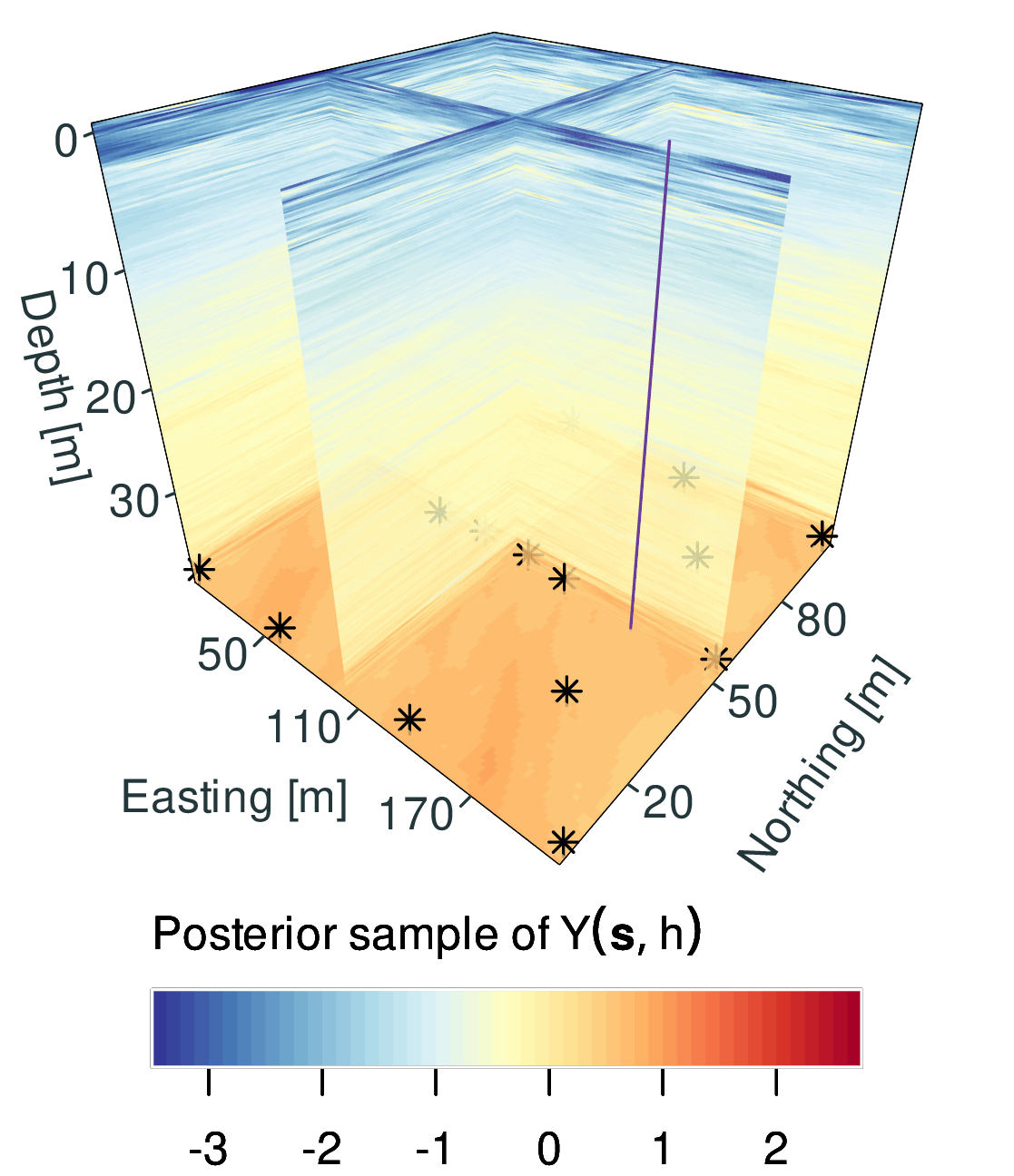}
    \includegraphics{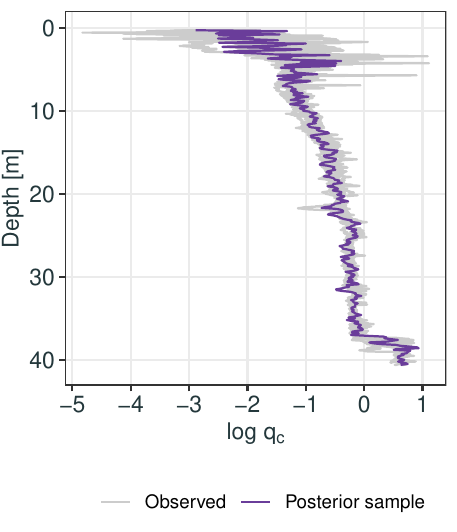} \\
    \includegraphics[width=0.49\linewidth]{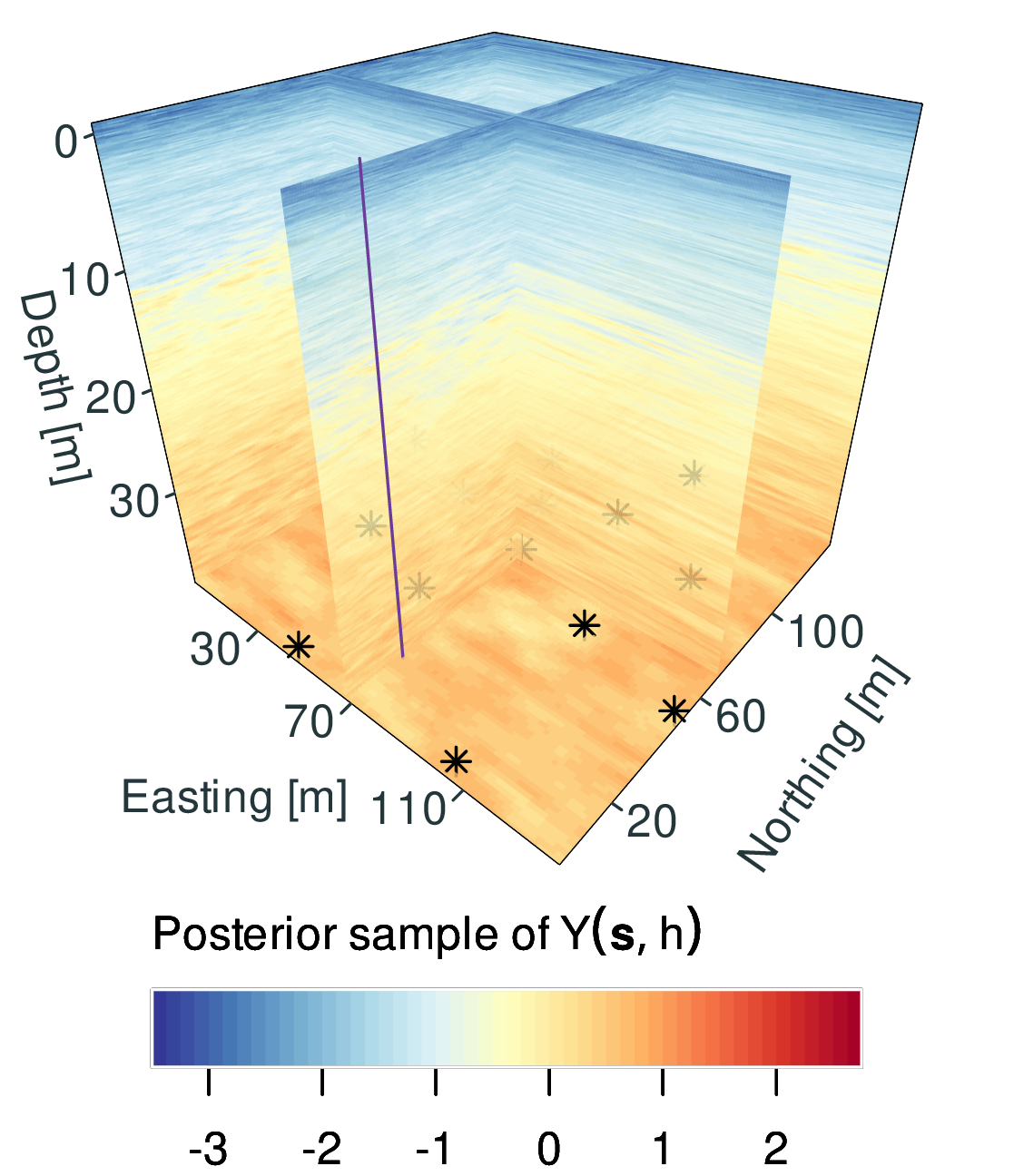}
    \includegraphics{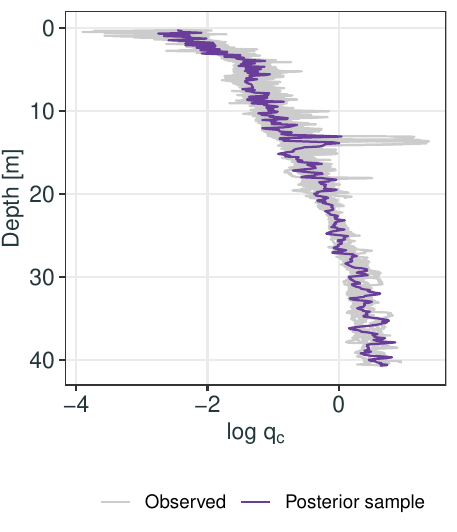}
  \end{center}

  \caption{
    Single samples from the posterior distribution of GeoWarp for the A1 site (top panels) and the B1 site (bottom panels). The left panels show slices of a full 3-D sample of $Y(\cdot, \cdot)$, while the right panels show single vertical samples of $\log q_c$ (equal to $Y(\cdot, \cdot)$ plus Gaussian white noise with variance $\sigma_\epsilon^2$) in purple alongside the CPTs from the site in gray. A purple line shows the location of the single vertical posterior sample, and stars show the horizontal location of the CPTs.
  }
  \label{fig:posterior_samples}
\end{figure}

\begin{table}[ht!]
  \caption{
    As in Table~\ref{tab:cross_validation_metrics}, with five parsimonious variants of GeoWarp added. For each site and metric the best-scoring model is marked in bold, as are the models with performance that is not significantly different from the best-scoring model (determined using a one-sided two-sample $t$-test at $\alpha = 0.05$).
  }
  \begin{center}
    \setlength{\tabcolsep}{5pt}
    \scriptsize
    \input{figures/cv-metrics-table-full}
  \end{center}

  \label{tab:cross_validation_metrics_full}
\end{table}

\begin{figure}[ht!]
  \begin{center}
    \includegraphics{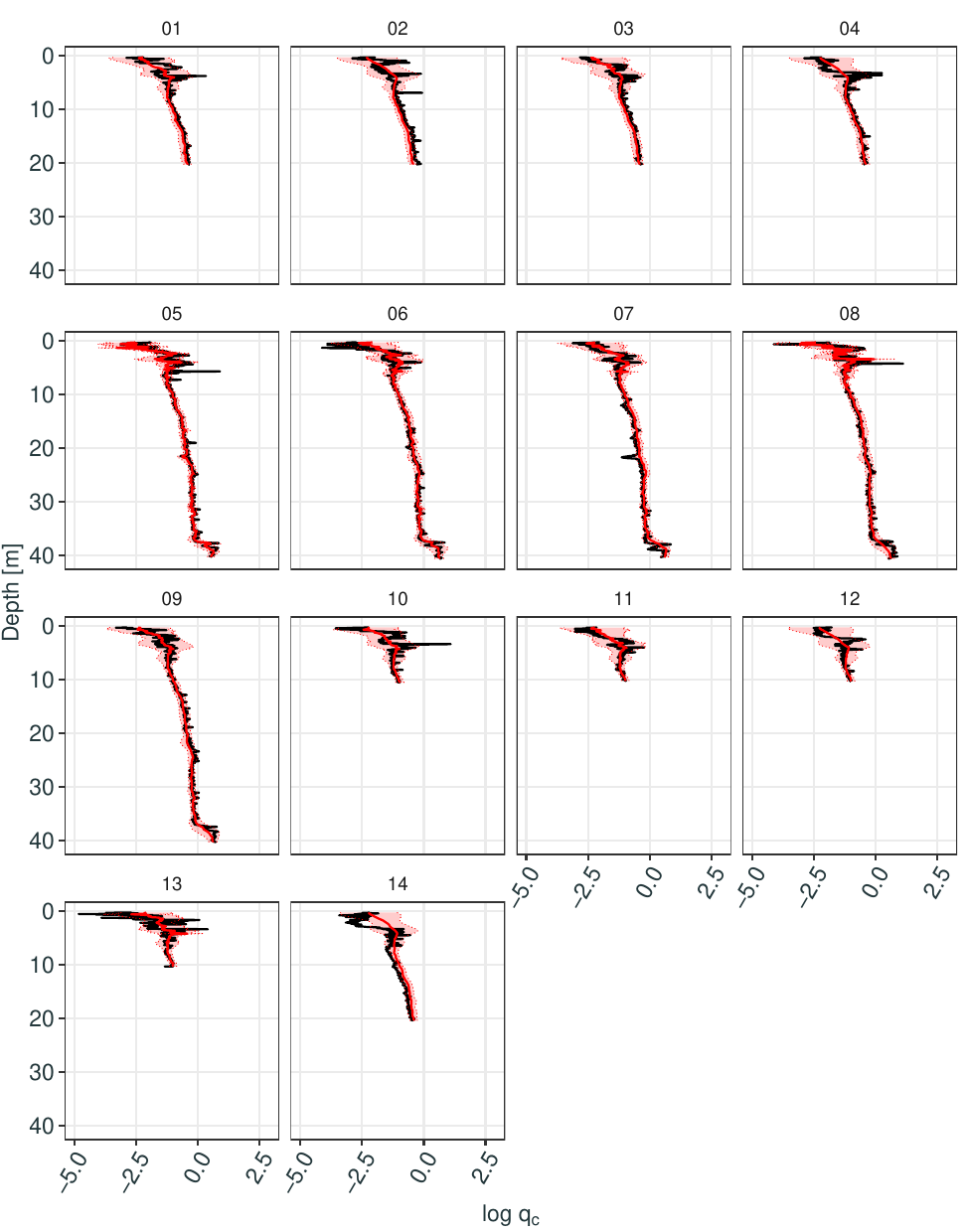}
  \end{center}

  \caption{
    GeoWarp cross-validation predictions for the A1 site. Each panel shows the data from the withheld CPT in black, and the posterior mean (red solid line) and pointwise 95\% posterior prediction interval (shaded area) predicted from the remaining CPTs.
  }
  \label{fig:cv_predictions_a1}
\end{figure}

\begin{figure}[ht!]
  \begin{center}
    \includegraphics{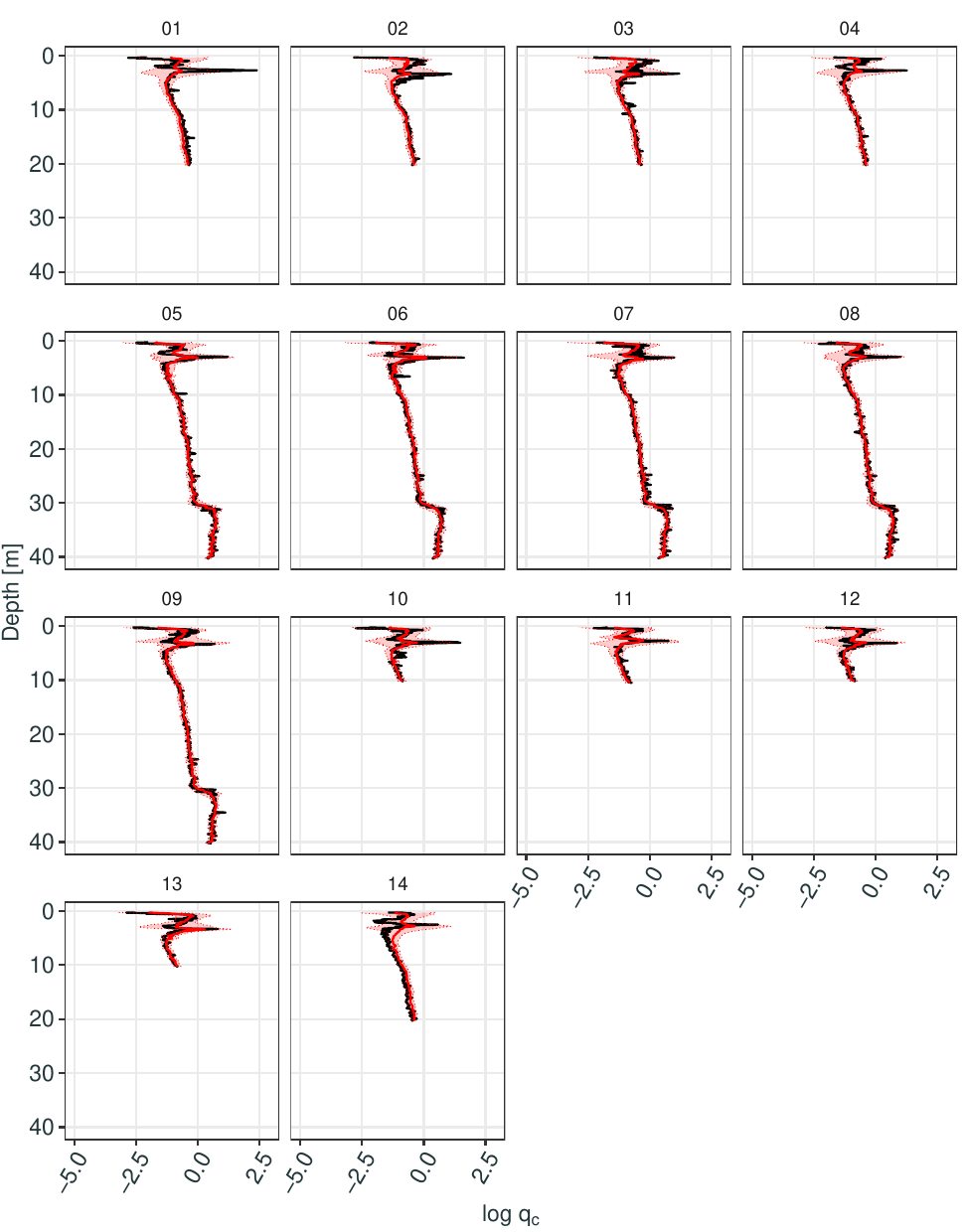}
  \end{center}

  \caption{As in Figure~\ref{fig:cv_predictions_a1}, but for the A2 site.}
  \label{fig:cv_predictions_a2}
\end{figure}

\begin{figure}[ht!]
  \begin{center}
    \includegraphics{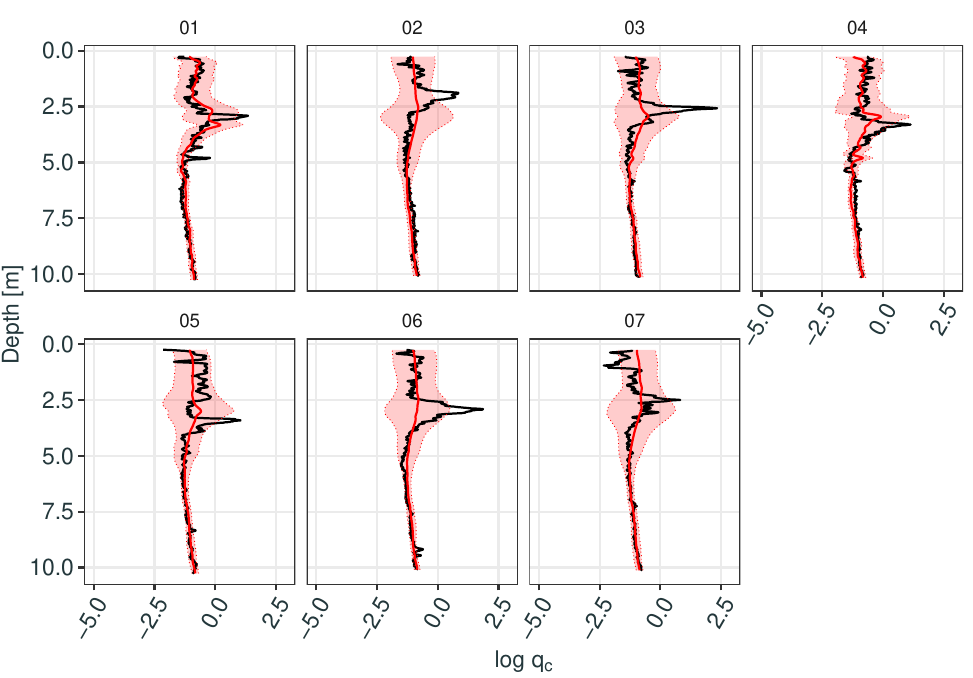}
  \end{center}

  \caption{As in Figure~\ref{fig:cv_predictions_a1}, but for the A3 site.}
  \label{fig:cv_predictions_a3}
\end{figure}

\begin{figure}[ht!]
  \begin{center}
    \includegraphics{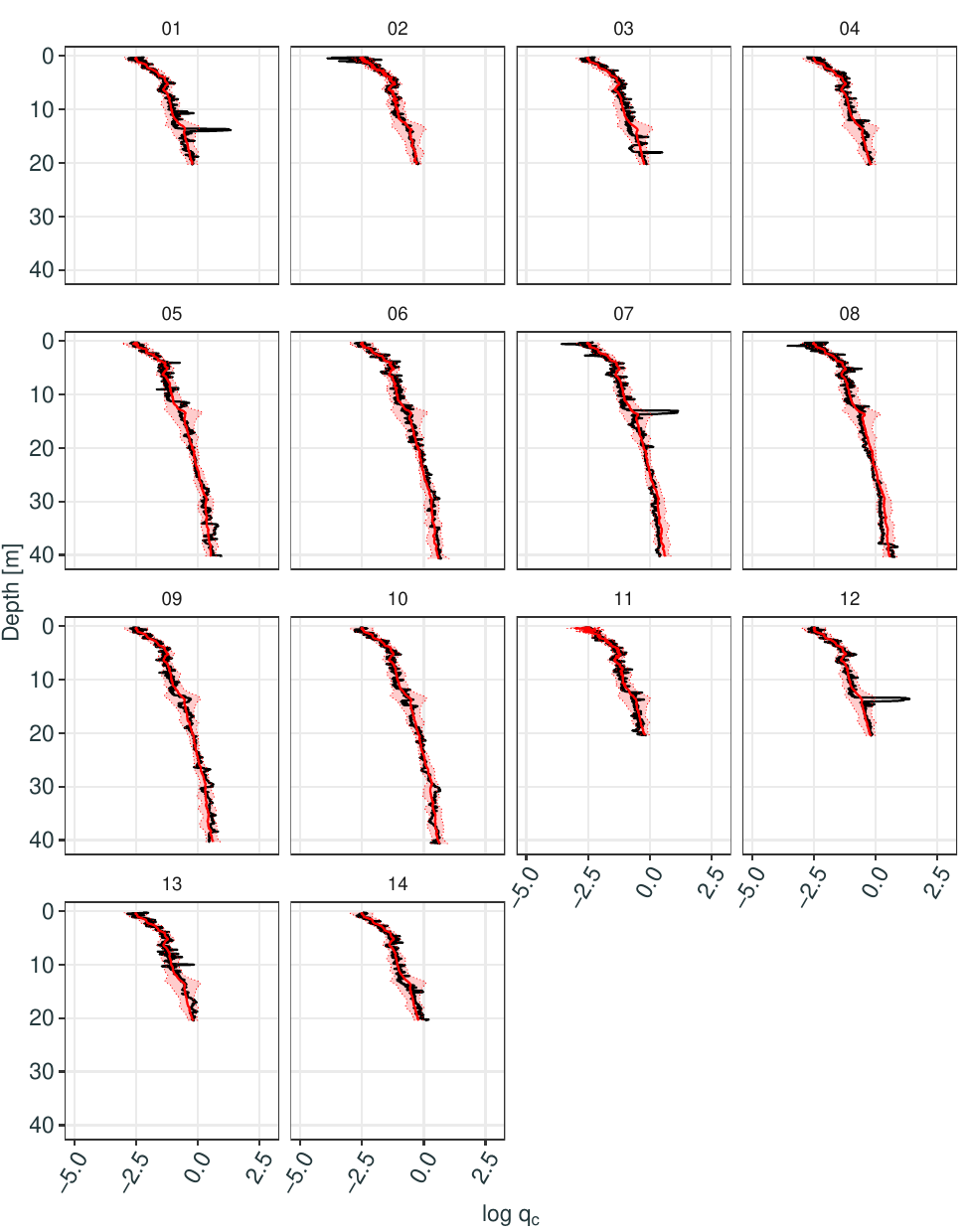}
  \end{center}

  \caption{As in Figure~\ref{fig:cv_predictions_a1}, but for the B1 site.}
  \label{fig:cv_predictions_b1}
\end{figure}

\begin{figure}[ht!]
  \begin{center}
    \includegraphics{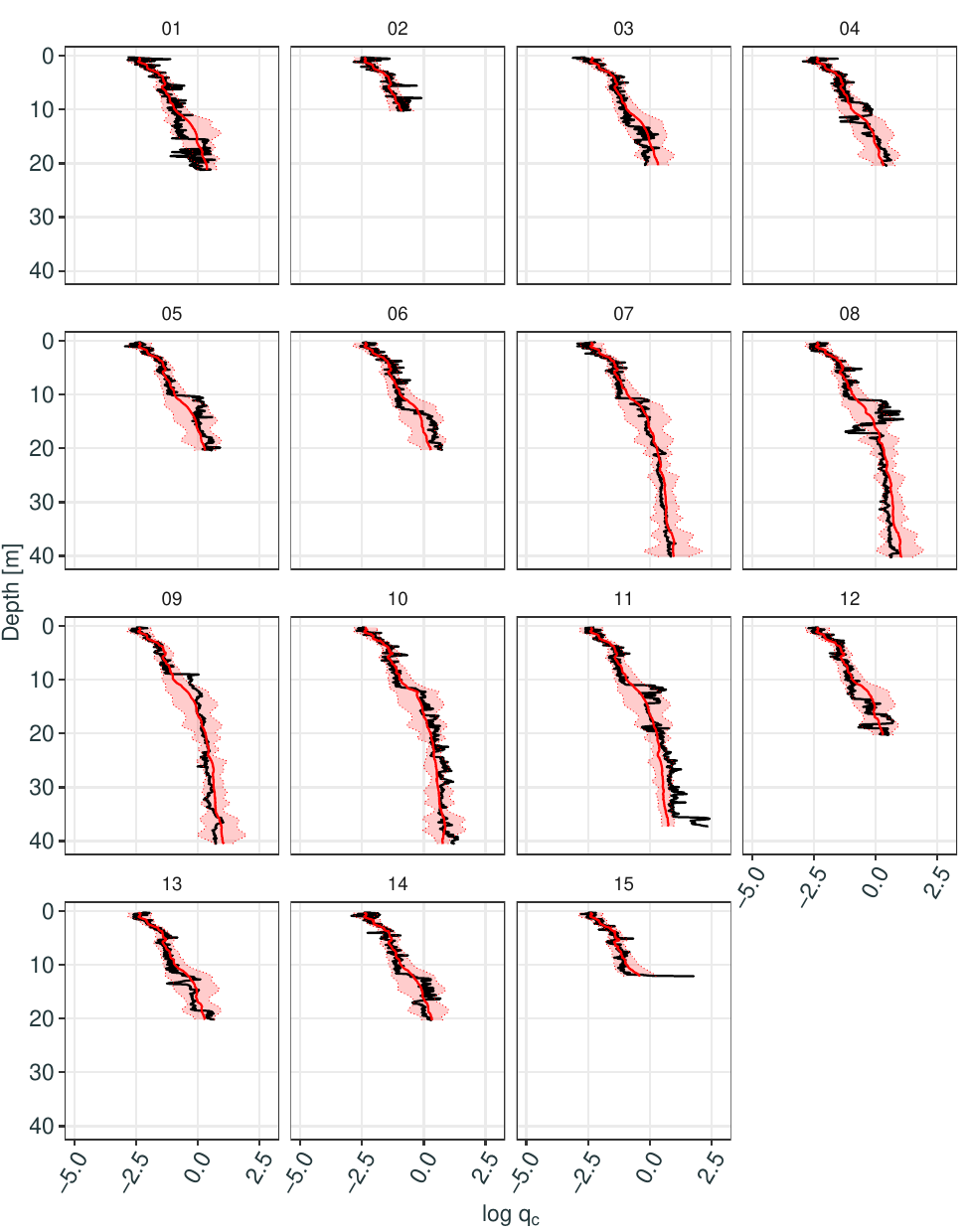}
  \end{center}

  \caption{As in Figure~\ref{fig:cv_predictions_a1}, but for the B2 site.}
  \label{fig:cv_predictions_b2}
\end{figure}

\begin{figure}[ht!]
  \begin{center}
    \includegraphics{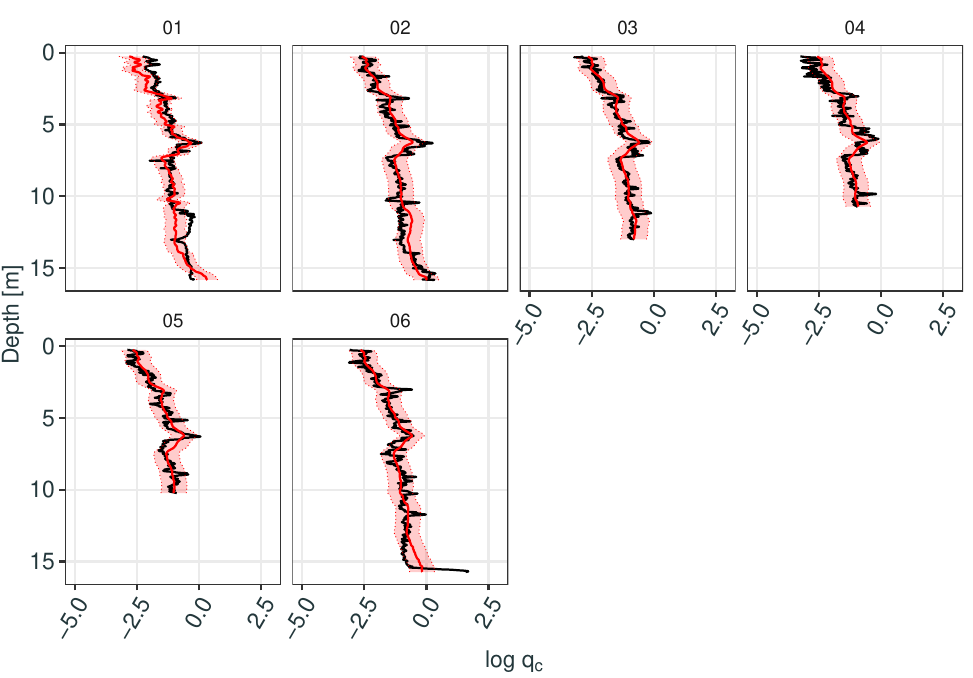}
  \end{center}

  \caption{As in Figure~\ref{fig:cv_predictions_a1}, but for the B3 site.}
  \label{fig:cv_predictions_b3}
\end{figure}

\begin{figure}[ht!]
  \begin{center}
    \includegraphics{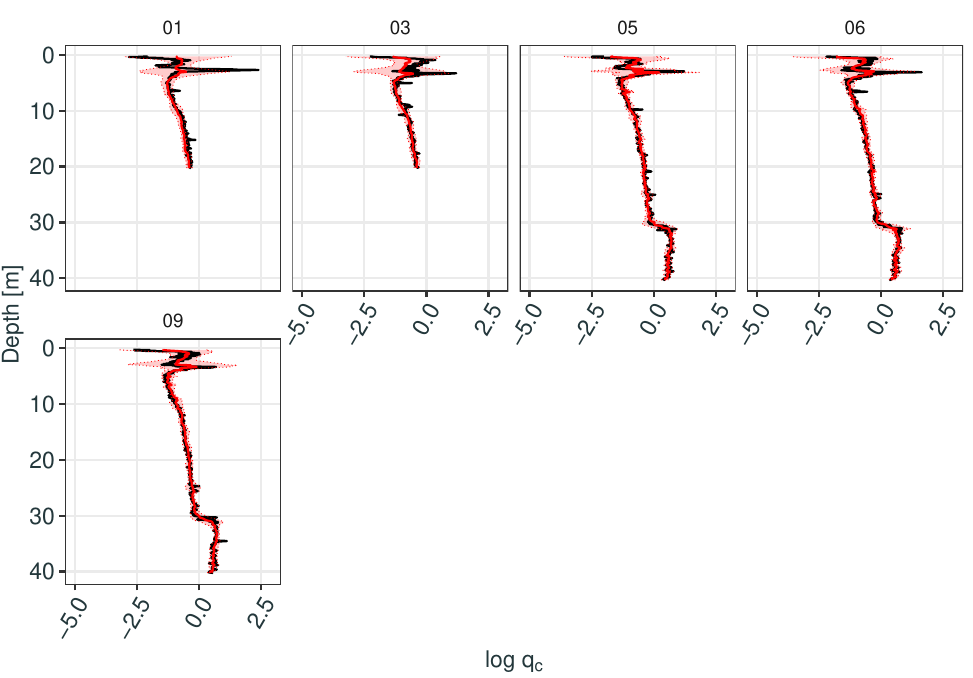}
  \end{center}

  \caption{As in Figure~\ref{fig:cv_predictions_a1}, but for the A2-T site.}
  \label{fig:cv_predictions_a2_t}
\end{figure}

\begin{figure}[ht!]
  \begin{center}
    \includegraphics{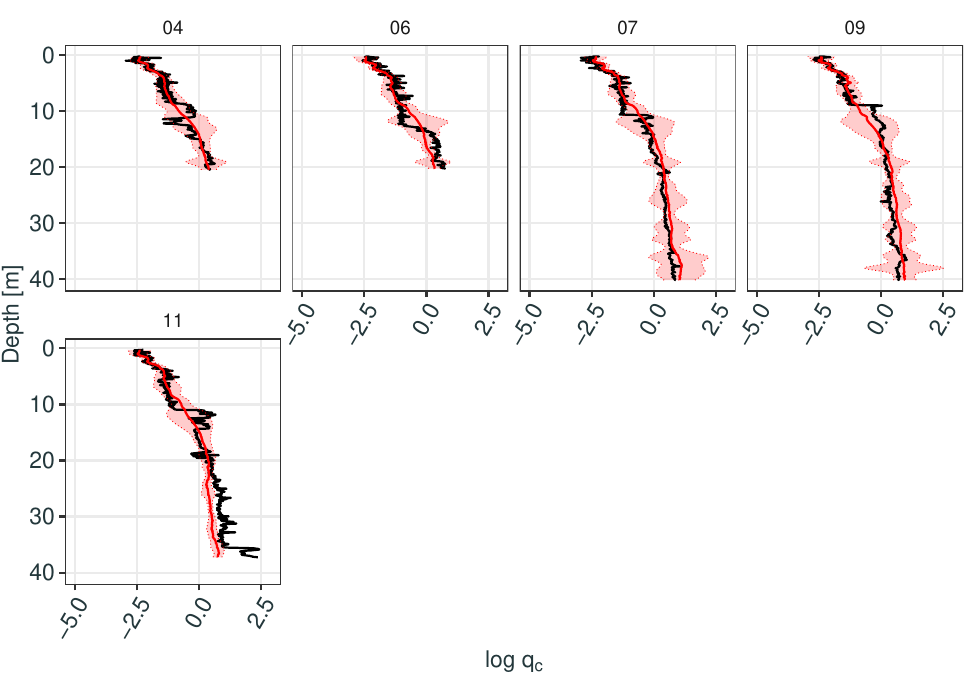}
  \end{center}

  \caption{As in Figure~\ref{fig:cv_predictions_a1}, but for the B2-T site.}
  \label{fig:cv_predictions_b2_t}
\end{figure}

\begin{figure}[ht!]
  \begin{center}
    \includegraphics{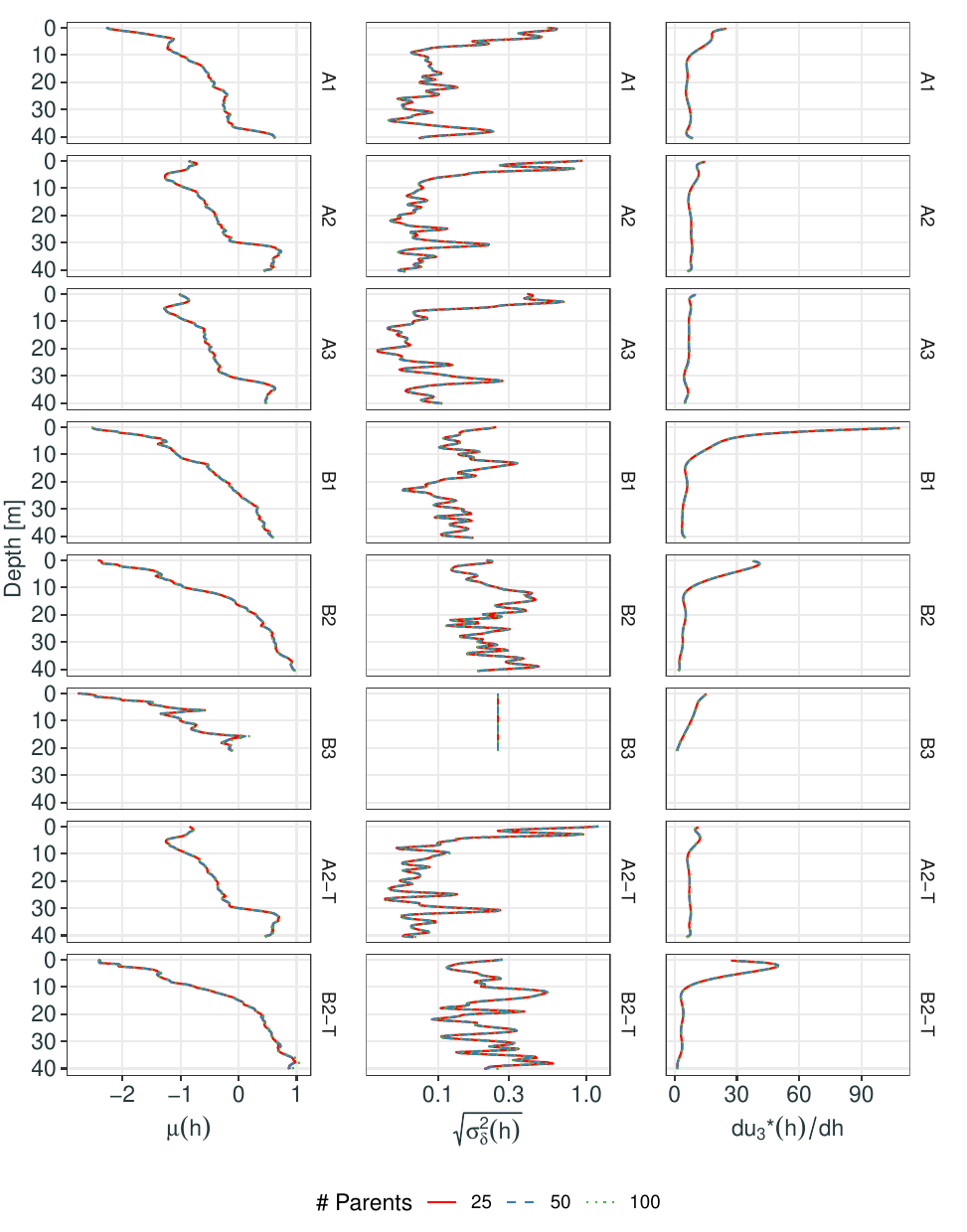}
  \end{center}

  \caption{
    Estimates of the mean profile ($\mu(h)$, left), the vertical standard deviation profiles ($\sqrt{\sigma_\delta^2(h)}$, middle, log scale), and the derivative of the vertical warping ($\textrm{d}u_3^*(h) / \textrm{d}h$, right), when using 25, 50, and 100 parents in the Vecchia approximation. Each row shows estimates for a different site. The similarity between the profiles means that the profiles are very similar and largely overlap.
  }
  \label{fig:vertical_profiles_n_parents}
\end{figure}

\begin{figure}[ht!]
  \begin{center}
    \includegraphics{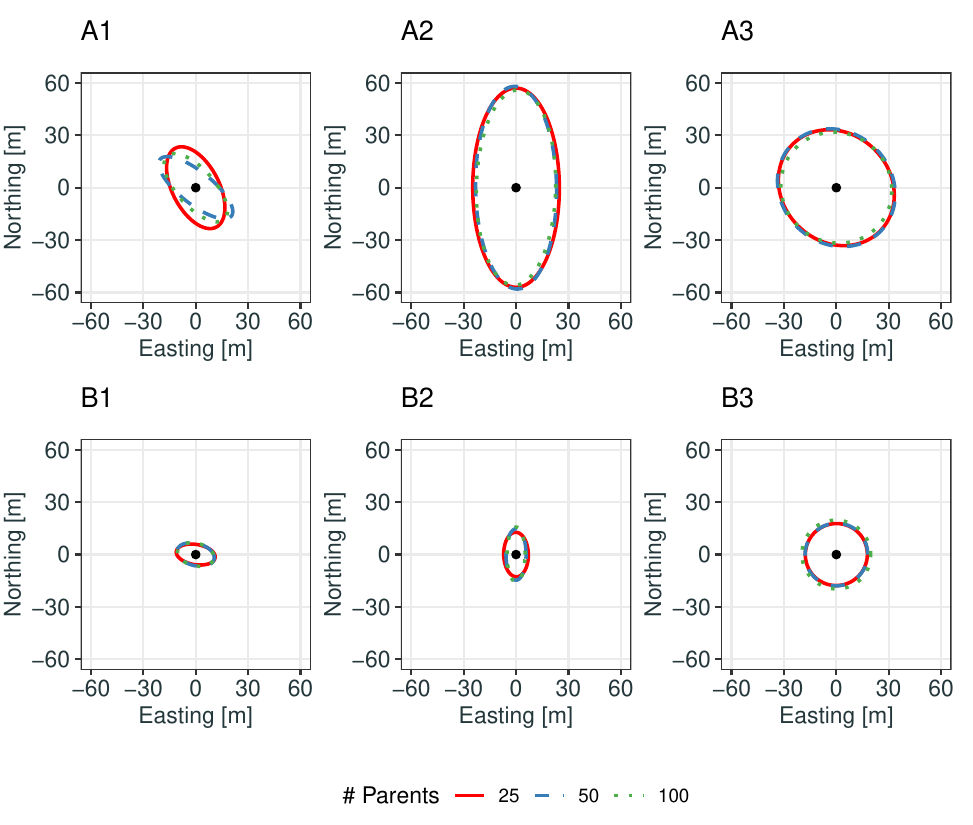}
  \end{center}

  \caption{
    As in Figure~\ref{fig:isocorrelation_horizontal}, but showing how the iso-correlation contours vary when using 25, 50, or 100 parents in the Vecchia approximation.
  }
  \label{fig:isocorrelation_horizontal_n_parents}
\end{figure}

\begin{figure}[ht!]
  \begin{center}
    \includegraphics{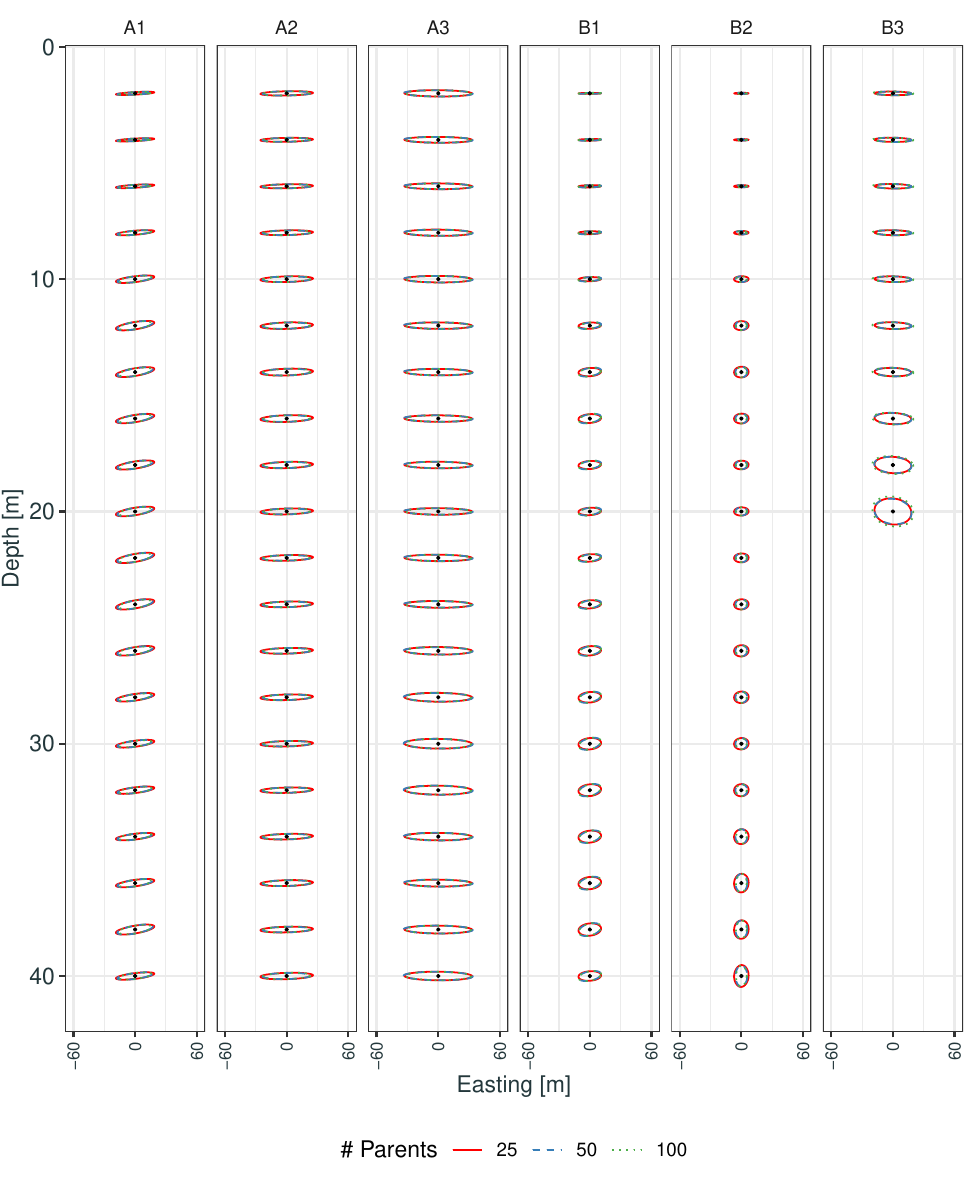}
  \end{center}

  \caption{
    Iso-correlation contours in the easting--depth plane centered on locations $(0, 0, 2)', \ldots, (0, 0, 40)'$ that show the coordinates for which the correlation in the deviation process with that location is equal to 0.5. Each column shows the contours for one of the six 3-D sites. Contours are shown for 25, 50, and 100 parents in the Vecchia approximation.
  }
  \label{fig:isocorrelation_vertical_n_parents_easting}
\end{figure}

\begin{figure}[ht!]
  \begin{center}
    \includegraphics{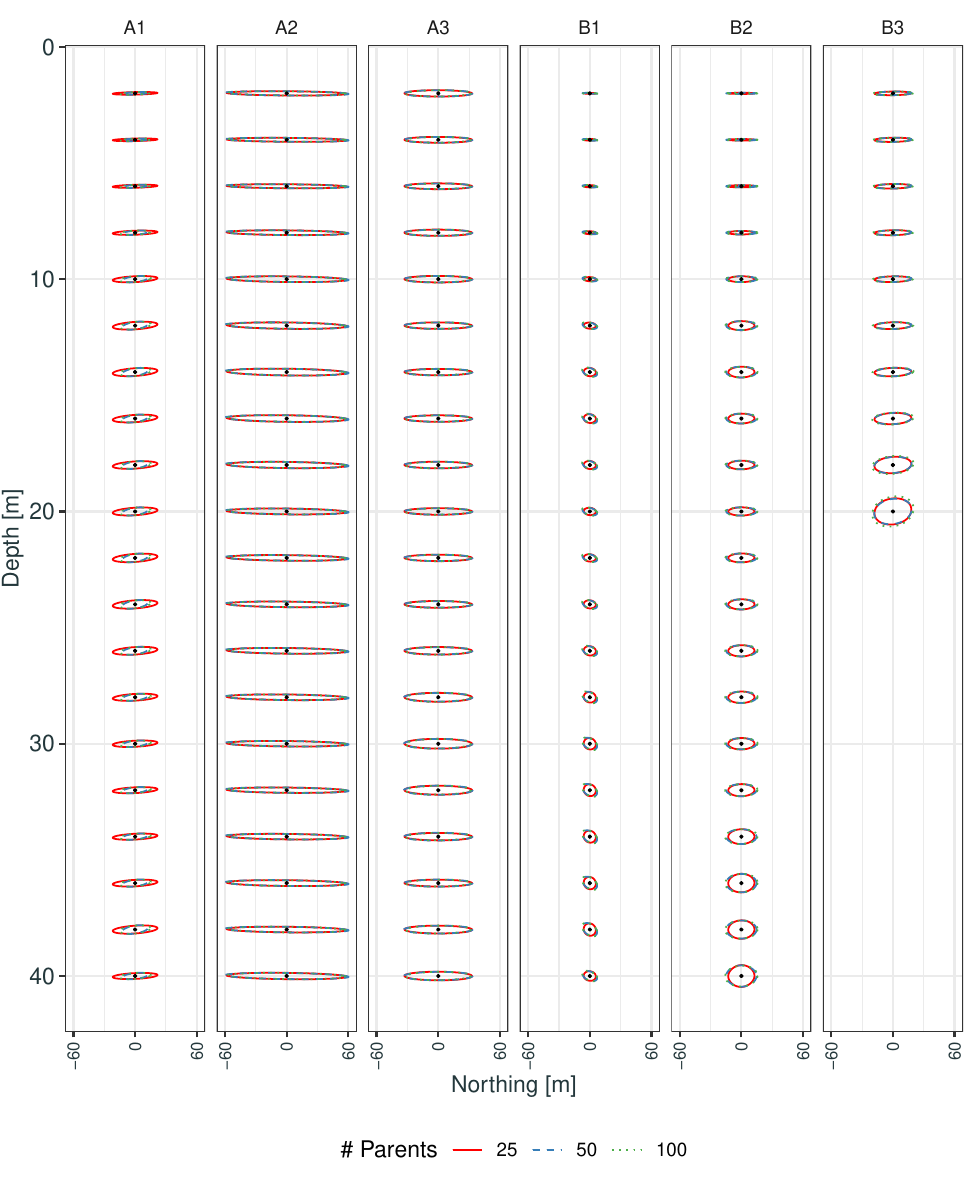}
  \end{center}

  \caption{
    As in Figure~\ref{fig:isocorrelation_vertical_n_parents_easting}, but showing iso-correlation contours in the northing--depth plane when using 25, 50, and 100 parents in the Vecchia approximation.
  }
  \label{fig:isocorrelation_vertical_n_parents_northing}
\end{figure}

\begin{figure}[ht!]
  \begin{center}
    \includegraphics{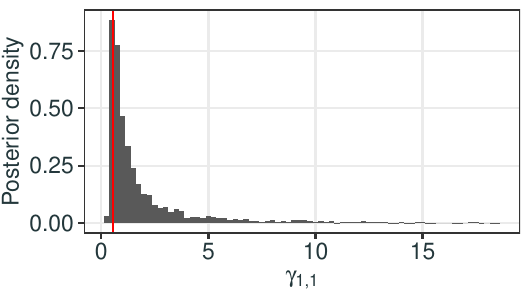}
  \end{center}

  \caption{
    Histogram of the estimated marginal posterior distribution of $\gamma_{1, 1}$ for the B3 site. The red line marks the MAP estimate of the same parameter.
  }
  \label{fig:gamma_horizontal_marginal_posterior}
\end{figure}

\begin{figure}[ht!]
  \begin{center}
    \includegraphics[width=16.5cm]{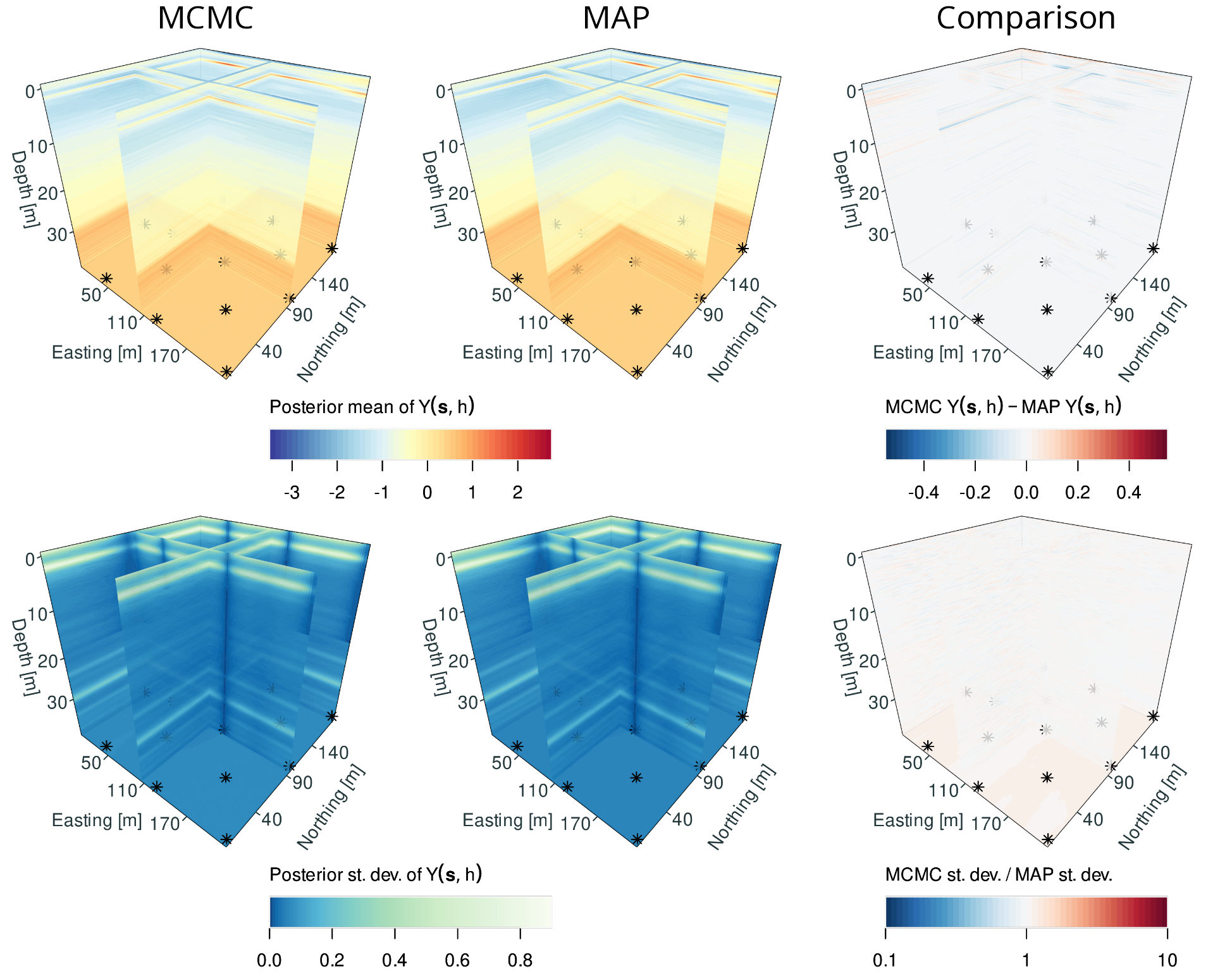}
  \end{center}

  \caption{
    Predictions over a 3-D grid at the A2 site. The panels in the top row show the MCMC posterior mean, the MAP posterior mean, and the difference between the two, respectively. The panels in the bottom row show the MCMC posterior standard deviation, the MAP posterior standard deviation, and the ratio between the two, respectively. Stars show the horizontal locations of the CPTs.
  }
  \label{fig:predicted_a2_mcmc}
\end{figure}

\begin{table}[ht!]
  \caption{
    Configuration and hyperparameter settings for the application of GeoWarp to NWS CPT data.
  }
  \centering
  \begin{tabular}{ll}
    \hline \hline
    \textbf{Configuration} & \textbf{Value} \\
    \hline
    Mean profile knot spacing & $\Delta_\mu = 0.1$ m \\
    Deviation process smoothness & $\nu = 3 / 2$ \\
    Variance knot spacing & $\Delta_\sigma = 1$ m \\
    Basis functions for AWUs & $L_1 = 2$, $L_2 = 2$, $L_3 = 20$ \\
    \hline
    Prior on $\gamma_{1, 1}, \gamma_{2, 1}$  & $\gamma_d^- = 1 / 200$, $\gamma_d^+ = 1 / 0.5$, for $d = 1, 2$ \\
    Prior on $\gamma_{3, 1}, \ldots, \gamma_{3, 20}$ & $a_3^\gamma = 1.01$, $b_3^\gamma = 0.01$ \\
    Prior on $\Rvec$  & $\rho_R = 6$ \\
    Prior on $\sigma_\epsilon^2$ & $a_\epsilon = 2.437$, $b_\epsilon = 0.544$ \\
    Prior on $\sigma_\beta^2$, $\sigma_\zeta^2$ & $a_\beta = a_\zeta = 0.166$, $b_\beta = b_\zeta = 8.932 \times 10^{-7}$ \\
    Prior on $\ell_\zeta$ & $\sigma_\ell^2 = 1$ \\
    \hline \hline
  \end{tabular}
  \label{tab:geowarp_configuration}
\end{table}

\begin{figure}[ht!]
  \begin{center}
    \includegraphics[width=16.5cm]{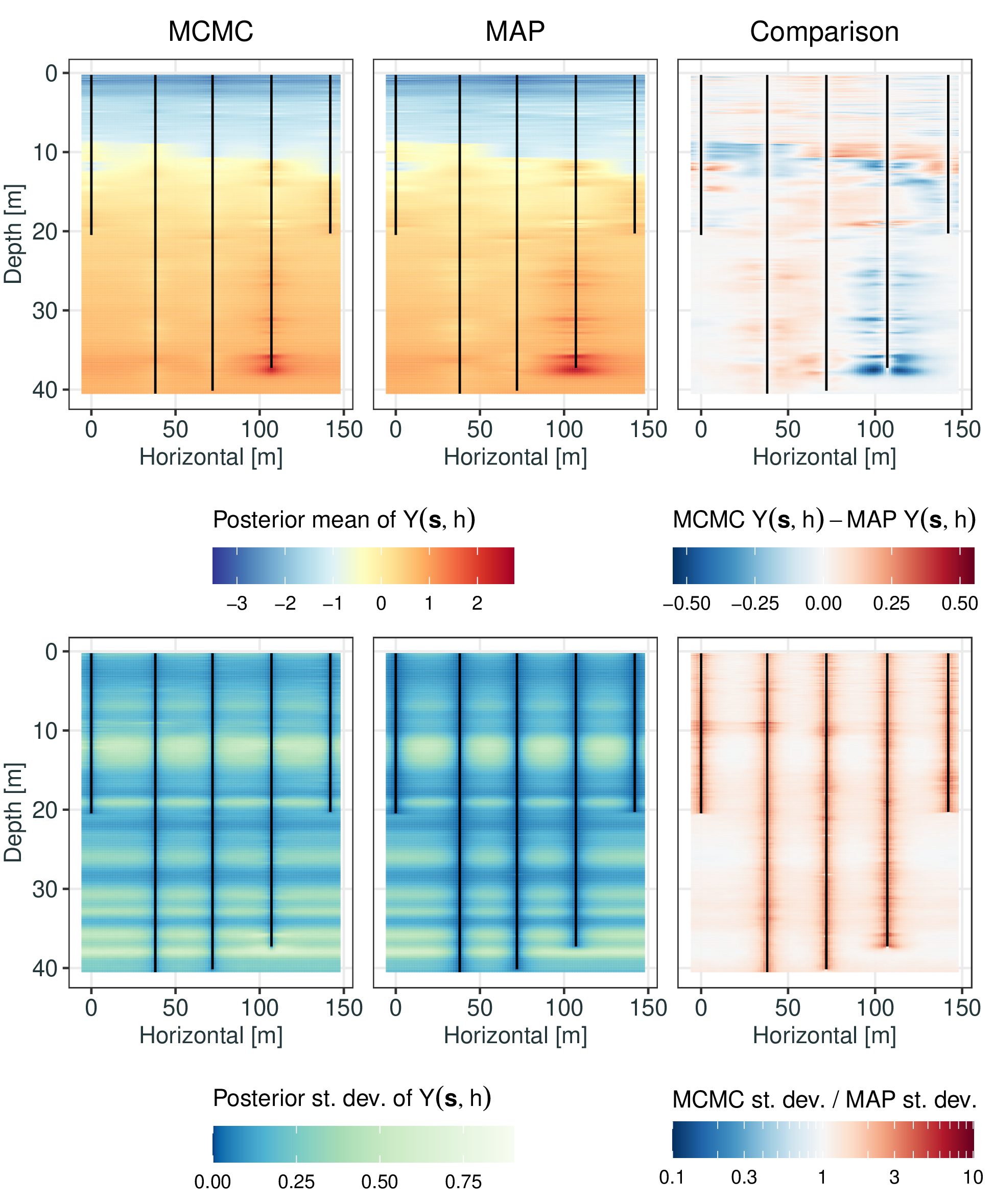}
  \end{center}

  \caption{
    As in Figure~\ref{fig:predicted_a2_mcmc}, but for the B2-T site over a 2-D grid.
  }
  \label{fig:predicted_b2t_mcmc}
\end{figure}

\begin{figure}[ht!]
  \begin{center}
    \includegraphics{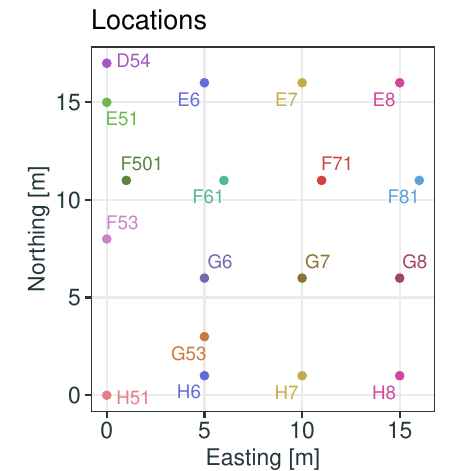} %
    \includegraphics{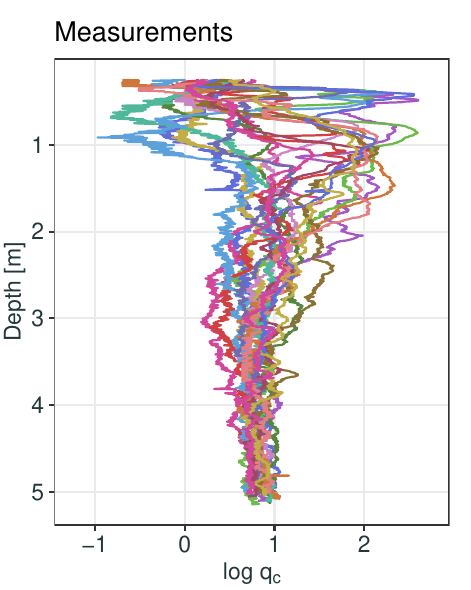}
  \end{center}
  \caption{
    Map of locations (left panel) and measurements of $\log q_c$ by depth (right panel) from CPTs taken at the South Parklands site discussed in Section~\ref{sec:jaksa_benchmark}. Labels and colors identify each CPT.
  }
  \label{fig:jaksa_data}
\end{figure}

\begin{table}[ht!]
  \caption{
    Cross-validated prediction scores for the South Parklands site described in Section~\ref{sec:jaksa_benchmark}. The best (lowest-scoring) model for each metric is marked in bold. The best model was significantly better than the other models for each metric (assessed using one-sided two-sample $t$-tests at $\alpha = 0.05$).
  }
  \begin{center}
    \setlength{\tabcolsep}{5pt}
    \input{figures/cv-metrics-table-jaksa}
  \end{center}

  \label{tab:cross_validation_metrics_jaksa}
\end{table}

\end{document}

%% file: figures/cv-metrics-table.tex
\begin{tabular}{rr|rrrrrr|>{\columncolor[gray]{0.95}}r|rr}
\hline \hline
& & A1 & A2 & A3 & B1 & B2 & B3 & All & A2-T & B2-T \\
\hline
\multirow{4}{*}{MSE}
& GeoWarp & \textbf{0.060} & \textbf{0.045} & \textbf{0.130} & \textbf{0.034} & \textbf{0.096} & \textbf{0.083} & \textbf{0.062} & \textbf{0.049} & \textbf{0.107} \\
& Linear & 0.118 & 0.122 & 0.210 & 0.088 & 0.211 & 0.157 & 0.139 & 0.114 & 0.242 \\
& Binned & \textbf{0.062} & 0.054 & \textbf{0.139} & \textbf{0.035} & 0.099 & \textbf{0.088} & 0.066 & 0.057 & 0.127 \\
& BCS & 0.125 & 0.226 & 1.021 & 0.072 & 0.153 & 2.008 & 0.231 & 0.145 & 0.157 \\
\hline
\multirow{4}{*}{CRPS}
& GeoWarp & \textbf{0.100} & \textbf{0.079} & \textbf{0.147} & \textbf{0.090} & \textbf{0.165} & \textbf{0.160} & \textbf{0.112} & \textbf{0.075} & \textbf{0.186} \\
& Linear & 0.173 & 0.178 & 0.230 & 0.155 & 0.261 & 0.216 & 0.194 & 0.168 & 0.284 \\
& Binned & 0.103 & 0.086 & 0.159 & 0.094 & 0.169 & 0.173 & 0.117 & 0.087 & 0.210 \\
& BCS & 0.192 & 0.242 & 0.627 & 0.147 & 0.234 & 0.835 & 0.238 & 0.203 & 0.223 \\
\hline
\multirow{4}{*}{Int05}
& GeoWarp & \textbf{1.104} & \textbf{0.848} & \textbf{1.790} & \textbf{1.010} & \textbf{1.949} & \textbf{1.593} & \textbf{1.269} & \textbf{0.780} & 2.905 \\
& Linear & 2.242 & 2.176 & 2.918 & 1.721 & 2.031 & 2.215 & 2.070 & 2.121 & \textbf{2.071} \\
& Binned & 1.299 & 1.001 & \textbf{1.860} & 1.115 & 2.221 & 2.787 & 1.474 & 1.344 & 4.245 \\
& BCS & 2.410 & 3.473 & 3.365 & 1.842 & 2.543 & 5.947 & 2.668 & 2.336 & 3.095 \\
\hline
\multirow{3}{*}{DSS}
& GeoWarp & \textbf{-2.686} & \textbf{-3.307} & \textbf{-1.956} & \textbf{-2.536} & \textbf{-0.605} & \textbf{-1.354} & \textbf{-2.215} & \textbf{-3.444} & 3.175 \\
& Linear & -1.119 & -1.091 & -0.497 & -1.410 & \textbf{-0.547} & -0.832 & -1.017 & -1.156 & \textbf{-0.397} \\
& BCS & -0.934 & -0.276 & 1.168 & -1.433 & \textbf{-0.524} & 1.723 & -0.658 & -0.791 & 0.334 \\
\hline
\multirow{3}{*}{DSS2}
& GeoWarp & \textbf{-4.381} & \textbf{-4.904} & \textbf{-3.495} & \textbf{-4.358} & \textbf{-3.157} & \textbf{-2.577} & \textbf{-4.107} & \textbf{-4.988} & \textbf{-1.342} \\
& Linear & -1.116 & -1.089 & -0.494 & -1.408 & -0.547 & -0.837 & -1.016 & -1.154 & -0.394 \\
& BCS & -1.612 & -1.235 & -0.080 & -2.230 & -1.493 & 0.469 & -1.538 & -1.805 & \textbf{-1.125} \\
\hline \hline
\end{tabular}

%% file: figures/cv-metrics-table-full.tex
\begin{tabular}{rr|rrrrrr|>{\columncolor[gray]{0.95}}r|rr}
\hline \hline
& & A1 & A2 & A3 & B1 & B2 & B3 & All & A2-T & B2-T \\
\hline
\multirow{9}{*}{MSE}
& GeoWarp & \textbf{0.060} & 0.045 & \textbf{0.130} & \textbf{0.034} & \textbf{0.096} & \textbf{0.083} & 0.062 & 0.049 & \textbf{0.107} \\
\cdashline{2-11}
& GW-NoWarp & 0.077 & 0.048 & \textbf{0.130} & 0.062 & 0.114 & \textbf{0.084} & 0.078 & 0.054 & 0.153 \\
\cdashline{2-11}
& GW-CV & \textbf{0.062} & 0.045 & \textbf{0.116} & \textbf{0.034} & \textbf{0.095} & \textbf{0.084} & 0.062 & 0.051 & 0.116 \\
& GW-NoWarp-CV & \textbf{0.059} & \textbf{0.040} & \textbf{0.122} & \textbf{0.034} & \textbf{0.095} & \textbf{0.082} & \textbf{0.060} & \textbf{0.042} & 0.114 \\
& GW-Vert-CV & \textbf{0.062} & 0.054 & \textbf{0.126} & \textbf{0.034} & \textbf{0.095} & \textbf{0.082} & 0.064 & 0.060 & 0.118 \\
& GW-WN-CV & \textbf{0.061} & 0.053 & 0.137 & \textbf{0.035} & 0.098 & \textbf{0.085} & 0.065 & 0.056 & 0.125 \\
\cdashline{2-11}
& Linear & 0.118 & 0.122 & 0.210 & 0.088 & 0.211 & 0.157 & 0.139 & 0.114 & 0.242 \\
& Binned & \textbf{0.062} & 0.054 & 0.139 & \textbf{0.035} & 0.099 & \textbf{0.088} & 0.066 & 0.057 & 0.127 \\
& BCS & 0.125 & 0.226 & 1.021 & 0.072 & 0.153 & 2.008 & 0.231 & 0.145 & 0.157 \\
\hline
\multirow{9}{*}{CRPS}
& GeoWarp & \textbf{0.100} & \textbf{0.079} & \textbf{0.147} & \textbf{0.090} & \textbf{0.165} & \textbf{0.160} & \textbf{0.112} & \textbf{0.075} & \textbf{0.186} \\
\cdashline{2-11}
& GW-NoWarp & 0.122 & 0.083 & \textbf{0.146} & 0.161 & 0.236 & 0.164 & 0.153 & 0.080 & 0.293 \\
\cdashline{2-11}
& GW-CV & 0.130 & 0.112 & 0.168 & \textbf{0.090} & \textbf{0.166} & \textbf{0.161} & 0.127 & 0.115 & 0.191 \\
& GW-NoWarp-CV & 0.114 & 0.094 & 0.167 & \textbf{0.090} & \textbf{0.165} & \textbf{0.160} & 0.119 & 0.090 & \textbf{0.186} \\
& GW-Vert-CV & 0.134 & 0.120 & 0.178 & \textbf{0.090} & \textbf{0.165} & \textbf{0.158} & 0.130 & 0.129 & 0.194 \\
& GW-WN-CV & 0.117 & 0.105 & 0.176 & 0.092 & 0.168 & \textbf{0.162} & 0.124 & 0.099 & 0.195 \\
\cdashline{2-11}
& Linear & 0.173 & 0.178 & 0.230 & 0.155 & 0.261 & 0.216 & 0.194 & 0.168 & 0.284 \\
& Binned & 0.103 & 0.086 & 0.159 & 0.094 & 0.169 & 0.173 & 0.117 & 0.087 & 0.210 \\
& BCS & 0.192 & 0.242 & 0.627 & 0.147 & 0.234 & 0.835 & 0.238 & 0.203 & 0.223 \\
\hline
\multirow{9}{*}{Int05}
& GeoWarp & \textbf{1.104} & \textbf{0.848} & \textbf{1.790} & \textbf{1.010} & 1.949 & \textbf{1.593} & \textbf{1.269} & \textbf{0.780} & 2.905 \\
\cdashline{2-11}
& GW-NoWarp & 1.549 & \textbf{0.868} & \textbf{1.765} & 2.423 & 3.474 & 1.758 & 2.102 & \textbf{0.817} & 4.736 \\
\cdashline{2-11}
& GW-CV & 1.770 & 1.567 & 2.282 & \textbf{1.000} & 1.983 & \textbf{1.581} & 1.594 & 1.679 & 2.494 \\
& GW-NoWarp-CV & 1.717 & 1.430 & 2.367 & \textbf{1.005} & \textbf{1.825} & \textbf{1.578} & 1.520 & 1.417 & 2.200 \\
& GW-Vert-CV & 1.814 & 1.698 & 2.356 & \textbf{1.003} & 1.983 & \textbf{1.543} & 1.635 & 1.841 & 2.549 \\
& GW-WN-CV & 1.744 & 1.662 & 2.744 & \textbf{1.026} & \textbf{1.854} & 1.692 & 1.607 & 1.694 & 2.472 \\
\cdashline{2-11}
& Linear & 2.242 & 2.176 & 2.918 & 1.721 & 2.031 & 2.215 & 2.070 & 2.121 & \textbf{2.071} \\
& Binned & 1.299 & 1.001 & \textbf{1.860} & 1.115 & 2.221 & 2.787 & 1.474 & 1.344 & 4.245 \\
& BCS & 2.410 & 3.473 & 3.365 & 1.842 & 2.543 & 5.947 & 2.668 & 2.336 & 3.095 \\
\hline
\multirow{8}{*}{DSS}
& GeoWarp & \textbf{-2.686} & \textbf{-3.307} & \textbf{-1.956} & \textbf{-2.536} & -0.605 & \textbf{-1.354} & \textbf{-2.215} & \textbf{-3.444} & 3.175 \\
\cdashline{2-11}
& GW-NoWarp & -1.992 & \textbf{-3.307} & \textbf{-2.070} & -1.137 & 1.736 & -0.940 & -1.137 & \textbf{-3.386} & 6.187 \\
\cdashline{2-11}
& GW-CV & -1.638 & -1.874 & -1.155 & -2.352 & -1.074 & \textbf{-1.358} & -1.710 & -1.754 & -0.154 \\
& GW-NoWarp-CV & -1.778 & -2.151 & -0.994 & -2.332 & \textbf{-1.270} & \textbf{-1.361} & -1.836 & -2.048 & \textbf{-0.579} \\
& GW-Vert-CV & -1.550 & -1.726 & -1.038 & -2.342 & -1.083 & \textbf{-1.420} & -1.655 & -1.496 & -0.096 \\
& GW-WN-CV & -1.745 & -1.876 & -0.661 & -2.298 & \textbf{-1.248} & -1.283 & -1.739 & -1.511 & -0.385 \\
\cdashline{2-11}
& Linear & -1.119 & -1.091 & -0.497 & -1.410 & -0.547 & -0.832 & -1.017 & -1.156 & -0.397 \\
& BCS & -0.934 & -0.276 & 1.168 & -1.433 & -0.524 & 1.723 & -0.658 & -0.791 & 0.334 \\
\hline
\multirow{8}{*}{DSS2}
& GeoWarp & \textbf{-4.381} & \textbf{-4.904} & \textbf{-3.495} & \textbf{-4.358} & -3.157 & \textbf{-2.577} & \textbf{-4.107} & \textbf{-4.988} & -1.342 \\
\cdashline{2-11}
& GW-NoWarp & -3.972 & \textbf{-4.894} & \textbf{-3.557} & -3.554 & -1.864 & -2.352 & -3.496 & \textbf{-4.956} & 0.359 \\
\cdashline{2-11}
& GW-CV & -3.820 & -4.096 & -3.085 & -4.253 & \textbf{-3.402} & \textbf{-2.578} & -3.825 & -4.024 & \textbf{-3.058} \\
& GW-NoWarp-CV & -3.271 & -3.828 & -2.879 & -3.669 & -3.282 & \textbf{-2.561} & -3.460 & -3.785 & -2.950 \\
& GW-Vert-CV & -3.779 & -4.018 & -3.040 & -4.251 & \textbf{-3.409} & \textbf{-2.616} & -3.800 & -3.909 & \textbf{-3.025} \\
& GW-WN-CV & -1.743 & -1.876 & -0.670 & -2.298 & -1.249 & -1.298 & -1.740 & -1.511 & -0.382 \\
\cdashline{2-11}
& Linear & -1.116 & -1.089 & -0.494 & -1.408 & -0.547 & -0.837 & -1.016 & -1.154 & -0.394 \\
& BCS & -1.612 & -1.235 & -0.080 & -2.230 & -1.493 & 0.469 & -1.538 & -1.805 & -1.125 \\
\hline \hline
\end{tabular}

%% file: figures/cv-metrics-table-jaksa.tex
\begin{tabular}{r|rrrrr}
\hline \hline
& MSE & CRPS & Int05 & DSS & DSS2 \\
\hline
GeoWarp
& \textbf{0.169}
& \textbf{0.202}
& 2.003
& \textbf{-1.199}
& \textbf{-3.334}
\\
GW-NoWarp
& 0.185
& 0.222
& 2.576
& -1.005
& -3.217
\\
GW-CV
& 0.176
& 0.226
& 2.554
& -0.631
& -3.005
\\
GW-NoWarp-CV
& 0.186
& 0.228
& 2.944
& -0.339
& -2.781
\\
GW-Vert-CV
& 0.194
& 0.238
& 2.552
& -0.612
& -2.993
\\
GW-WN-CV
& 0.189
& 0.232
& 2.596
& -0.635
& -0.632
\\
Linear
& 0.221
& 0.254
& 2.726
& -0.492
& -0.487
\\
Binned
& 0.190
& 0.210
& \textbf{1.803}
& 
& 
\\
BCS
& 0.235
& 0.269
& 2.834
& -0.370
& -0.818
\\
\hline \hline
\end{tabular}